\documentclass[hyper]{JHEP3}  

\usepackage{amsmath,amssymb,amsfonts,amsxtra, mathrsfs, makeidx,graphics,graphicx,amsthm,epstopdf,epsfig}

\newcommand{\todo}[1]{{\bf ?????!!!! #1 ?????!!!!}\marginpar{$\Longleftarrow$}}

\newcommand{\be}{\begin{equation}}
\newcommand{\ee}{\end{equation}}
\newcommand{\beq}{\begin{equation}}
\newcommand{\eeq}{\end{equation}}
\newcommand{\ba}{\begin{array}}
\newcommand{\ea}{\end{array}}
\newcommand{\bea}{\begin{eqnarray}}
\newcommand{\eea}{\end{eqnarray}}
\newcommand{\ben}{\begin{enumerate}}
\newcommand{\een}{\end{enumerate}}
\newcommand{\bean}{\begin{eqnarray*}}
\newcommand{\eean}{\end{eqnarray*}}
\newcommand{\eref}[1]{(\ref{#1})}

\newcommand{\nn}{\nonumber}

\newcommand{\tr}{\mathop{\rm Tr}}

\newcommand{\BC}{\mathbb{C}}
\newcommand{\BR}{\mathbb{R}}
\newcommand{\BZ}{\mathbb{Z}}
\newcommand{\sC}{\mathscr{C}}
\newcommand{\sD}{\mathscr{D}}
\newcommand{\sH}{\mathscr{H}}
\newcommand{\comment}[1]{}

\newcommand{\CT}{{\cal T}}

\newcommand{\CMm}{{\cal M}^{\mathrm{mes}}}
\newcommand{\gm}{ g^{\mathrm{mes}}}

\newcommand{\MSN}{{\cal M^{\mathrm{MSN}}}}

\newcommand{\gMSN}{ g^{\mathrm{MSN}}}

\newcommand{\CN}{{\cal N}}

\newcommand{\CC}{{\cal C}}
\newcommand{\BF}{\mathbb{F}}

\newcommand{\perm}{\mathrm{perm}}

\newcommand{\ud}{\mathrm{d}}

\newcommand{\PL}{\mathrm{PL}}

\newcommand{\f}{{\cal F}^{\flat}}
\newcommand{\gf}{g^{\f}}
\newcommand{\firr}[1]{{}^{{\rm Irr}}\!{\cal F}^{\flat}_{#1}}

\title{Phases of M2-brane Theories}

\author{John Davey, Amihay Hanany, Noppadol Mekareeya and Giuseppe Torri\\
Theoretical Physics Group, The Blackett Laboratory \\
Imperial College London, Prince Consort Road\\ 
London,  SW7 2AZ,  UK \\
Email: {\tt j.davey07, a.hanany, n.mekareeya07, giuseppe.torri08@imperial.ac.uk}}

\abstract{We investigate different toric phases of 2+1 dimensional quiver gauge theories arising from M2-branes probing toric Calabi--Yau 4 folds. 
A brane tiling for each toric phase is presented.  We apply the `forward algorithm' to obtain the toric data of the mesonic moduli space of vacua and exhibit the equivalence between the vacua of different toric phases of a given singularity.  The structures of the Master space, the mesonic moduli space, and the baryonic moduli space are examined in detail.  We compute the Hilbert series and use them to verify the toric dualities between different phases.  The Hilbert series, R-charges, and generators of the mesonic moduli space are matched between toric phases.}

\preprint{Imperial/TP/09/AH/01}

\begin{document}
\section{Introduction}
Supersymmetric Chern-Simons (CS) theories in 2+1 dimensions have recently attracted great interest as theories for multiple M2-branes in various backgrounds.  The excitement was triggered by the independent works of Bagger--Lambert \cite{BL} and Gustavsson \cite{gus}.  A key role was played by 3-algebras which, at first sight, do not have a usual field theory structure.  Later it was understood that the theory can be recast as an ordinary field theory \cite{VanRaamsdonk:2008ft}.  A $U(N) \times U(N)$ CS theory at level $(k, -k)$ with bi-fundamental matter fields was subsequently proposed by Aharony, Bergman, Jafferis and Maldacena (ABJM) \cite{Aharony:2008ug} as a model describing $N$ M2-branes in the $\BC^4/\BZ_k$ orbifold background.  After the proposal of the ABJM theory, a number of generalisations have been explored \cite{Benna:2008zy,Imamura:2008nn,Terashima:2008ba, Jafferis:2008qz, Fuji:2008yj,Hosomichi:2008jb,Kim:2008gn,Aharony:2008gk,Ooguri:2008dk, Imamura:2008dt, Imamura:2008ji, Imamura:2009ur, Lambert:2008et}.  

In particular, the $\CN = 2$ CS theory with a general quiver structure is studied in \cite{Martelli:2008si, Ueda:2008hx, Hanany:2008cd}. It is shown how D-term conditions and the moduli space are modified compared to the 3+1 dimensional $\CN = 1$ gauge theory with the same quiver diagram. {\bf Brane tilings} \cite{Hanany:2005ve,Franco:2005rj}, \cite{Franco:2005sm, Feng:2005gw, Broomhead:2008an} are convenient tools to establish the relation between 3+1 dimensional gauge theories and their moduli spaces which are Calabi--Yau 3 folds.  As discussed in \cite{Hanany:2008cd, Hanany:2008fj}, we can conveniently use brane tilings\footnote{There have also been studies on brane crystals \cite{Lee:2006hw, Lee:2007kv, Kim:2007ic, Imamura:2008qs}, which are three-dimensional bipartite graphs, to establish the relation between 2+1 dimensional gauge theories and their moduli spaces which are Calabi--Yau 4 folds.  However, in this paper, we focus only on brane tilings.} (with a few modifications from the 3+1 dimensional case) to study 2+1 dimensional CS theories as well.  In this paper, we refer to each gauge theory by its brane tiling.

An interesting aspect of 2+1 dimensional CS theories on which we focus in this paper is {\bf toric duality}.  It corresponds to a situation in which one singular Calabi--Yau variety has more than one quiver gauge theory (which we refer to as a \emph{(toric) phase} or a \emph{model}) that has this manifold as its mesonic moduli space of vacua.  Toric dualities have been studied in detail in the setup of D3-branes at singularities \cite{Feng:2000mi,Feng:2001xr,Feng:2002zw, Feng:2002fv, Feng:2001bn, Franco:2003ea, Franco:2003ja, Franco:2002mu, Feng:2002kk, Forcella:2008ng}.  Recently, there has been progress along this line in the case of M2-branes, \emph{e.g.} connections between models have been mentioned in \cite{Hanany:2008fj, Franco:2008um} and a number of models have been classified and systematically studied in \cite{taxonomy}. 

It should be emphasised that all models we study are brane tilings but \emph{not} the general class of quiver gauge theories, since every brane tiling gives rise to a quiver but not every quiver gives rise to a brane tiling. All known M2-brane theories so far are brane tiling models.

In this paper, we study supersymmetric CS theories arising from M2-branes probing various toric Calabi--Yau 4 folds.  For each Calabi--Yau variety, we discuss different toric phases and represent each of them by a brane tiling.  We then apply the `forward algorithm' \cite{Feng:2000mi} to obtain the toric data of the mesonic moduli space and exhibit the equivalence between the vacua of different toric phases.  The global symmetry of each model can be found using its toric data (charge matrices and the toric diagram).  The global symmetries of any two toric phases are thus expected to be the same.  We subsequently construct the Hilbert series of the Master space and the mesonic moduli space from which the R-charges and generators of the mesonic moduli space can be determined.  The mesonic Hilbert series, R-charges and generators are matched between toric phases.

Before discussing the models in detail, we summarise some useful results on the 2+1 dimensional CS theory in Section \ref{summary}.

\paragraph{Note added:}  During the completion of this work, we became aware of two relevant papers: One by Amariti, Forcella, Girardello and Mariotti \cite{Davide}, and one by Franco, Klebanov and Rodriguez-Gomez \cite{Seba}.

\section{A Summary of the 2+1 Dimensional Supersymmetric Chern--Simons Theory} \label{summary}
This paper deals with the study of 2+1 dimensional quiver Chern--Simons (CS) theories with $\CN=2$ supersymmetry (four supercharges).   
The theories consist of a product of gauge groups. There are no kinetic terms for the gauge fields but instead there are CS terms.
The matter fields consist of bi-fundamental and adjoint matter.  
Let the quiver CS theory have gauge group with $G$ factors, and a total of $E$ fields, we then have the gauge group $\prod_{a=1}^G U(N_a)$ and the Lagrangian, written in $\CN=2$ superspace notation:
\begin{equation} \label{lagrange}
\mathcal{L}= -\int d^4 \theta\left( \sum\limits_{X_{ab}} X_{ab}^\dagger e^{-V_a} X_{ab} e^{V_b}
-i \sum\limits_{a=1}^G k_a \int\limits_0^1 dt V_a \bar{\mathcal{D}}^{\alpha}(e^{t V_a} \mathcal{D}_{\alpha} e^{-tV_a})
\right) + 
\int d^2 \theta W(X_{ab}) + \mathrm{c.c.}
\end{equation}
where $a$ indexes the factors in the gauge group, $X_{ab}$ are the superfields accordingly charged, $V_a$ are the vector multiplets, $\mathcal{D}$ is the superspace derivative, $W$ is the superpotential and $k_a$ are the CS levels which are integers; an overall trace is implicit since all the fields are matrix-valued. 

The first and third terms in \eref{lagrange} are respectively usual matter and superpotential terms.  
It is useful to write the second term, which corresponds to the CS terms, explicitly in component notation.
The 2+1 dimensional $\CN=2$ vector multiplet $V_a$ consists of a gauge field $A_a$, a scalar field $\sigma_a$, a two-component Dirac spinor $\chi_a$, and an auxiliary scalar field $D_a$, all transforming in the adjoint representation of the gauge group $U(N_a)$.  
This can be viewed as a dimensional reduction of the 3+1 dimensional ${\CN}=1$ vector multiplet. 
In particular, $\sigma_a$ arise from the zero modes of the components of the vector fields in the direction along which we reduce.
In component notation, the CS terms, in Wess--Zumino (WZ) gauge, are given by
\bea \label{csterms}
S_{\mathrm{CS}}\, = \, \sum_{a=1}^G \frac{k_a}{4\pi}\int  \mathrm{Tr} \,\left( A_a \wedge \ud A_a + \frac{2}{3} A_a\wedge A_a\wedge A_a - \bar\chi_a \chi_a +
2D_a \sigma_a \right)~.
\eea

\paragraph{The vacuum equations.} From \eref{lagrange}, we obtain the following vacuum equations:
\begin{eqnarray}
\nn \partial_{X_{ab}} W &=& 0~, \\
\nn \mu_a(X) := \sum\limits_{b=1}^G X_{ab} X_{ab}^\dagger - 
\sum\limits_{c=1}^G  X_{ca}^\dagger X_{ca} + [X_{aa}, X_{aa}^\dagger] &=&  4k_a\sigma_a~, \\
\label{DF} \sigma_a X_{ab} - X_{ab} \sigma_b &=& 0 \ .
\end{eqnarray}
The first set of \eref{DF} are referred to as the \emph{F-term equations}.  The others are in analogy to the  \emph{D-term equations} of $\CN=1$ gauge theories in 3+1 dimensions, with the last equation being a new addition.  
We refer to the space of all solutions for \eref{DF} as the {\bf mesonic moduli space} and denote it as $\CMm$.   

\paragraph{Parity invariance.} 
The parity operator commutes with the supersymmetry generators. 
Since all terms in \eref{csterms} come from the second term of the supersymmetric Lagrangian \eref{lagrange}, it follows that all terms in \eref{csterms} transform in the same way under parity.  From the first two terms of \eref{csterms}, we see that the gauge fields $A_a$ and the derivative transform in the same way under parity.  Hence, the first two terms in \eref{csterms} (as well as the third term, which is a Dirac bilinear in 2+1 dimensions) are negative under parity.  The fourth term in \eref{csterms} must be negative under parity.  Note that the usual equation of motion of $D_a$ tells us that $D_a$ is bilinear in scalars $X^\dagger_{ab}$ and $X_{ab}$.  Since $X_{ab} \rightarrow X_{ab}$ under parity, $D_a \rightarrow D_a$ under parity. Thus, it follows that $\sigma_a \rightarrow -\sigma_a$ under parity. Since $k_a \rightarrow -k_a$, it follows that the vacuum equations \eref{DF} are invariant under parity.  Below, we shall demonstrate this fact geometrically using the toric diagram.

\paragraph{Connection to M2-branes.} For the rest of the paper, we assume that
\begin{itemize}
\item All gauge groups are $U(N)$ with $N$ having the physical interpretation as the number of M2-branes in the stack on which the gauge theory is living;
\item The superpotential $W$ satisfies the {\bf toric condition} \cite{Feng:2002zw}: Each chiral multiplet appears precisely twice in $W$. Once with a positive sign and once with a negative sign.  Under such assumptions, the moduli space is conjectured to receive no quantum corrections due to supersymmetry and due to conformal invariance in the IR.
\end{itemize}
As a consequence, for $N=1$, the space transverse to the single M2-brane is a toric non-compact Calabi-Yau cone, and it is conjectured to be \emph{the mesonic moduli space} $\CMm$ discussed in the previous paragraph.  Hence, $\CMm$ is a \emph{4 dimensional} toric Calabi--Yau cone.  In which case, we can apply the forward algorithm \cite{Feng:2000mi} which takes gauge theory information (quiver, superpotential and CS levels) as input and gives toric data of the moduli space as output.  We may as well consider the mesonic moduli space for higher $N$ which, as a result of the first assumption, is simply the $N$-th symmetric product of the one for $N=1$ case\footnote{The Hilbert series can be obtained using the \emph{plethystic exponential} \cite{pleth, Hanany:2008qc, master}.}.  However, we note that the former is no longer toric \cite{master} and the forward algorithm is not applied.  In this paper, we focus only on the case of $N=1$. 

\subsection{The Moduli Space of Abelian Theories}  
The gauge group is simply $U(1)^G$ and we henceforth refer to this case as the \emph{abelian case}. 
\paragraph{Conditions on the CS levels.}  
From the second equation of \eref{DF}, since each quiver field has a start and an end and hence appears precisely twice in the sum, once with a positive sign and once with a negative sign, it follows that
\bea
\sum_a k_a \sigma_a = 0~. \label{ks}
\eea 
The third equation of \eref{DF} sets all $\sigma_a$ to a single field, say $\sigma$.  From \eref{ks}, we see that for $\sigma \neq 0$, we must impose the following constraints on the CS levels:
\bea
\left(k_1, \ldots, k_G \right)  \neq 0~, \qquad \quad \sum_{a=1}^{G} k_a = 0~.  \label{k-con}
\eea
Note that if the last equality is not satisfied, then $\sigma$ is identically zero and \eref{DF} reduces to the usual vacuum equations for 3+1 dimensional gauge theories.  In which case, the mesonic moduli space is 3 dimensional.  Thus, \eref{k-con} are indeed necessary conditions for the mesonic moduli space to be 4 dimensional, as we require.
For simplicity, we also take
\bea
 \gcd(\{k_a\}) = 1
\eea
so that we do not have to consider orbifold actions on the moduli space.  However, it is easy to generalise to the case of higher $\gcd(\{k_a\})$, and several explicit examples are given in \cite{Hanany:2008cd, Hanany:2008qc}.

\paragraph{Baryonic charges.} The moduli space $\CMm$ is a symplectic quotient of the space of solutions to the F-terms prescribed by the first equation modulo the gauge conditions prescribed by the D-terms.  Because of the condition that all $k_a$ sum to 0 imposed in \eref{k-con}, there is an overall $U(1)$ (corresponding to the position of the M2-brane) which has to be factored out. Furthermore, there is another $U(1)$ which must be factored out. This is because, from the second equation of \eref{DF}, the presence of CS couplings induces Fayet--Iliopoulos (FI)-like terms on the space of D-terms :
\bea
\zeta_a = 4k_a \sigma~. \label{zetaFI}
\eea
We emphasise that these FI-like terms are not the same as the FI parameters for a 3+1 dimensional theory.  This is because the latter are parameters in the Lagrangian, whereas for the former, $\sigma$ is an auxiliary field - not a parameter. 
From \eref{zetaFI},  we see that the vector $\zeta_a$ aligns along a direction set by the CS integers $k_a$.  
It picks \emph{one} direction out of the $(G-1)$ baryonic directions which are present in the 3+1 dimensional theory\footnote{The reader is reminded from \cite{master} that for a 3+1 dimensional theory, the mesonic moduli space is a Calabi--Yau 3-fold, and there are $(G-1)$ baryonic directions.}. 
This direction becomes \emph{mesonic} in the 2+1 dimensional theory and fibres over the Calabi-Yau 3-fold to give a \emph{mesonic moduli space as a Calabi-Yau 4-fold}. 
The remaining $(G-2)$ directions stay baryonic in the 2+1 dimensional theory.  
Thus, in summary, there are $(G-2)$ baryonic charges \emph{coming from the D-terms}.  
We emphasise a subtle point here: Although there are indeed $G-2$ baryonic directions coming from the D-terms, this does not imply that all possible baryonic directions of the particular Calabi-Yau 4-fold are given by these $G-2$ directions.  It only provides a lower bound. There are \emph{at least} $G-2$ such baryonic directions and a different formulation may give more than this number. Such a situation occurs, for example, in Phase II of the $\CC \times \BC$ theory and Phase II of the $D_3$ theory.  Below, we discuss how to count all baryonic charges using the toric diagram.  

\paragraph{The Master space (for $N=1$).} The \emph{Master space}, $\f$, is defined to be the space of solutions of the F-terms \cite{master}.  It is a toric variety for the abelian case. It is of the dimension $4+(G-2) = G+2 \ $.  The mesonic moduli space can be obtained by imposing D-terms:
\begin{equation}\label{symp}
\CMm = \f // U(1)^{G-2} \ .
\end{equation}
Note that the $G-2$ baryonic charges are in the null space of the matrix
\begin{equation}\label{C}
C =\left(\begin{matrix}
1 & 1 & 1 & \ldots & 1 \\ k_1 & k_2 & k_3 & \ldots & k_G
\end{matrix}\right) \ .
\end{equation}
This can be seen as follows.  If the charge vector $\mathbf{q} = (q_1, \ldots, q_G)$ is in the null space of $C$, then $C \cdot \mathbf{q}^t = 0$, \emph{i.e.} the $G$ charges are subject to 2 relations:
\bea
\sum_{a=1}^G  q_a = 0~, \qquad \quad
\sum_{a=1}^G  k_a q_a = 0~. 
\eea
The first equation, which fixes the total charge to be zero, implies that $\mathbf{q}$ is perpendicular to the vector $(1, \ldots, 1)_{1 \times G}$, and the second equation implies that $\mathbf{q}$ is perpendicular to the direction set by the CS integers $(k_1, \ldots k_G)$ .  Since the vectors $(1, \ldots, 1)_{1 \times G}$ and $(k_1, \ldots k_G)$ are orthogonal due to \eref{k-con}, the independent components of $\mathbf{q}$ are indeed the $G-2$ baryonic charges. 

\subsubsection{Brane Tilings, Perfect Matchings, and Toric Diagrams}
\paragraph{Brane tilings.} The toric condition, which requires that each field appears exactly twice with opposite 
signs, naturally gives rise to a bipartite graph on $T^2$ which is also known as a \emph{brane tiling}. A bipartite graph is a graph consisting of vertices of two colours, say, white and black, and every edge connects two vertices with different colours.  The tiling may also be drawn on the plane $\BR^2$ provided that one keeps in mind the periodicity of the smallest unit (called the {\it fundamental domain} and represented in the red frame in the pictures in subsequent sections).  
Each face of the tiling corresponds to a gauge group and each edge corresponds to a bi-fundamental field.  The superpotential can be obtained easily from the tiling in the way that we shall discuss below.  In this subsection, we use indices $\wp, \varrho, \ldots$ for nodes, $a, b, \ldots$ for faces, and $i, j, \ldots$ for edges. The field $\Phi_i \equiv X_{ab}$ transforms under $U(1)_a$ and $U(1)_b$ gauge groups corresponding to the two faces $a$ and $b$ sharing the edge $i$.  The bipartiteness gives rise to a natural orientation of each edge $i$ corresponding to the field $\Phi_i$.  It is indicated by an arrow crossing the edge from the face $a$ to the face $b$:  In this paper, we adopt the convention that \emph{the arrow `circulates' clockwise around the white node and counterclockwise around the black nodes}.  We can therefore uniquely assign the $U(1)_a$ charge $d_{ai}$ to the edge $i$ corresponding to the field $\Phi_i = X_{ab}$ as follows:  
\bea \label{incidence}
d_{ai} = \left \{ 
\begin{array}{rl}
+1 & \quad \text{for an outgoing arrow from the face $a$}~, \\
-1 & \quad \text{for an incoming arrow to the face $a$}~, \\
0 &  \quad \text{if the edge $i$ is not a side of the face $a$}~.  
\end{array} \right .
\eea
We call the $G \times E$ matrix $d$ an \emph{incidence matrix}.   We also assign integers $n_i$ to the edge $i$ such that the CS level $k_a$ of the gauge group $a$ is given by\footnote{This way of representing $k_a$ is introduced in \cite{Hanany:2008cd} and is also used in \cite{Imamura:2008qs}.}
\bea
k_a = \sum_i d_{ai} n_{i}~. \label{kn}
\eea
Due to bipatiteness of the tiling, we see that the relation $\sum_{a} k_a = 0$ is satisfied as required.   The superpotential can be written as
\bea
W = \sum_\wp \mathrm{sign}(\wp) \prod_{j_\wp} \Phi_{j_\wp}~,
\eea
where the product is taken over the edges $j_\wp$ around the node $\wp$, and $\mathrm{sign}(\wp)$ is +1 if $\wp$ is a white node\footnote{The reader should note the similarity between white nodes and British roundabouts. They both have a positive effect and you go round them clockwise.} and $-1$ if $\wp$ is a black node.

\paragraph{Brane realisation.}  As discussed in detail in \cite{Imamura:2008qs}, a brane tiling for the 2+1 dimensional CS theory can be regarded a D4-NS5 system in Type IIA theory on $\BR^{1,7}\times T^2$. The NS5-brane fills an $\BR^{1,3}$ subspace of $\BR^{1,7}$ and is on a complex curve on $\BR^2\times T^2$ such that the NS5-brane forms a collection of tiles that wrap the $T^2$, with the NS brane forming the edges of the tiles. In the remaining two coordinates on $\BR^{1,7}$ the brane system sits in a fixed position. The D4-branes span an $\BR^{1,2}$ subspace of $\BR^{1,3}$ and are wrapping the tiles in the $T^2$ directions, having boundaries that end on the NS5-brane. The gauge groups are realised on the D4-branes, giving rise to a $U(N)$ gauge group per $N$ D4-branes that span the tile. The edges are separating two tiles and open strings stretched between them give rise to chiral multiplets in bi-fundamental representations. Let $A$ be the gauge field on the D4-brane, and let $\phi$ be the 0-form gauge field on the NS5-brane.  This 0-form gauge field couples to the field strength $\ud A$ on the boundary of a D4-brane via the usual WZ coupling $\phi~\ud A \wedge \ud A$.  Integrating by parts, we may write down the boundary term in the D4-brane action as
\begin{equation}
S_{\text{boundary}}=\frac{1}{2\pi}\int_{\partial{\rm D4}} A\wedge \ud A\wedge \ud \phi~.
\end{equation}
This induces the CS coupling which is given by
\begin{equation}
k_a=\oint  \ud \phi~, \label{kdphi}
\end{equation}
where the integration is taken over the boundary of the face $a$ (\emph{i.e.,} along the boundary of the corresponding D4-brane). 
The one-form field strength $\ud \phi$ along the edge $i$ can be identified with the integer $n_i$. Being a field strength it is quantized and therefore $k_a$ are integers. Linear combinations of the edge contributions $n_i$ are integers and we therefore expect that each edge of the tiling gives an integer contribution with the orientation determining the sign.  Thus, \eref{kdphi} is indeed equivalent to the relation \eref{kn}.

\paragraph{Kasteleyn matrices.} Many important properties of the tiling are governed by the \emph{Kasteleyn matrix} $K(x, y, z)$, which is a weighted, signed adjacency matrix of the graph with (in our conventions) the rows indexed by the black nodes, and the columns indexed by the white nodes.  The entry $K_{\wp \varrho}$ of the Kasteleyn matrix is zero if there is no connection between the black node $\wp$ and the white node $\varrho$.  Otherwise, $K_{\wp \varrho}$ can be written as
\bea
K_{\wp \varrho} (x,y,z) =  \sum_{ \{ j_{\wp \varrho} \} } \Phi_{j_{\wp \varrho}} z^{n_{j_{\wp \varrho}}} w_{j_{\wp \varrho}}(x,y)~, \label{Kasteleyn}
\eea
where $j_{\wp \varrho}$ represent an edge connecting the black node $\wp$ to the white node $\varrho$, $\Phi_{j_{\wp \varrho}}$ is the field associated with this edge, and $w_{j_{\wp \varrho}}(x,y)$ is $x$ or $y$ (or $x^{-1}$ or $y^{-1}$, depending on the orientation of the edge) if the edge ${j_{\wp \varrho}}$ crosses the fundamental domain \cite{Hanany:2005ve,Franco:2005rj} and $w_{j_{\wp \varrho}}(x,y) = 1$ if it does not.  A number of examples are given in subsequent sections.  

\paragraph{Perfect matchings.} A \emph{perfect matching} is a subset of edges in the tiling, or equivalently a subset of elementary fields, that covers each node exactly once. As discussed in \cite{master}, the coherent component of the Master space of a toric quiver theory is generated by perfect matchings of the associated tiling.  We can obtain the perfect matchings from the Kasteleyn matrix $K(x,y,z)$ as follows: \emph{The quiver fields in the $\alpha$-th term of the permanent\footnote{The permanent is similar to the determinant: the signatures of the permutations are not taken into account and all terms come with a $+$ sign. One can also use the determinant but then certain signs must be introduced \cite{Hanany:2005ve,Franco:2005rj}.} of the Kasteleyn matrix are the elements of the $\alpha$-th perfect matching $p_\alpha$, i.e.}  
\bea \label{permk}
\mathrm{perm}~K = \sum_{\alpha=1}^c p_\alpha ~ x^{u_\alpha} y^{v_\alpha} z^{w_\alpha}~.
\eea
The coordinates $(u_\alpha, v_\alpha, w_\alpha)$, with $\alpha=1, \ldots, c$, are points in a 3d \emph{toric diagram} of the 2+1 dimensional theory.  From \eref{Kasteleyn}, we see that $w_\alpha$ is a linear combination of the integers $n_i$.  Indeed, if we set $z=1$, we then recover the 2d Newton polygon which gives a 2d toric diagram of the 3+1 dimensional theory. Note that there is also another way of constructing the toric diagram; this will be mentioned in a paragraph below.  In Appendix \ref{permkastel}, we prove that the permanent of the Kasteleyn matrix indeed gives rise to coordinates of the points in the toric diagram.

\paragraph{The perfect matching matrix.} We collect the correspondence between the perfect matchings and the quiver fields in an $E \times c$ matrix (where $E$ is the number of quiver fields and $c$ is the number of perfect matchings) called the \emph{perfect matching matrix} $P$.  If $E=c$ (\emph{i.e.} the fundamental domain contains precisely one pair of black and white nodes), we can relabel $p_\alpha$ so that $P$ becomes an identity matrix.  On the other hand, if $E \neq c$, then the null space of the matrix $P$ is non-trivial, and there exists a $(c-G-2) \times c$ matrix $Q_F$ whose rows are basis vectors (which are taken to be orthogonal) of the nullspace of $P$:
\bea
Q_F = \ker(P)~. \label{qfkerp}
\eea
Therefore, by construction, we find the relation
\bea
P \cdot Q_F^t =0~. \label{relpm}
\eea 
This matrix equation gives the relations between the perfect matchings $p_\alpha$.   Hence, the coherent component $\firr{}$ of the Master space can be viewed as the space $\BC^c$ generated by the perfect matchings modded out by the relations encoded in $Q_F$:
\bea
\firr{} = \BC^c//Q_F~. \label{sympq}
\eea
Hence, the matrix $Q_F$ can be regarded as the \emph{charge matrix associated with the F-terms}. The coherent component $\firr{}$ is $c - (c-G-2) = G+2$ dimensional, as expected.  Note that the sum of entries in each row of $Q_F$ vanishes.  This is equivalent to saying that $(1, 1, \ldots, 1)_{1 \times G}$ is in the null space of $Q_F$, or in other words, is spanned by the row vectors of $P^t$ (see \eref{relpm}).  It can be seen that the sum of all rows of $P^t$ is proportional to $(1, \ldots, 1)_{1\times G}$, and hence the statement in the previous sentence follows.


\paragraph{Baryonic charges of perfect matchings.}  Let us determine the baryonic charges of the perfect matchings.  In order to do so, we remind the reader of the definition \eref{incidence} of the incidence matrix $d$, which maps the fields into their quiver charges.  Furthermore, we recall the definition of the perfect matching matrix $P$, which maps the perfect matchings to the fields.  Let  $\widetilde{Q}$ be a $G \times c$ matrix which maps the perfect matchings into their quiver charges.  Then,
\bea
d_{G \times E} = \widetilde{Q}_{G \times c} \cdot (P^t)_{c \times E}~, \label{qtilde}
\eea
where the subscripts denote the sizes of matrices.  Recall that the $G-2$ baryonic charges are in the null space of $C$ given by \eref{C}.  We can define a $(G-2) \times G$ matrix $\ker(C)$ whose rows are orthogonal basis vectors of the null space of $C$.  This matrix projects the space of quiver charges onto the null space of $C$.  Hence, \emph{the baryonic charges of the perfect matching} are given by the $(G-2) \times c$ matrix:
\bea
(Q_D)_{(G-2) \times c} = \ker{(C)}_{(G-2) \times G} \cdot \widetilde{Q}_{G \times c}~. \label{QD}
\eea 
In analogy to $Q_F$, the mesonic moduli space can be written as
\bea
\CMm = \firr{} // Q_D = \left( \BC^c//Q_F \right) // Q_D~. \label{quoteFD}
\eea 
The matrix $Q_D$ can be regarded as the \emph{charge matrix associated with the D-terms}.  Note that the sum of entries in each row of $Q_D$ vanishes, since $(1, 1, \ldots, 1)_{1 \times G}$ is in the null space of $\ker(C)$ as discussed in the comment below \eref{C}.  If the number of perfect matchings $c$ is equal to the number of quiver fields $E$ (\emph{i.e.} there is precisely one pair of black and white nodes in the fundamental domain), then $P$ can be arranged to be the identity matrix and hence
\bea
(Q_D)_{(G-2) \times c}  = \ker{(C)}_{(G-2) \times G} \cdot d_{G \times E} \qquad \quad \text{(for $c=E$)} ~. \
\eea

\paragraph{The toric diagram.}  There are 2 methods of constructing the toric diagram:
\begin{itemize}
\item The first method was mentioned in the preceding paragraph. In particular, the coordinates $(u_\alpha, v_\alpha, w_\alpha)$ of the $\alpha$-th point in the toric diagram are respectively given by the power of $x, y, z$ in \eref{permk}.  
\item The second method is to make use of the charge matrices $Q_F$ and $Q_D$ via \eref{quoteFD}.  We construct a $(c-4) \times c$ matrix $Q_t$ as follows:
\bea \label{Qtdef}
(Q_t)_{(c-4) \times c} =
\left( \begin{array}{c}
(Q_D)_{(G-2)\times c} \\
(Q_F)_{(c-G-2) \times c} 
\end{array} \right)~.
\eea
Then, let us define a $4 \times c$ matrix 
\bea
G_t = \ker(Q_t) \label{Gt}
\eea
whose rows are basis vectors of the null space of $Q_t$. The matrix $G_t$ projects the space of perfect matchings onto the null space of $Q_t$.  Note that columns of length $4$ of $G_t$ signify a 4-fold.
Since $(1, \ldots, 1)_{1 \times G}$ lives in both the null spaces of $Q_F$ and $Q_D$, it follows that we can always pick a row of $G_t$ to be $(1, \ldots, 1)_{1 \times G}$.  This implies that the end points of these $c$ 4-vectors  lie in a 3 dimensional hyperplane.  Therefore, we may remove the first row of $G_t$ and obtain a $3 \times c$ matrix $G'_t$. The columns of $G'_t$ give the coordinates of points in the toric diagram, which represent the toric 4-fold by an integer polytope in 3 dimensions.
\end{itemize}
We emphasise that the 3d toric diagram is defined up to a $GL(3,\BZ)$ transformation. Below we demonstrate for every toric phase that  two methods indeed give the same toric diagram up to such a transformation.

\paragraph{The mesonic symmetries.} 
 In the context of the AdS/CFT correspondence, $\CN=2$ superconformal gauge theories in 2+1 dimensions are dual to M-theory on $\mathrm{AdS}_4 \times \mathrm{SE}^7$ (where $\mathrm{SE}^7$ denotes a Sasaki--Einstein 7-manifold).
There are 4 global $U(1)$ symmetries which come from the metric and are isometries of the Sasaki--Einstein 7-manifold. 
The toric condition implies that the isometry group is $U(1)^4$ or an enhancement of $U(1)^4$ to a non-abelian group.    
This isometry group is called the {\bf mesonic symmetry} and can be determined by the $Q_t$ matrix.  
In particular, the existence of a non-abelian $SU(k)$ factor (with $k>1$) in the mesonic symmetry is implied by the number $k$ of repetitions of columns in the $Q_t$ matrix. 
Since the mesonic symmetry has a total rank 4, we can classify all possible mesonic symmetries according to the partitions of 4 as follows:
\begin{itemize}
\item $SU(4) \times U(1)$~,
\item $SU(3) \times SU(2) \times U(1)$~,
\item $SU(3) \times U(1) \times U(1)$~,
\item $SU(2) \times SU(2) \times SU(2) \times U(1)$~,
\item $SU(2) \times SU(2) \times U(1) \times U(1)$~,
\item $SU(2) \times U(1) \times U(1) \times U(1)$~,
\item $U(1) \times U(1) \times U(1) \times U(1)$~.
\end{itemize}
If it turns out that there is precisely one $U(1)$ factor in the mesonic symmetry, we can immediately identify it with the R-charge.  Otherwise, there is a minimisation problem to be solved in order to determine which linear combination of these $U(1)$ charges gives the right R-charge in the IR \cite{Hanany:2008fj}.  In some simple cases, we can bypass this calculation using a symmetry argument. 

 \paragraph{The baryonic symmetries.}  Each external point in the toric diagram corresponds to a 5-cycle in the Sasaki--Einstein 7-manifold. Not all of these 5-cycles are homologically independent but one can choose a basis of homologically stable 5-cycles inside the Sasaki--Einstein 7-manifold. Every 5-cycle in this basis gives rise to a massless gauge field in $\text{AdS}_4$, coming from Kaluza--Klein reduction of the M-theory 6-form (dual to the 3-form) on the 5-cycle.  These massless gauge fields are dual to the {\bf baryonic} $U(1)$ symmetries in the gauge theory.  The number of such homologically stable 5-cycles, which is thus equal to \emph{the number of baryonic charges} $N(\mathscr{B})$, is equal to the number of external points $N(\mathscr{E})$ in the toric diagram minus 4:
\bea
N(\mathscr{B}) = N(\mathscr{E}) - 4~. \label{numberbary}
\eea
The global symmetry of the theory is a product of mesonic and baryonic symmetries.

\paragraph{Parity invariance of the Calabi--Yau 4-fold.} We mention above that the vacuum equations and the mesonic moduli space are invariant under parity.  This fact can also be seen from the toric diagram perspective as follows. Since under a parity transformation $k_a \rightarrow -k_a$, it follows from \eref{kn} that $n_i \rightarrow -n_i$ (the $d_{ai}$ do not change sign as we are not dealing with charge conjugation).  It follows from the discussion after \eref{permk} that, for each point in the toric diagram, the third coordinate $w_\alpha \rightarrow  -w_\alpha$, whereas the first and second coordinates $u_\alpha, v_\alpha$ remain unchanged.  This is however a $GL(3, \BZ)$ action on the coordinates. We thus arrive at our conclusion.

\paragraph{A summary of the forward algorithm.} We summarise the forward algorithm in the following diagram (as in \cite{taxonomy}):
\begin{equation}\label{forward}
\begin{array}{lllllll}
\fbox{
\mbox{
\begin{tabular}{l}
INPUT 1: \\
~~Quiver
\end{tabular}
}}
& \rightarrow & d_{G \times E}	& \rightarrow	&
	(Q_D)_{(G-2) \times c} = 
	 \ker{(C)}_{(G-2) \times G} \cdot \widetilde{Q}_{G \times c}\quad (d_{G \times E} = \widetilde{Q}_{G \times c} \cdot (P^t)_{c \times E}) \\
&&\\[-0.3cm]
\vspace{-0.5cm}&&& \nearrow &&&\\
\fbox{
\mbox{
\begin{tabular}{l}
INPUT 2: \\
~~CS Levels
\end{tabular}
}}
& \rightarrow & C_{2 \times G}	&&&&\\[-0.3cm]
&&& \nearrow &&&\\
\fbox{\mbox{
\begin{tabular}{l}
INPUT 3: \\
~~Superpotential
\end{tabular}
}}
& \rightarrow & P_{E \times c}	& \rightarrow
		& (Q_F)_{(c-G-2)\times c} = \ker (P) \\
&&&&~~~~~~\downarrow\\
&&&& 
\hspace{-1in}
(Q_t)_{(c-4) \times c} =
\left( \begin{array}{c}
(Q_D)_{(G-2)\times c} \\
(Q_F)_{(c-G-2) \times c} 
\end{array} \right) \rightarrow
\fbox{\mbox{
\begin{tabular}{l}
OUTPUT:  \\
~~$(G_t)_{4 \times c} = {\ker}(Q_t) $\\
\end{tabular}
}}
\end{array}
\end{equation}

\paragraph{Notation and nomenclature.}  We denote the $i$-th bi-fundamental field transforming in the fundamental (antifundamental) representation of the gauge group $a$ (gauge group $b$) by $X^{i}_{ab}$ and similarly $\phi^i_a$ denotes the $i$-th adjoint field in the gauge group $a$ (when there is only a single arrow the $i$-index is dropped).  We refer to gauge theories in subsequent sections by their mesonic moduli space (\emph{e.g.,} the $\BC^4$ theory), and in each subsection we name toric phases according to the features of their tilings (\emph{e.g.,} Phase I of the $\BC^4$ theory is called the `chessboard model' as its tiling is similar to the chessboard).  We use the shorthand notation listed in Table \ref{t:nomen} for our nomenclature, \emph{e.g.} the two double-bonded one-hexagon model is denoted by $\mathscr{D}_2 \mathscr{H}_1$.
\begin{table}[h]
 \begin{center} 
   {\small
  \begin{tabular}{|c||c|}
  \hline
  Shorthand notation & Object referred to \\
  \hline
  $\mathscr{C}$ & chessboard \\
  $\mathscr{D}_n$ & $n$ double bonds \\   
  $\mathscr{H}_n$ & $n$ hexagons  \\
  $\mathscr{S}_n$ & $n$ squares  \\   
  $\partial_n$ & $n$ diagonals  \\   
  $\mathscr{O}_n$ & $n$ octogons  \\   
  \hline
  \end{tabular}}
  \end{center}
\caption{Shorthand notation for the nomenclature of the brane tilings used in paper.}
\label{t:nomen}
\end{table}

\section{Phases of the $\BC^4$ Theory}
It was shown in \cite{Hanany:2008fj,  Franco:2008um, taxonomy} that there are two different gauge theories which have $\BC^4$ as a mesonic moduli space. We recall the first, more discussed one and then move over to the much less discussed theory. Since the two gauge theories have the same moduli space, we may test their correspondence on a deeper level - compare gauge invariant operators, compare scaling dimensions, and check if they are indeed dual to each other. This is the subject of the following subsections.

\subsection{Phase I: The Chessboard Model (The ABJM Theory)} \label{sec:ABJM}
\begin{figure}[ht]
\begin{center}
  \vskip 1cm
  \hskip -6cm
  \includegraphics[totalheight=1.5cm]{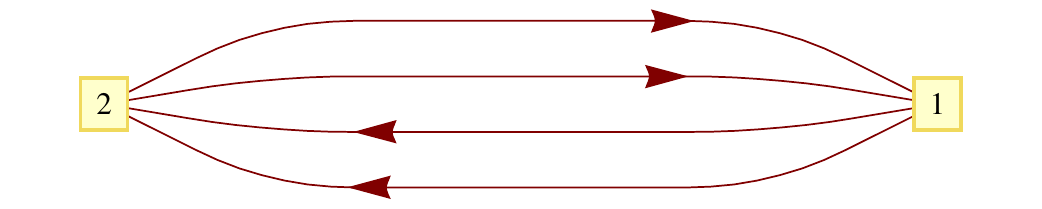}
  \vskip -3.0cm
  \hskip 8cm
  \includegraphics[totalheight=5.5cm]{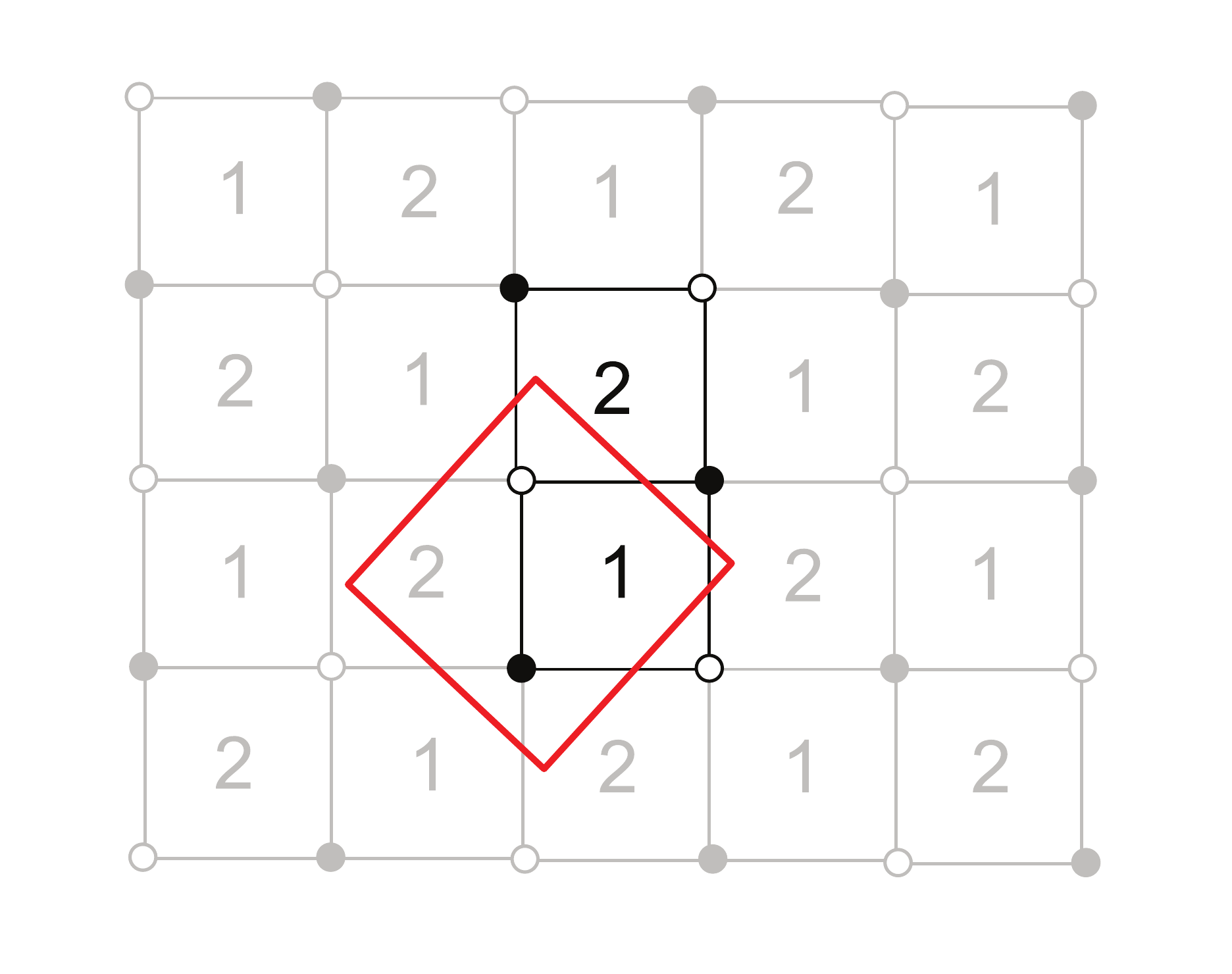}
 \caption{[Phase I of $\BC^4$] (i) Quiver diagram for the $\mathscr{C}$ model. \ (ii) Tiling for the $\mathscr{C}$ model.}
  \label{f:con}
\end{center}
\end{figure}

The chessboard model (which we shall refer to as $\mathscr{C}$) contains two gauge groups $U(N)_1 \times U(N)_2$  and bi-fundamental fields $X_{12}^i$ and $X_{21}^i$ (with $i=1,2$).  The superpotential is given by 
\bea
W = \tr(X^1_{12} X^1_{21} X^2_{12} X^2_{21} - X^1_{12} X^2_{21} X^2_{12} X^1_{21})~. 
\eea
According to \eref{k-con}, we take the Chern--Simons levels to be $k_1=-k_2=1$. 
The quiver diagram and tiling are drawn in Figure \ref{f:con}.  In 3+1 dimensions, the chessboard tiling actually gives rise to the conifold theory (which we shall refer to as $\CC$); however, for the 2+1 dimensional theory, there is an additional structure, namely each edge in the tiling bears an integer $n_i$ according to \eref{kn}.  In the following paragraph, we see that the mesonic moduli space of the 2+1 dimensional chessboard model indeed differs from the mesonic moduli space of the 3+1 dimensional conifold theory but still coincides with its master space.
\begin{figure}[ht]
\begin{center}
   \includegraphics[totalheight=7cm]{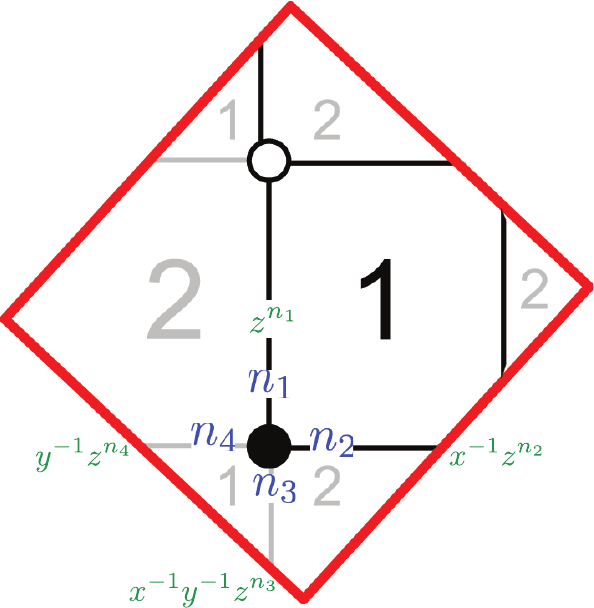}
 \caption{[Phase I of $\BC^4$] The fundamental domain of the tiling for the $\mathscr{C}$ model: Assignments of the integers $n_i$ to the edges are shown in blue and the weights for these edges are shown in green.}
  \label{f:fdphase1c4}
\end{center}
\end{figure}

\paragraph{The toric diagram.}  We demonstrate two methods of constructing the toric diagram. 
\begin{itemize}
\item {\bf The Kasteleyn matrix.} We assign the integers $n_i$ to the edges according to Figure \ref{f:fdphase1c4}.  From \eref{kn}, we find that 
\bea
\text{Gauge group 1~:} \qquad k_1 &=& 1 = n_1 - n_2 + n_3 - n_4~, \nn \\
\text{Gauge group 2~:} \qquad k_2 &=& -1 = -n_1 + n_2 - n_3 + n_4 ~. 
\eea  
We choose
\bea
n_3= 1,\quad n_1=n_2=n_4=0~.
\eea
We can now determine the Kasteleyn matrix for this phase of the theory. Since the fundamental domain contains only one white node and one black node, the Kasteleyn matrix is $1\times 1$ and, therefore, coincides with its permanent:
\bea \label{Kabjm}
K &=& X^1_{12} z^{n_1} + X^1_{21} x^{-1} z^{n_2} + X^2_{12} x^{-1} y^{-1} z^{n_3} + X^2_{21} y^{-1} z^{n_4} \nn \\
&=& X^1_{12} + X^1_{21} x^{-1}  +  X_{12}^2 x^{-1} y^{-1} z +  X_{21}^2  y^{-1}  \qquad \ \text{(for $n_3= 1,~ n_1=n_2=n_4=0$)}~. \nn \\
\eea
The powers of $x, y, z$ in each term of \eref{Kabjm} give the coordinates of each point in the toric diagram. 
We collect these points in the columns of the following $G_K$ matrix:
\bea
G_K = \left(
\begin{array}{cccc}
 -1 & 0 & -1 & 0 \\
 0 & -1 & -1 & 0 \\
 0 & 0 & 1 & 0
\end{array}
\right)~.
\eea

\item {\bf The charge matrices.} 
From \eref{Kabjm}, the perfect matchings can therefore be taken as
\bea
p_1 = X^1_{12},\quad p_2 = X_{12}^2, \quad p_3 = X_{21}^2, \quad p_4 = X^1_{21}~.
\eea
Since there is a one-to-one correspondence between the perfect matchings and the quiver fields, $Q_F =0$.  Since the number of gauge groups is $G=2$, there is $G-2=0$ baryonic charge from the D-terms and hence $Q_D=0$.  Thus, we have $Q_t = 0$.    
From \eref{Gt}, we find that {\footnotesize $G_t = \left( \begin{array}{cccc} 1&1&1&1\\1&0&0&0\\0&1&0&0\\0&0&1&0 \end{array} \right)$}.
After removing the first row, the columns give the coordinates of points in the toric diagram:
\bea
G'_t = \left( \begin{array}{cccc} 1&0&0&0\\0&1&0&0\\0&0&1&0 \end{array} \right)~. \label{gtabjm}
\eea
We see that the toric diagram is merely 4 corners of a tetrahedron (Figure \ref{f:torphase1c4}).   This is in fact the toric diagram of $\BC^4$ \cite{Hanany:2008fj, taxonomy}.
\begin{figure}[ht]
\begin{center}
  \includegraphics[totalheight=3.0cm]{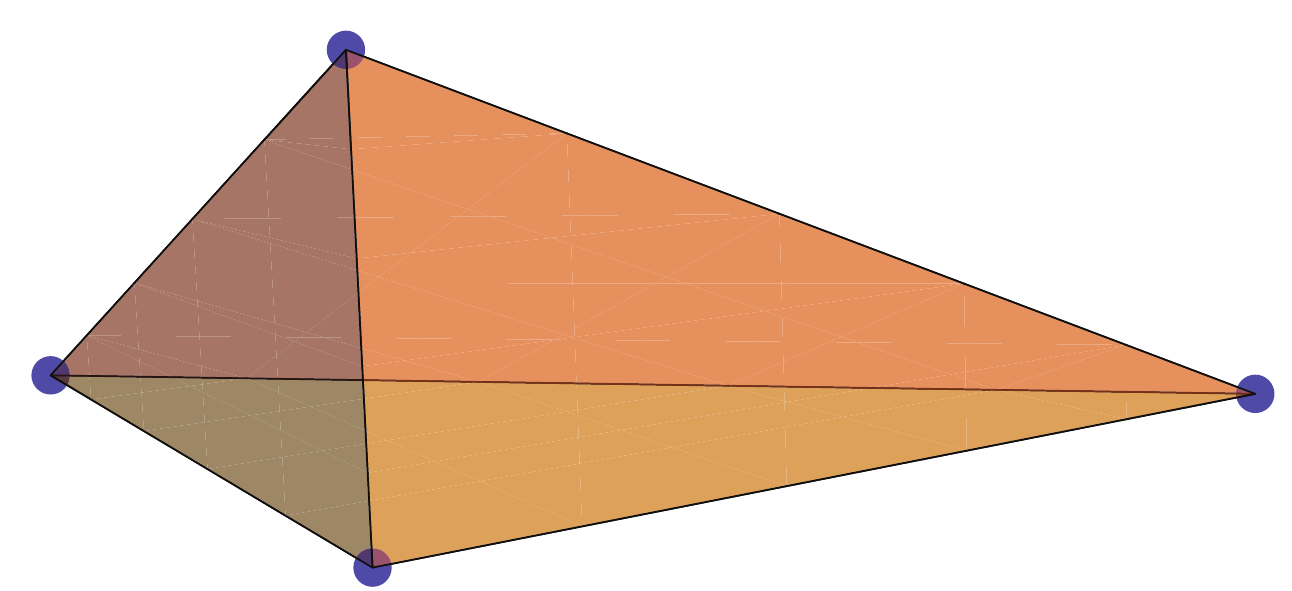}
 \caption{The toric diagram of the $\BC^4$ theory.}
  \label{f:torphase1c4}
\end{center}
\end{figure}
\end{itemize}
\noindent Note that the toric diagrams constructed from $G_K$ and $G'_t$ are the same up to a transformation 
{\footnotesize $\CT = \left( \begin{array}{ccc} -1&0&-1\\0&-1&-1\\0&0&1 \end{array} \right) \in GL(3, \BZ)$},
where we have $G_K = \CT \cdot G'_t$.

\begin{figure}[ht]
\begin{center}
  \includegraphics[totalheight=3.0cm]{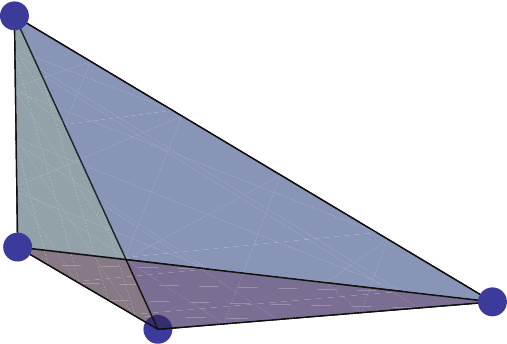}
 \caption{The lattice of generators of the $\BC^4$ theory.}
  \label{f:latc4}
\end{center}
\end{figure}

\paragraph{The moduli space.} For the abelian case, the fields are simply complex numbers and so the superpotential vanishes.  Therefore, the Master space is $\f_{\sC} = \BC^4$.  From Figure \ref{f:torphase1c4}, there are 4 external points in the toric diagram.  It follows that the number of baryonic charges is $4-4 = 0$, and hence the mesonic moduli space coincides with the Master space: 
\bea
\CMm_{\sC} = \f_{\sC} = \BC^4~.  \label{abjmmod}
\eea
Since all four columns of the $Q_t$ matrix are the same, the mesonic symmetry of this model is $SU(4)\times U(1)$.   Note that this $U(1)$ is not the full R-symmetry, which is actually $Spin(8)$.  However, since it assigns equal weight to all fields, it can be identified with the scaling dimension $1/2$.  
The four fields transform as the fundamental representation of the $SU(4)$.   The Hilbert series is given by 
\bea
\gm_1(t,x_1,x_2,x_3; \sC) = \frac{1}{(1-t x_1)\left(1-\frac{t x_2}{x_1}\right)\left(1- \frac{t x_3}{x_2}\right)\left(1-\frac{t}{x_3}\right)} &=& \sum^{\infty}_{k=0}[k,0,0] t^k\ , \label{n1con} 
\eea
where $t$ is the fugacity counting scaling dimensions and $x_1,x_2$ and $x_3$ are fugacities for the $SU(4)$ weights.
Let us compute the plethystic logarithm of the Hilbert series:
\bea
\PL[\gm_1(t,x_1,x_2,x_3; \sC)] = t\left(x_1 + \frac{x_2}{x_1}+\frac{x_3}{x_2}+\frac{1}{x_3}\right) = [1,0,0]t \ . \label{plabjm}
\eea
\paragraph{The generators.} We can see that the mesonic moduli space is generated by four operators:
\bea
X_{12}^1, \quad X_{12}^2, \quad X_{21}^1, \quad X_{21}^2~.  \nn
\eea  
We can represent these generators in a lattice (Figure \ref{f:latc4}) by plotting the powers of $x_1, x_2, x_3$ of the character in \eref{plabjm}. Note that the lattice of generators is the dual of the toric diagram (nodes are dual to faces and edges are dual to edges).  For the $\BC^4$ theory, the toric diagram is a tetrahedron (4 nodes, 6 edges and 4 faces), which is a self-dual lattice.  Therefore, the lattice of generators is the same as the toric diagram.

\paragraph{The 2+1 dimensional  chessboard model $\sC$ V.S. the 3+1 dimensional conifold theory $\CC$.}  The Master space of the 3+1 dimensional conifold theory (see \cite{master}) coincides with the Master space of the 2+1 dimensional chessboard model (see \eref{abjmmod}):
\bea
 \f_{\sC} = \f_{\CC} = \BC^4~.
\eea
However, the mesonic moduli spaces of these two theories are different.  The space $\CMm_{\CC}$ of the conifold theory is a Calabi-Yau 3-fold whose affine coordinates are given by a hypersurface $\{xy - wz = 0\} \subset \BC^4$.  The Hilbert series is given by
\bea
\gm_1(t; \CC) = \frac{1-t^2}{(1-t)^4} = \frac{1+t}{(1-t)^3}~.
\eea
On the other hand, according to \eref{abjmmod}, the space $\CMm_{\sC}$ of the 2+1 dimensional chessboard theory is simply $\BC^4$.

\subsection{Phase II: The One Double-Bonded One-Hexagon Model}
\begin{figure}[ht]
\begin{center}
  \vskip 1cm
  \hskip -7cm
  \includegraphics[totalheight=2.5cm]{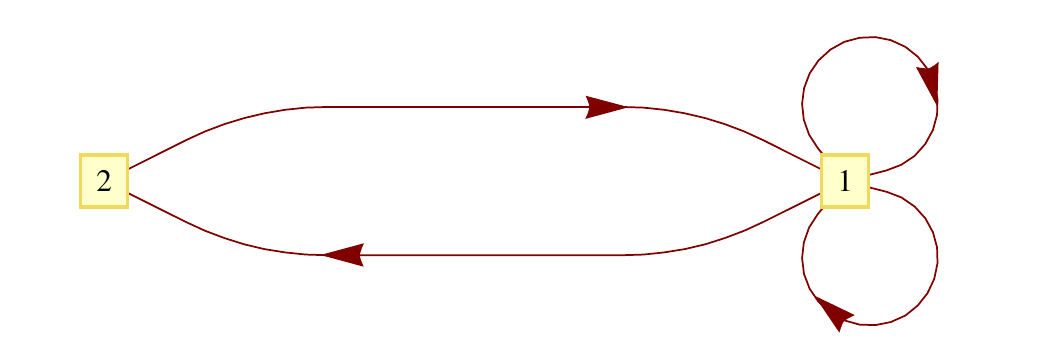}
  \vskip -4.0cm
  \hskip 8.5cm
  \includegraphics[totalheight=5.0cm]{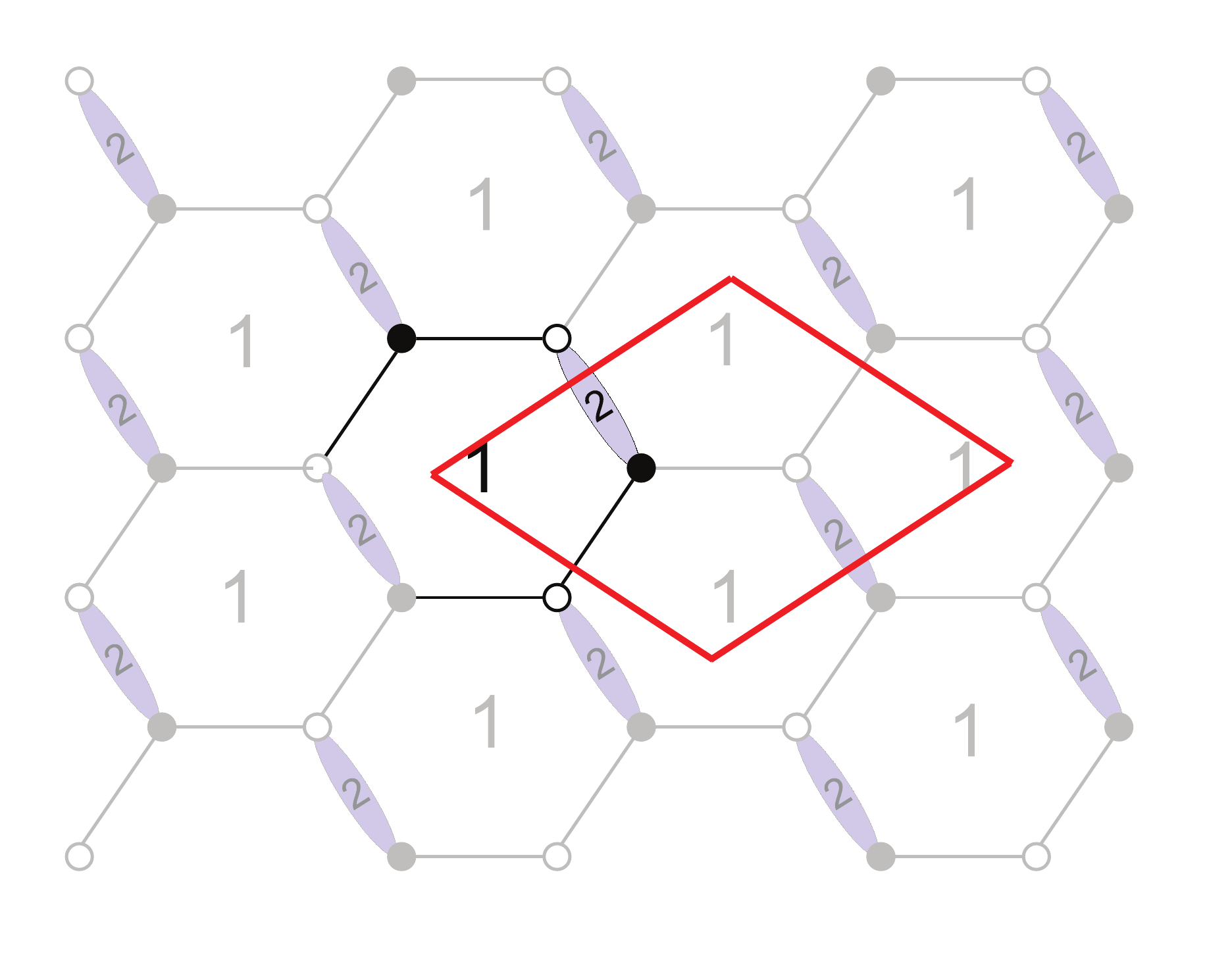}
 \caption{[Phase II of $\BC^4$] (i) Quiver diagram for the $\sD_1\sH_1$ model. \ (ii) Tiling for the $\sD_1\sH_1$ model.}
  \label{f:phase2c4}
\end{center}
\end{figure}

This model (which we shall refer to as $\sD_1\sH_1$) contains two gauge groups $U(N)_1 \times U(N)_2$.  There are 2 bi-fundamental fields $X_{12}$ and $X_{21}$ as well as 2 adjoint fields transforming in one of the two gauge groups.  Without loss of generality, we take this gauge group to be $U(N)_1$ and denote the adjoint fields by $\phi_1^1$ and $\phi_1^2$.   The superpotential is given by
\bea
W = \tr(X_{21}[\phi_1^1,\phi_1^2]X_{12})~.
\eea
According to \eref{k-con}, we take the Chern--Simons levels to be $k_1=-k_2=1$.
The quiver diagram and tiling\footnote{The tiling for this theory was introduced in \cite{Hanany:2008fj}.} are drawn in Figure \ref{f:phase2c4}. 

\begin{figure}[ht]
\begin{center}
   \includegraphics[totalheight=6.0cm]{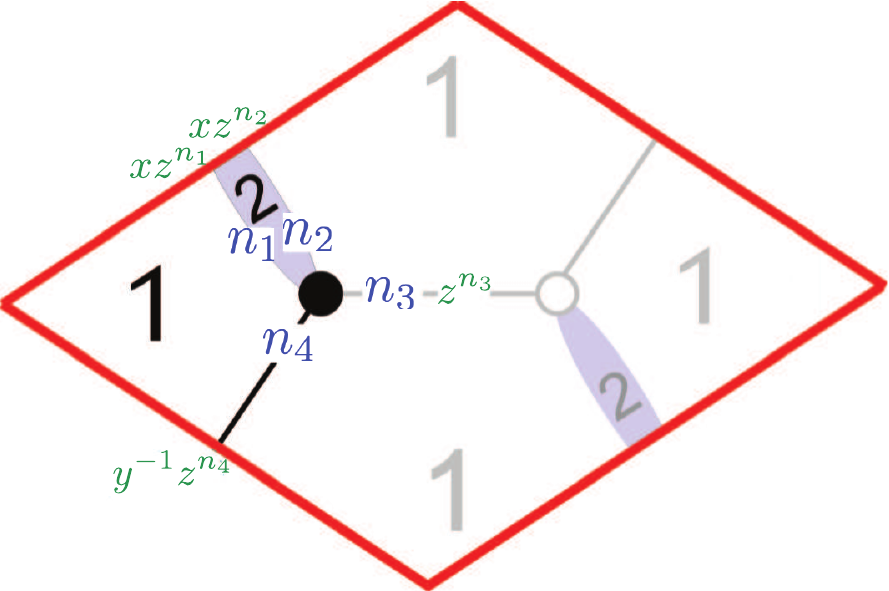}
 \caption{[Phase II of $\BC^4$] The fundamental domain of tiling for the $\sD_1\sH_1$ model : Assignments of the integers $n_i$ to the edges are shown in blue and the weights for these edges are shown in green.}
  \label{f:fdphase2c4}
\end{center}
\end{figure}

\paragraph{The toric diagram.} We demonstrate two methods of constructing the toric diagram. 
\begin{itemize}
\item {\bf The Kasteleyn matrix.} We assign the integers $n_i$ to the edges according to Figure \ref{f:fdphase2c4}.  From \eref{kn}, we find that 
\bea
\text{Gauge group 1~:} \qquad k_1 &=& 1 = -n_1 + n_2 ~, \nn \\
\text{Gauge group 2~:} \qquad k_2 &=& -1 = n_1 - n_2 ~.
\eea  
We choose
\bea
n_2= 1,\quad n_1=n_3=n_4=0~.
\eea
We can now construct the Kasteleyn matrix for this model. Since the fundamental domain contains only one black node and one white node, the Kasteleyn matrix is a $1\times 1$ matrix and, therefore, coincides with its permanent:
\bea \label{permKph2c4}
K &=&  \phi_{1}^1 z^{n_3} + \phi_{1}^2 y^{-1} z^{n_4} + X_{21} x z^{n_1} + X_{12} x z^{n_2} \nn \\
&=& \phi_{1}^1 + \phi_{1}^2 y^{-1} + X_{21} x  + X_{12} x z \qquad \text{(for $n_2=1,~n_1=n_3=n_4=0$)}~. 
\eea
The powers of $x, y, z$ in each term of $K$ give the coordinates of each point in the toric diagram.  
We collect these points in the columns of the following $G_K$ matrix:
\bea
G_K = \left(
\begin{array}{cccc}
 1 & 0 & 1 & 0 \\
 0 & -1 & 0 & 0 \\
 0 & 0 & 1 & 0
\end{array}
\right)~.
\eea

\item {\bf The charge matrices.} 
From \eref{permKph2c4}, the perfect matchings can therefore be taken as
\bea
p_1 = X_{12}, \quad p_2 = \phi_{1}^2, \quad p_3 = X_{21}, \quad p_4 = \phi_{1}^1~.
\eea
Since there is a one-to-one correspondence between the perfect matchings and the fields, $Q_F =0$.  Since the number of gauge groups is $G=2$, there is $G-2=0$ baryonic charge from the D-terms and hence $Q_D=0$.  Thus, we have $Q_t = 0$.  Therefore, we have the same $G'_t$ as in \eref{gtabjm}.  The toric diagram is 4 corners of a tetrahedron as in Figure \ref{f:torphase1c4}.  Thus, we have shown that the toric diagram of phase II is indeed identical to that of phase I.  
\end{itemize}
\noindent Note that the toric diagrams constructed from $G_K$ and $G'_t$ are the same up to a transformation 
{\footnotesize $\CT =\left(
\begin{array}{ccc}
 1 & 0 & 1 \\
 0 & -1 & 0 \\
 0 & 0 & 1
\end{array}
\right) \in GL(3, \BZ)$},
where we have $G_K = \CT \cdot G'_t$.



\paragraph{The moduli space.}  Since all four columns of the $Q_t$ matrix are the same, the mesonic symmetry of this model is $SU(4)\times U(1)$.   Note that this $U(1)$ is not the full R-symmetry, which is actually $Spin(8)$.  However, since it assigns equal weight to all fields, it can be identified with the scaling dimension $1/2$.  The four fields transform as the fundamental representation of the $SU(4)$. 
It follows that 
\bea
\CMm_{\sD_1\sH_1} = \f_{\sD_1\sH_1} =\BC^4~,
\eea
with the Hilbert series given by \eref{n1con}. The plethystic logarithm, of course, coincides with that of the chessboard model and the generators  are therefore
\bea
X_{12},\quad X_{21},\quad \phi^{1}_{2},\quad \phi^{2}_{2} \ .
\eea
Note that there is a one-to-one correspondence between the generators of this model and those of Phase I.


\section{Phases of the $\CC \times \BC$ Theory} 

\subsection{Phase I: The One Double-Bonded Chessboard Model}
\begin{figure}[ht]
\begin{center}
  \hskip -7cm
  \includegraphics[totalheight=5.0cm]{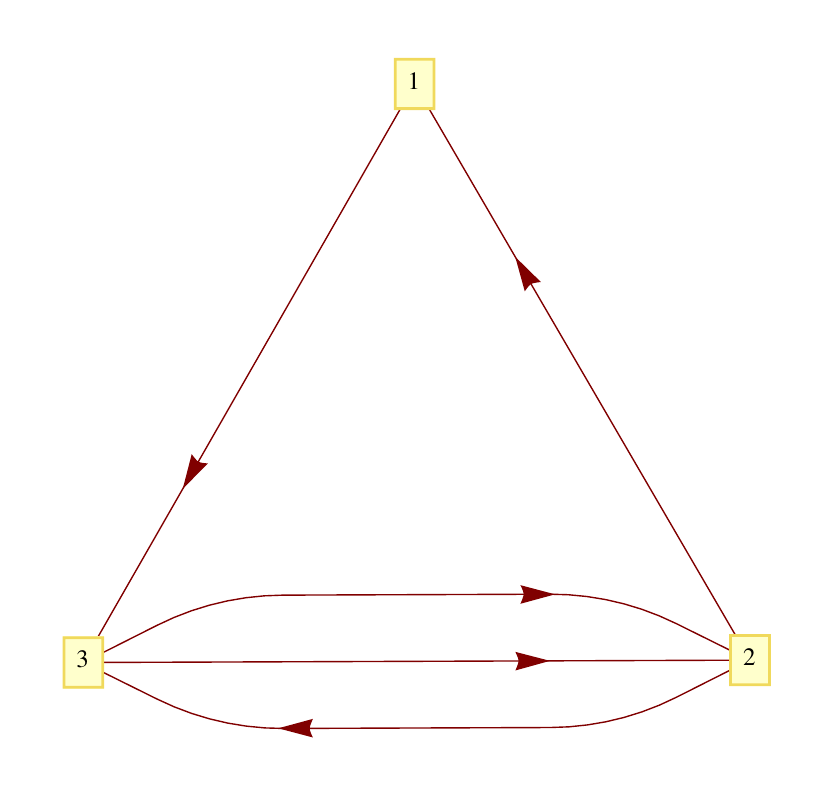}
   \vskip -5.0cm
  \hskip 8.0cm
  \includegraphics[totalheight=5.0cm]{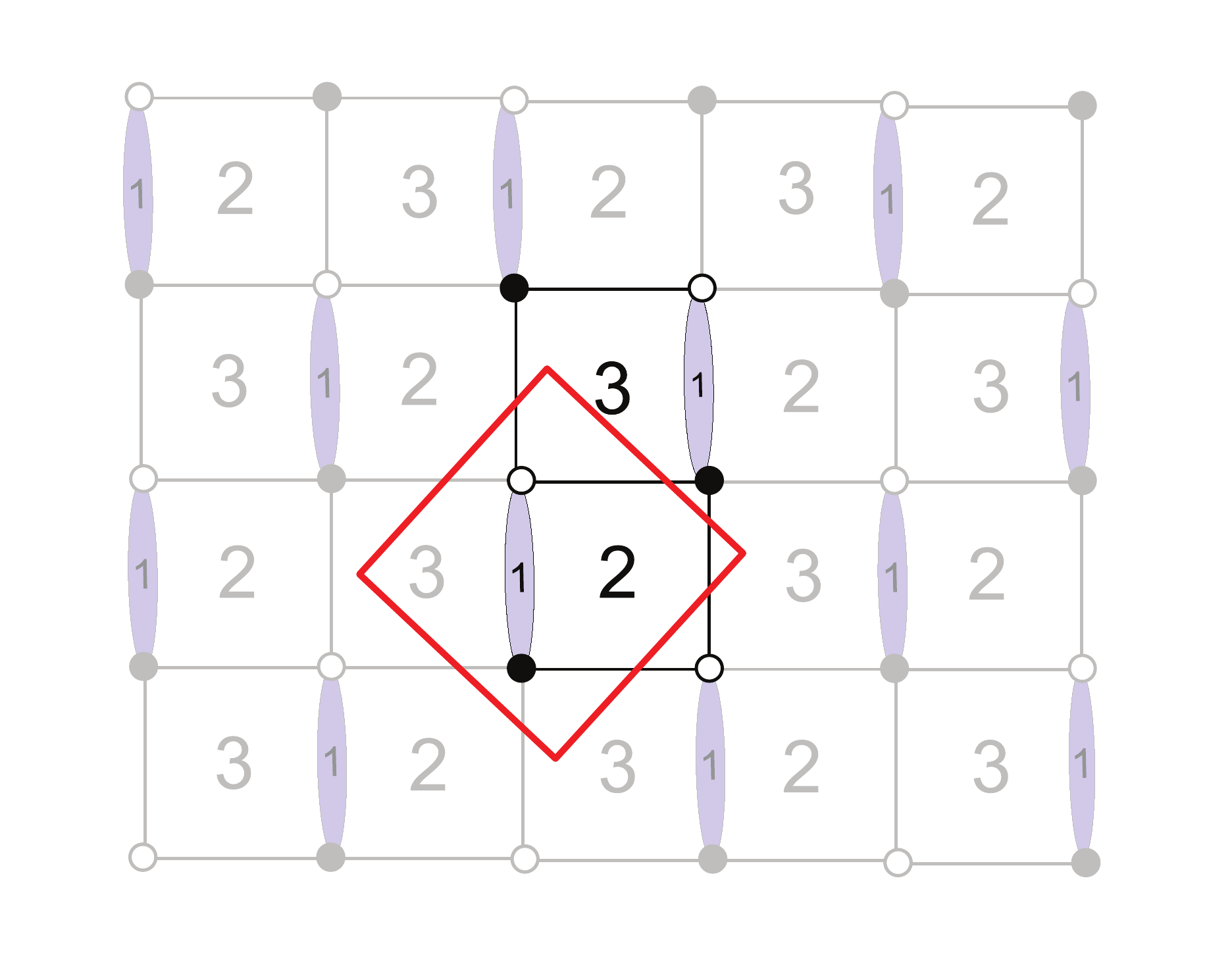}
 \caption{[Phase I of $\CC \times \BC$] (i) Quiver diagram of the $\sD_1\sC$  model.\ (ii) Tiling of the $\sD_1\sC$  model.}
  \label{f:phase2conxc}
\end{center}
\end{figure}

This model (which we shall refer to as $\sD_1\sC$) was first introduced in \cite{taxonomy} as part of a classification procedure for all models that have 2 terms in the superpotential.  It has 3 gauge groups and five chiral multiplets which we will denote as $X_{13}, X_{23}, X_{21}, X_{32}^1, X_{32}^2$, with a superpotential:
\bea
W= \tr \left( X_{21} X_{13} X^1_{32} X_{23} X^2_{32} - X_{21} X_{13} X^2_{32} X_{23} X^1_{32}\right)~. 
\eea
The quiver diagram and tiling are given in Figure \ref{f:phase2conxc}.
We choose the CS levels to be $k_1 = 1,~k_2 = -1,~ k_3 =0$.

\begin{figure}[ht]
\begin{center}
   \includegraphics[totalheight=8.0cm]{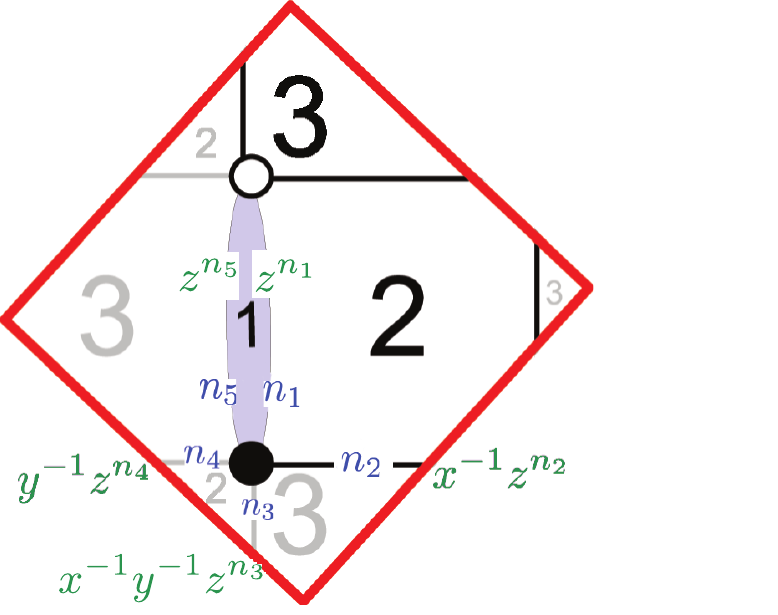}
 \caption{[Phase I of $\CC \times \BC$]. The fundamental domain of tiling for the $\sD_1\sC$  model: Assignments of the integers $n_i$ to the edges are shown in blue and the weights for these edges are shown in green.}
  \label{f:fdphase2conxc}
\end{center}
\end{figure}

\paragraph{The toric diagram.}   We demonstrate two methods of constructing the toric diagram. 
\begin{itemize}
\item {\bf The Kasteleyn matrix.} We assign the integers $n_i$ to the edges according to Figure \ref{f:fdphase2conxc}.  From \eref{kn}, we find that 
\bea
\text{Gauge group 1~:} \qquad k_1 &=& -1  = -n_1+n_5 ~, \nn \\
\text{Gauge group 2~:} \qquad k_2 &=& 1 = -n_2 +n_1 - n_4 + n_3 ~,  \nn \\
\text{Gauge group 3~:} \qquad k_3 &=& 0 = -n_3 + n_2 + n_4 -n_5~.
\eea  
We choose
\bea
n_1= 1,\qquad n_i=0 \;\;\text{otherwise}~.
\eea
We can construct the Kasteleyn matrix, which for this case is just a $1\times 1$ matrix and, therefore, coincides with its permanent:
\bea  \label{permKph1conxc}
K &=& X_{13} z^{n_5} + X_{21} z^{n_1} + X^1_{32} x^{-1} z^{n_2} + X_{23} x^{-1} y^{-1} z^{n_3} + X^2_{32} y^{-1} z^{n_4}  \nn \\
&=& X_{13} + X_{21} z + X_{32}^1 x^{-1}  + X_{23} x^{-1} y^{-1}  + X_{32}^2 y^{-1} \qquad \text{(for $n_1 = 1$ and $n_i =0$ otherwise)} ~. \nn\\
\eea
The powers of $x, y, z$ in each term of $K$ give the coordinates of each point in the toric diagram.
We collect these points in the columns of the following $G_K$ matrix:
\bea
G_K = \left(
\begin{array}{ccccc}
 -1 & 0 & -1 & 0 & 0 \\
 -1 & 0 & 0 & -1 & 0 \\
 0 & 0 & 0 & 0 & 1
\end{array}
\right)~.
\eea

\item {\bf The charge matrices.}  
From \eref{permKph1conxc}, the perfect matchings can therefore be taken as
\bea \label{pmph1conxc}
p_1 = X^{1}_{32},\;\; p_2 = X^{2}_{32}, \;\; p_3 = X_{13}, \;\; p_4 = X_{23},\;\;  p_5 =  X_{21} \ . \qquad
\eea
Since there is a one-to-one correspondence between the quiver fields and the perfect matchings, it follows that 
\bea
Q_F=0~. \label{QFD1C}
\eea
From \eref{QD}, we find that 
\bea
Q_D = (1,1,-1,-1,0)~.  \label{QDD1C}
\eea  
Note that since the CS coefficient $k_3=0$, we can immediately identify the baryonic charges with the quiver charges under the gauge group 3, and hence arrive at \eref{QDD1C}.
The total charge matrix is given by
\bea
Q_t = Q_D = (1,1,-1,-1,0)~.  \label{QtD1C}
\eea
We obtain the matrix $G_t$ from \eref{Gt}, and after removing the first row, the columns give the coordinates of points in the toric diagram:  
\bea
G'_t = \left(
\begin{array}{ccccc}
 1 & 0 & 1 & 0 & 0 \\
 1 & 0 & 0 & 1 & 0 \\
 0 & 0 & 0 & 0 & 1
\end{array}
\right)~. \label{Gptd2c}
\eea
We see that the toric diagram is merely 5 corners of a pyramid (Figure \ref{f:torconxc}).   This is in fact the toric diagram of $\CC \times \BC$ \cite{Hanany:2008fj}.
\begin{figure}[h]
\begin{center}
  \includegraphics[totalheight=4.0cm]{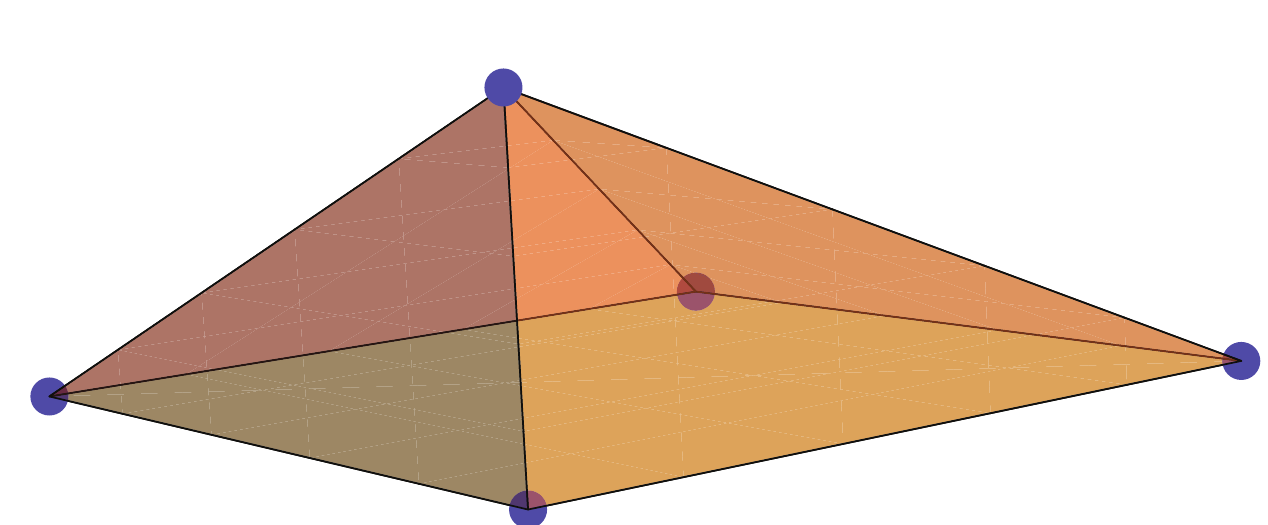}
 \caption{The toric diagram of the $\CC \times \BC$ theory.}
  \label{f:torconxc}
\end{center}
\end{figure}
\end{itemize}

\noindent Note that the toric diagrams constructed from $G_K$ and $G'_t$ are the same up to a transformation 
{\footnotesize $\CT =\left( \begin{array}{ccc} -1&0&0\\0&-1&0\\0&0&1 \end{array} \right)  \in GL(3, \BZ)$},
where we have $G_K = \CT \cdot G'_t$.

\paragraph{The Master space.}  
Since the Master space is generated by the perfect matchings (subject to the relation \eref{QFD1C}), it follows that 
\bea
\f_{\sD_1\sC} = \BC^5~.  \label{fflatD1C}
\eea
Since there are two pairs of repeated columns in the $Q_t$ matrix, the mesonic symmetry of the theory is $SU(2) \times SU(2) \times U(1)_q \times U(1)_R$. 
From Figure \ref{f:torconxc}, there are 5 external points in the toric diagram. From \eref{numberbary}, we thus have $5-4 =1$ baryonic charge.  This baryonic charge comes from the D-terms, and its assignment to the perfect matchings is given by the $Q_D$ matrix. The global symmetry of the theory is a product of mesonic and baryonic symmetries: $SU(2) \times SU(2) \times U(1)_q \times U(1)_R \times U(1)_B$. The presence of two mesonic $U(1)$ charges implies that there is a minimisation problem to be solved in order to determine which linear combination of these charges gives the right R-charge in the IR (see \cite{Hanany:2008fj} for details of the computation).  A consistent charge assignment to the perfect matchings is given in Table \ref{t:chargeconxc}. 
\begin{table}[h]
 \begin{center} 
  \begin{tabular}{|c||c|c|c|c|c|c|}
  \hline
  \;& $SU(2)_1$&$SU(2)_2$&$U(1)_q$&$U(1)_B$&$U(1)_R$&  fugacity \\
  \hline \hline
  $p_1$&$1$&0&1&1&$3/8$ & $t^3 q b x_1$ \\
  \hline
  $p_2$&$-1$&0&1&1&$3/8$ & $t^3 q b/x_1$ \\
  \hline
  $p_3$&0&$1$&1&$-1$ &$3/8$ & $t^3 q x_2/b$\\
  \hline
  $p_4$&0&$-1$&1&$-1$&$3/8$ & $t^3 q/ (b x_2)$\\
  \hline
  $p_5$&0&0&$-4$&0&$1/2$ & $t^4/q^4$\\
  \hline
  \end{tabular}
  \end{center}
\caption{Charges under the global symmetry of the $\CC \times \BC$ theory. Here $t$ is the fugacity associated with the $U(1)_{R}$ charges.  The power of $t$ counts R-charges in the unit of $1/8$, $q$ is the fugacity associated with the $U(1)_{q}$ charges, and $x_1,~x_2$ are respectively the $SU(2)_1,~SU(2)_2$ weights.}
\label{t:chargeconxc}
\end{table}
Instead of doing this computation, we can give arguments for the correct result as follows. The perfect matching which parametrises $\BC$ is expected to be a free field and therefore have R-charge $1/2$. The remaining 4 perfect matchings are completely symmetric and the requirement of R-charge 2 to the superpotential divides $3/2$ equally among them, resulting in R-charge of $3/8$ per each. The baryonic charge is determined by the charge matrix $Q_D$ \eref{QDD1C} which gives the linear relations between the 4 perfect matchings and the remaining $U(1)$ is determined by demanding that the superpotential has charge 0.
From \eref{fflatD1C}, it is immediate to write down the Hilbert series of the Master space using the charge assignment in Table \ref{t:chargeconxc}:
\bea
\gf_1(t_1,t_2,x_1,x_2,b; \sD_1\sC) = \frac{1}{\left(1-t_1 b x_1 \right)\left(1-\frac{t_1 b}{x_1}\right) \left(1-\frac{t_1 x_2}{b}\right) \left(1-\frac{t_1}{b x_2}\right)\left(1-t_2\right)}~, \label{hsfflatd1c}
\eea
where $t_1 = t^3 q$ and $t_2 = t^4/q^4$. 

\begin{figure}[h]
\begin{center}
  \includegraphics[totalheight=3.0cm]{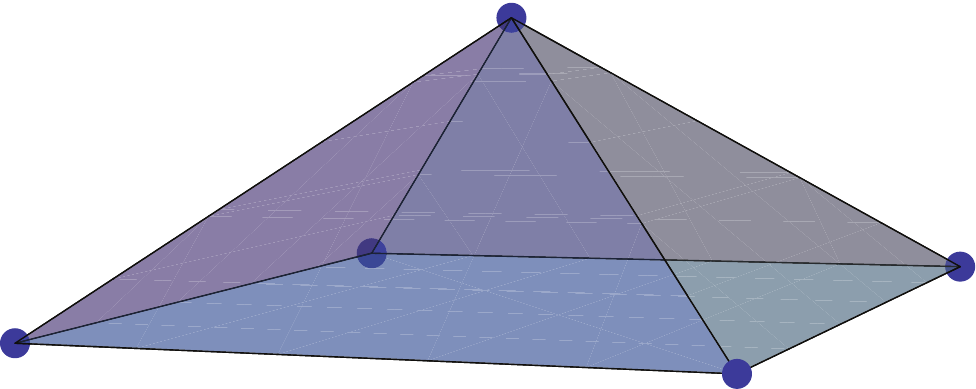}
 \caption{The lattice of generators of the $\CC \times \BC$ theory.}
  \label{f:latconxc}
\end{center}
\end{figure}

\paragraph{The mesonic moduli space.} From \eref{quoteFD}, the mesonic moduli space is given by
\bea
\CMm_{\sD_1\sC} =  \BC^{5}//Q_D = \BC^{5}//(1,1,-1,-1,0) ~. \label{mesonicd1c}
\eea 
Therefore, the Hilbert series of this space can be obtained by integrating \eref{hsfflatd1c} over the baryonic fugacity $b$:
\bea
\gm_1(t_1,t_2,x_1,x_2; \sD_1\sC) &=& \oint_{|b|=1} \frac{db}{2\pi i b}\frac{1}{\left(1-t_1 x_1 b\right)\left(1-\frac{t_1 b}{x_1}\right)\left(1-\frac{t_1 x_2}{b}\right)\left(1-\frac{t_1}{x_2 b}\right)\left(1-t_2\right)} \nn \\
&=& \frac{1-t_1^4}{\left(1-t^2_1 x_1 x_2\right)\left(1-\frac{t^2_1 x_2}{x_1}\right)\left(1-t_2\right)\left(1-\frac{t^2_1 x_1}{x_2}\right)\left(1-\frac{t^2_1}{x_1 x_2}\right)} \nn\\
&=& \frac{1}{1-t_2} \times  \frac{1-t_1^4}{\left(1-t^2_1 x_1 x_2\right)\left(1-\frac{t^2_1 x_2}{x_1}\right)\left(1-\frac{t^2_1 x_1}{x_2}\right)\left(1-\frac{t^2_1}{x_1 x_2}\right)} \nn \\
&=& \sum^{\infty}_{i=0} t_2^{i} \sum^{\infty}_{n=0}[n;n]t_1^{2n}~. \label{meshsd1c}
\eea
It is apparent from the third equality that the mesonic moduli space is indeed $\CC \times \BC$.
The unrefined Hilbert series is
\bea
\gm_1 (t,t,1,1; \sD_1\sC) = \frac{1+t^2}{(1-t)(1-t^2)^3}~.
\eea
The order of the pole $t=1$ indicates that the space $\CMm_{\sD_1\sC}$ is 4 dimensional, as expected.
The plethystic logarithm  of the Hilbert series is
\bea \label{pln1ph2conC}
\PL[\gm_1(t_1,t_2,x_1,x_2; \sD_1\sC)] &=& \left(x_1+\frac{1}{x_1}\right)\left(x_2+\frac{1}{x_2}\right)t^2_1 + t_2  - t^4_1 \nn \\
&=& [1; 1] t^2_1 + t_2 - t^4_1 \ . \label{mespld1c}
\eea 
\paragraph{The generators.} We see that the generators of the mesonic moduli space are
\bea
M^{1}_{1} &=& X_{13}X^{1}_{32} = p_1 p_3~,\quad M^{2}_{1} = X_{13}X^{2}_{32} = p_2 p_3~,\quad M^{1}_{2}= X_{23} X^{1}_{32} = p_1 p_4~, \nn \\ 
M^{2}_{2}&=& X_{23} X^{2}_{32} = p_2 p_4~,\quad X_{21} = p_5~. \label{genph1conxc}
\eea
Note that we require gauge invariance with respect to the gauge group 3, and so the indices corresponding to the gauge group 3 are contracted.
Among these generators, there is a relation:
\bea
\left(X_{13}X^{1}_{32}\right)\left(X^{2}_{32}X_{23}\right)=\left(X_{13}X^{2}_{32}\right)\left(X^{1}_{32}X_{23}\right)~,
\eea
or in a more concise notation:
\bea
\det ~ \! \! M = 0 \ .
\eea
We can represent the generators \eref{genph1conxc} in a lattice (Figure \ref{f:latconxc}) by plotting the powers of the weights of the characters in \eref{mespld1c}.
Note that the lattice of generators is the dual of the toric diagram (nodes are dual to faces and edges are dual to edges).  For the $\CC \times \BC$ theory, the toric diagram is a pyramid (5 nodes, 8 edges and 5 faces), which is a self-dual lattice.  Therefore, the lattice of generators is the same as the toric diagram.

\subsection{Phase II: The Two-Hexagon Model}
\label{C2Z2}
\begin{figure}[ht]
\begin{center}
  \vskip 1cm
  \hskip -6cm
  \includegraphics[totalheight=1.2cm]{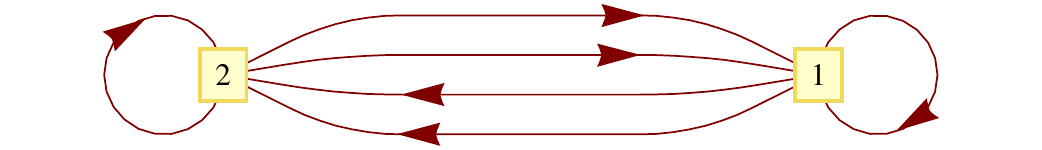}
  \vskip -3.0cm
  \hskip 9cm
  \includegraphics[totalheight=4.5cm]{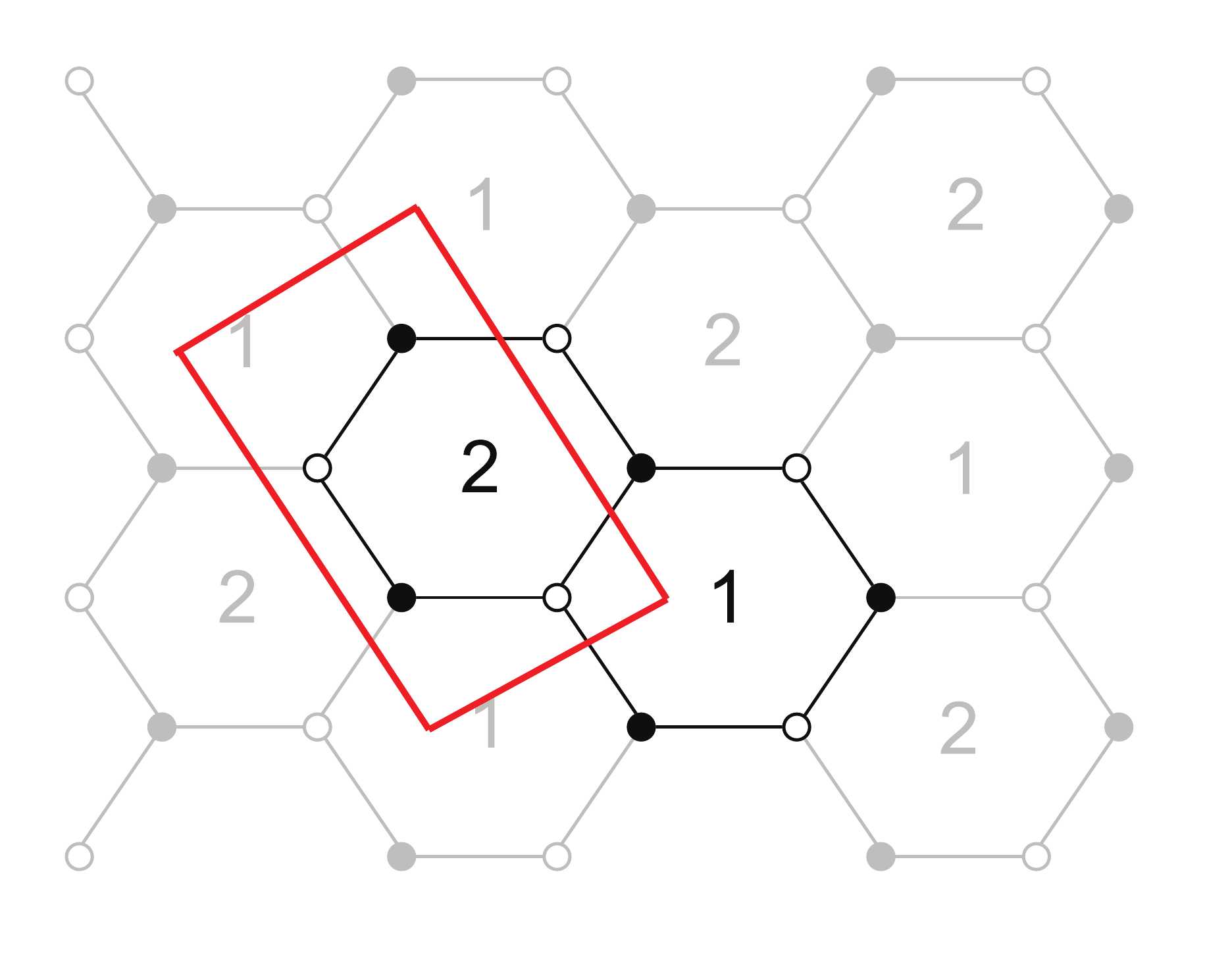}
 \caption{[Phase II of $\CC \times \BC$] (i) Quiver diagram for the $\sH_2$ model. \ (ii) Tiling for the $\sH_2$ model.}
  \label{f:phase1conxc}
\end{center}
\end{figure}

This model, studied in \cite{Hanany:2008cd, Hanany:2008fj} (which we shall refer to as $\sH_2$) has two gauge groups and six chiral multiplets denoted as $\phi_1, \phi_2, X_{12}^1, X_{12}^2, X_{21}^1, X_{21}^2$.  The quiver and tiling are drawn in Figure \ref{f:phase1conxc}. Note that in 3+1 dimensions this tiling corresponds to the $\BC^2/\BZ_2 \times \BC$ theory. The superpotential is given by
\begin{equation} \label{spph1conc}
W = \tr \left( \phi_1 (X_{12}^2 X_{21}^1 - X_{12}^1 X_{21}^2 ) + {\phi}_2 (X_{21}^2 X_{12}^1 - X_{21}^1 X_{12}^2) \right) \ .
\end{equation}
According to \eref{k-con}, we take the Chern--Simons levels to be $k_1=-k_2=1$. \\

\paragraph{The Master space.}  From \eref{spph1conc}, we see that the Master space  of the $\sH_2$ model \cite{master} is 
\begin{equation}\label{N2id}
\f_{\sH_2}=\mathbb{V}(X_{12}^1X_{21}^2-X_{12}^2X_{21}^1, (\phi_1-\phi_2) X_{12}^1,  (\phi_1-\phi_2) X_{12}^2, (\phi_1 -\phi_2) X_{21}^2, (\phi_1-\phi_2) X_{21}^1) \ .
\end{equation}
It is clear that $\f_{\sH_2}$ is reducible and decomposes into two irreducible components as $\f_{\sH_2}= \firr{\sH_2}\hbox{  } \cup \hbox{  }L_{\sH_2}~,$ where
\bea\label{N2red}
\firr{\sH_2} &=& \mathbb{V}(\phi_1 - \phi_2, X_{12}^1X_{21}^2 - X_{12}^2X_{21}^1) \qquad \text{(Higgs branch)} \ , \nn \\
L_{\sH_2} &=& \mathbb{V}(X_{12}^1, X_{12}^2, X_{21}^1, X_{21}^2) \qquad \qquad \quad \; \;  \text{(Coulomb branch)}\ .
\eea
We see that the \emph{coherent component} is  
\bea
\firr{\sH_2} = \CC \times \mathbb{C} ~, 
\eea
where the $\mathbb{C}$ is parametrised by $\phi_1 = \phi_2$ and the conifold singularity $\CC$ is described by the chiral fields $\{X_{12}^1,X_{12}^2,X_{21}^1,X_{21}^2\}$ with the constraint $X_{12}^1 X_{21}^2 = X_{12}^2 X_{21}^1 $. The component $L_{\sH_2} = \mathbb{C}^2$ is parametrised by the fields $\{\phi_1,\phi_2 \}$. These two branches meet on the complex line parametrised by $\phi_1=\phi_2$:
\bea
 \firr{\sH_2}\hbox{  } \cap \hbox{  }L_{\sH_2} = \BC~.
\eea

\begin{figure}[ht]
\begin{center}
   \includegraphics[totalheight=8.0cm]{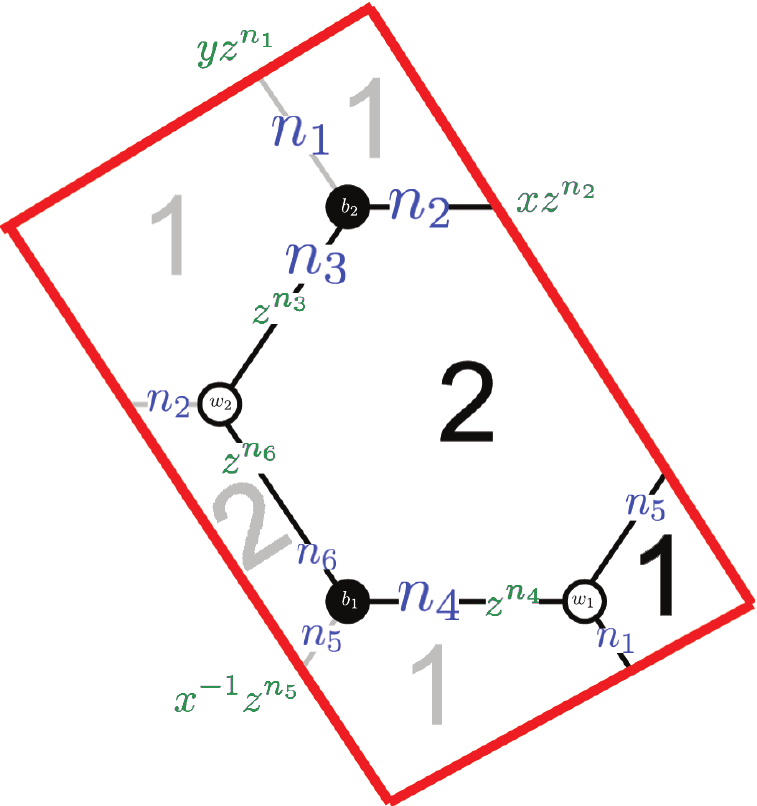}
 \caption{[Phase II of $\CC \times \BC$]. The fundamental domain of tiling for the $\sH_2$ model: Assignments of the integers $n_i$ to the edges are shown in blue and the weights for these edges are shown in green.}
  \label{f:fdphase1conxc}
\end{center}
\end{figure}

\paragraph{The Kasteleyn matrix.} We assign the integers $n_i$ to the edges according to Figure \ref{f:fdphase1conxc}.  From \eref{kn}, we find that 
\bea
\text{Gauge group 1~:} \qquad k_1 &=& 1  = -n_2+n_3+n_4 - n_5 ~, \nn \\
\text{Gauge group 2~:} \qquad k_2 &=& -1 = n_2 - n_3 -n_4 + n_5 ~.
\eea  
We choose
\bea
n_3= 1,\quad n_i=0 \; \text{otherwise}~.
\eea
We can now construct the Kasteleyn matrix:
\be
K =   \left(
\begin{array}{c|cc}
\; & w_1 & w_2 \\
\hline
b_1 & X^{1}_{21} x^{-1} z^{n_5}+ X^{2}_{12} z^{n_4}  & \ \phi_2 z^{n_6}  \\
 b_2 & \phi_1 y z^{n_1}      & \  X^{2}_{21} x z^{n_2}+ X^{1}_{12} z^{n_3}
\end{array}
\right) ~.
\ee
The permanent of this matrix is
\bea
\perm~K &=&   X^{1}_{21}X^{2}_{21} z^{n_2+n_5}  + X^{2}_{12} X^{2}_{21} x z^{n_2+n_4}  + X^{1}_{21}X^{1}_{12} x^{-1} z^{n_3+n_5} + X^{1}_{12} X^{2}_{12} z^{n_3+n_4} + \phi_{1}\phi_{2} y z^{n_1+n_6} \nn \\
&=& X^{1}_{21}X^{2}_{21}  + X^{2}_{12} X^{2}_{21} x   + X^{1}_{21}X^{1}_{12} x^{-1} z + X^{1}_{12} X^{2}_{12} z + \phi_{1}\phi_{2} y  \nn \\  
&&\text{(for $n_3 = 1$ and $n_i =0$ otherwise)} ~ .
\label{permph1conxc}
\eea

\paragraph{The perfect matchings.} From \eref{permph1conxc}, we write each perfect matching as a collection of fields (on the coherent component) as follows:
\bea \label{pmph2conxc}
p_1 = \{ X^{1}_{12}, X^{2}_{12} \},\;\; p_2 = \{X^{2}_{21}, X^{2}_{12} \} , \;\; p_3 = \{X^{1}_{12}, X^{1}_{21} \}, \;\; p_4 = \{ X^{1}_{21}, X^2_{21} \}, \;\;  p_5=  \{ \phi_{1}, \phi_{2} \} \ . \qquad
\eea
We see below that this choice of the perfect matchings is precisely equal to the perfect matching of Phase I.
In turn, we find the parameterisation of fields in terms of perfect matchings:
\bea
X^{1}_{12} = p_1 p_3,  \quad X^{1}_{21} = p_2 p_3, \quad X^{2}_{12} = p_1 p_4, \quad X^{2}_{21} = p_2 p_4, \quad \phi_1 = \phi_2 = p_5~. 
\eea
The correspondence is summarised in the perfect matching matrix:
\beq
P=\left(\begin{array} {c|ccccc}
 \;& p_1&p_2&p_3&p_4&p_5\\
  \hline 
  X^{1}_{12}&1&0 &1&0&0\\
   X^{2}_{12}&1&0 &0&1&0\\
  X^{1}_{21}&0&1&1&0&0\\
  X^{2}_{21}&0&1 &0&1&0\\
  \phi_{1}&0&0&0&0&1\\
  \phi_{2}&0&0&0&0&1
\end{array}
\right).
\eeq
Basis vectors of of the null space of $P$ are given in the rows of the following matrix:
\bea
Q_F = (1,1,-1,-1,0)~. \label{q1}
\eea
Hence, from \eref{relpm}, we see that the relations between the perfect matchings are given by
\bea 
p_1+p_2 - p_3 - p_4 = 0~. \label{relpmph2conxc}
\eea
Since the coherent component of the Master space is generated by the perfect matchings (subject to the relation \eref{relpmph2conxc}), it follows from \eref{sympq} that 
\bea
\firr{\sH_2} = \BC^5// Q_F = \BC^5//(1,1,-1,-1,0)~.  \label{quot1}
\eea

\paragraph{The mesonic moduli space.}  Since the number of gauge groups is $G=2$, it follows that there is $G-2=0$ baryonic charge from the D-terms, \emph{i.e.} 
\bea
Q_D = 0~.  \label{qdh2}
\eea
From \eref{quoteFD}, the mesonic moduli space is identical to the Master space and is given by
\bea
\CMm_{\sH_2} = \firr{\sH_2} = \BC^5//(1,1,-1,-1,0)~. \label{quot2}
\eea
Comparing this equation to \eref{mesonicd1c}, we find that the mesonic moduli space of this model is indeed identical to that of Phase I:
\bea
\CMm_{\sH_2} = \CMm_{\sD_1\sC}  = \CC \times \BC~. \label{ph2conxc}
\eea

\paragraph{The toric diagram.} We demonstrate two methods of constructing the toric diagram. 
\begin{itemize}
\item {\bf The charge matrices.} From \eref{q1} and \eref{qdh2}, we see that the total charge matrix $Q_t$ is given by
\bea
Q_t = (1,1,-1,-1,0)~,
\eea
which is identical to that of Phase I.  Hence, the $G'_t$ matrix coincides with that of Phase I and is given by \eref{Gptd2c}.  Thus, we arrive at the toric diagram in Figure \ref{f:torconxc}.  This indeed confirms the relation \eref{ph2conxc}.

\item {\bf The Kasteleyn matrix.} 
The powers of $x, y, z$ in each term of \eref{permph1conxc} give the coordinates of each point in the toric diagram.  
We collect these points in the columns of the following $G_K$ matrix:
\bea
G_K = \left(
\begin{array}{ccccc}
 0 & 0 & -1 & 1 & 0 \\
 0 & 0 & 0 & 0 & 1 \\
 1 & 0 & 1 & 0 & 0
\end{array}
\right)~.
\eea
Note that the toric diagrams constructed from the $G_K$ matrix and the $G'_t$ matrix (given by \eref{Gptd2c}) are the same up to a transformation 
{\footnotesize $\CT =\left( \begin{array}{ccc} -1&1&0\\0&0&1\\1&0&0 \end{array} \right)  \in GL(3, \BZ)$},
where we have $G_K = \CT \cdot G'_t$.
\end{itemize}

\paragraph{The baryonic charge.}  From Figure \ref{f:torconxc}, there are 5 external points in the toric diagram.  From \eref{numberbary}, we thus have $5-4 =1$ baryonic charge in this model.  We emphasise that this baryonic charge does \emph{not} come from the D-terms, as $Q_D = 0$.  Since $Q_F$ is the only non-zero charge matrix available in the theory, from \eref{quot2}, it is natural to assign the baryonic charge $U(1)_B$ to each perfect matchings according to the $Q_F$ matrix.  

\paragraph{The global symmetry.}  
Since there are two pairs of repeated columns in the $Q_t$ matrix, the mesonic symmetry of the theory is $SU(2) \times SU(2) \times U(1)_q \times U(1)_R$. 
The global symmetry of the theory is a product of mesonic and baryonic symmetries: $SU(2) \times SU(2) \times U(1)_q \times U(1)_R \times U(1)_B$, which is identical to that of Phase I.  The R-charges of perfect matchings can be determined as follows.  As discussed above, the perfect matching $p_5 = \phi_1 = \phi_2$ parametrises $\BC$, and so it is expected to be a free field with an R-charge $1/2$.  The remaining 4 perfect matchings are completely symmetric and the requirement of R-charge 2 to the superpotential divides 3/2 equally among them, resulting in R-charge of 3/8 per each.   We can therefore assign global charges to the perfect matchings as in Table \ref{t:chargeconxc}.

\paragraph{The Hilbert series.} From the above discussion, we see that the Hilbert series of the mesonic moduli space of this model and its plethystic logarithm are given respectively by (\ref{meshsd1c}) and (\ref{mespld1c}).
The latter indicates that the mesonic moduli space is a complete intersection generated by the fields 
\bea
X^{1}_{12} = p_1 p_3,  \quad X^{1}_{21} = p_2 p_3, \quad X^{2}_{12} = p_1 p_4, \quad X^{2}_{21} = p_2 p_4, \quad \phi_1 = p_5, \quad \phi_2 = p_5~,
\eea
subject to the relations:
\bea
X^{1}_{12}X^{2}_{21} = X^{2}_{12}X^{1}_{21}~,\qquad \phi_{1} = \phi_{2} \ . \label{relph1conxc}
\eea
Note that, in terms of the perfect matchings, the generators of this model are precisely the same as those of Phase I.

\subsection{Phase III: The Two Double-Bonded One-Hexagon Model}
\begin{figure}[ht]
\begin{center}
 \vskip 1cm
  \hskip -7cm
  \includegraphics[totalheight=1.5cm]{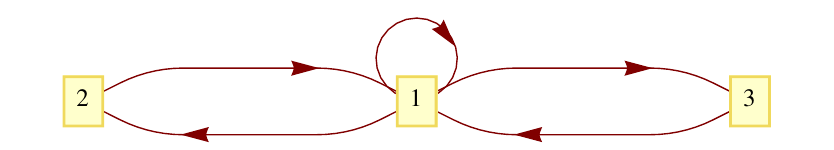}
   \vskip -3cm
  \hskip 8cm
  \includegraphics[totalheight=5.0cm]{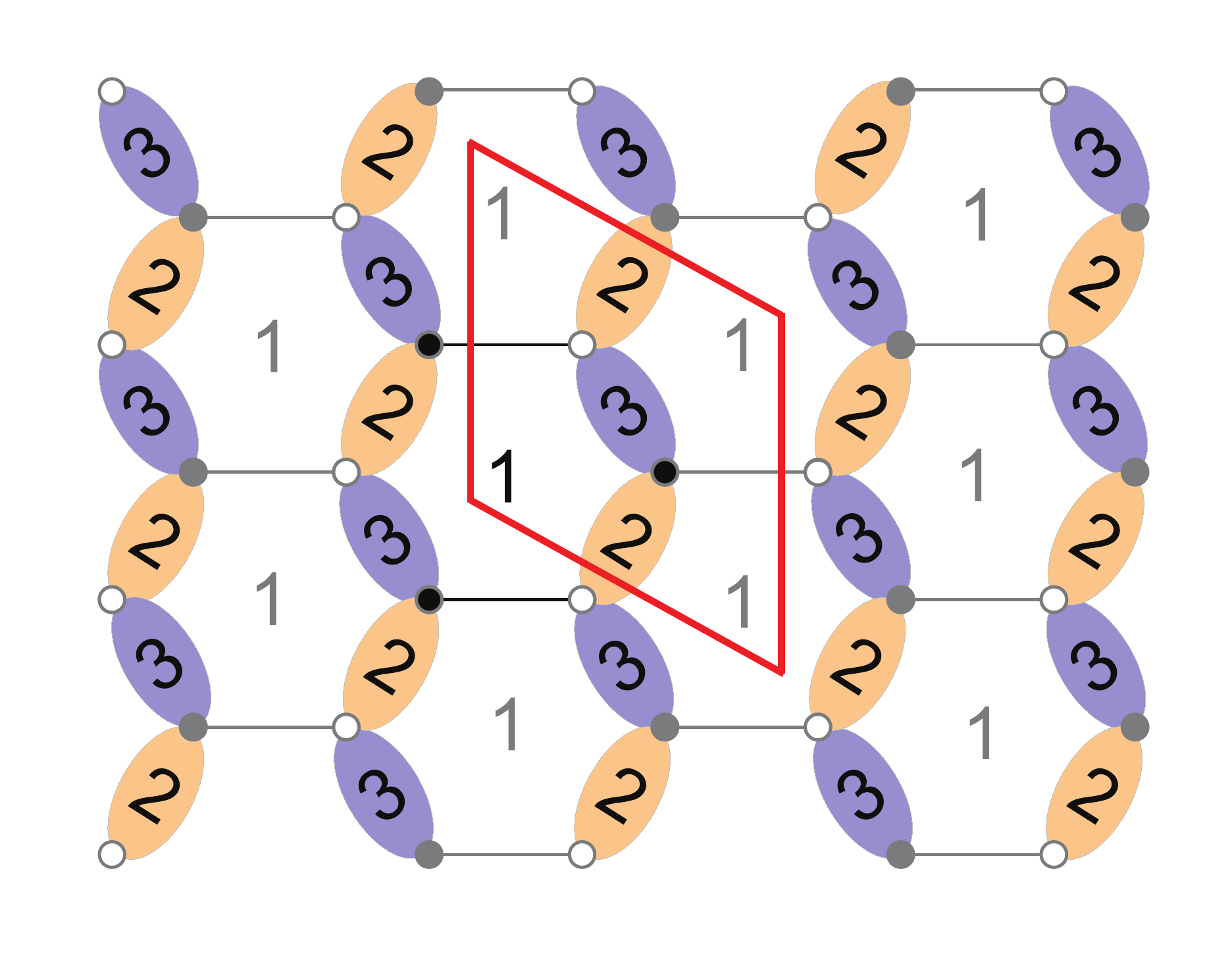}
 \caption{[Phase III of $\CC \times \BC$]  (i) Quiver diagram of the $\sD_2\sH_1$  model.\ (ii) Tiling of the $\sD_2\sH_1$  model.}
  \label{f:phase3conxc}
\end{center}
\end{figure}

This model (which we shall refer to as $\sD_2\sH_1$) was introduced in \cite{taxonomy} as part of a classification procedure for all models that have 2 terms in the superpotential.  It has 3 gauge groups and five chiral multiplets which we will denote as $X_{12}, X_{21}, X_{13}, X_{31}, \phi_1$, with a superpotential:
\bea
W= \tr \left( \phi_1 X_{12} X_{21} X_{13} X_{31} - \phi_1 X_{13} X_{31} X_{12} X_{21} \right)~. 
\eea
The quiver diagram and tiling are given in Figure \ref{f:phase3conxc}.
We choose the CS levels to be $k_1 = 0,~k_2 = 1,~ k_3 =-1$.

\begin{figure}[ht]
\begin{center}
   \includegraphics[totalheight=8.0cm]{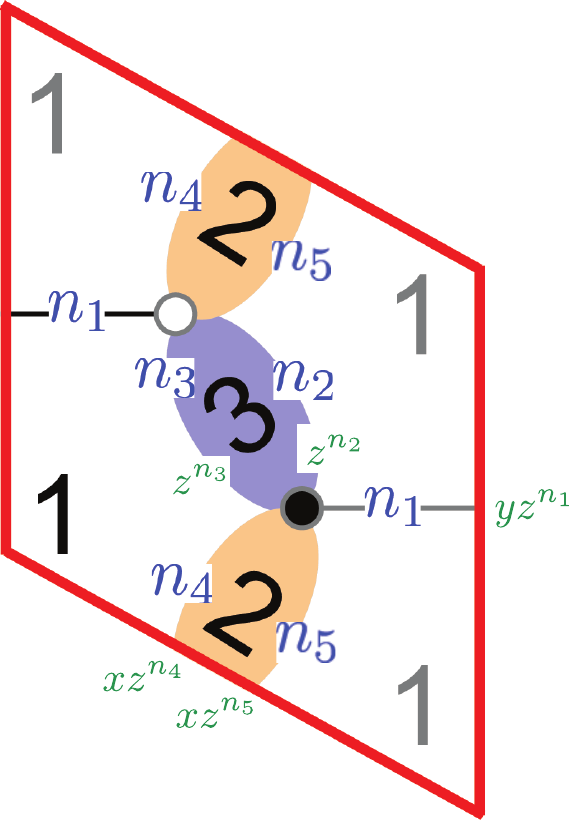}
 \caption{[Phase III of $\CC \times \BC$]  The fundamental domain of tiling for the $\sD_2\sH_1$  model: Assignments of the integers $n_i$ to the edges are shown in blue and the weights for these edges are shown in green.}
  \label{f:fdphase3conxc}
\end{center}
\end{figure}

\paragraph{The toric diagram.}   We demonstrate two methods of constructing the toric diagram. 
\begin{itemize}
\item {\bf The Kasteleyn matrix.}  We assign the integers $n_i$ to the edges according to Figure \ref{f:fdphase3conxc}.  From \eref{kn}, we find that 
\bea
\text{Gauge group 1~:} \qquad k_1 &=& 0  = n_2 - n_3 + n_4 -n_5 ~, \nn \\
\text{Gauge group 2~:} \qquad k_2 &=& 1 = -n_4 +n_5 ~,  \nn \\
\text{Gauge group 3~:} \qquad k_3 &=& -1 = -n_2 + n_3~.
\eea  
We choose
\bea
n_2 = n_5 = 1,\qquad n_i=0 \;\;\text{otherwise}~.
\eea
We can construct the Kasteleyn matrix, which for this case is just a $1\times 1$ matrix and, therefore, coincides with its permanent:
\bea \label{Kph3conxc}
K &=& \phi_1 y z^{n_1} + X_{13} z^{n_2} + X_{31} z^{n_3} + X_{12} x z^{n_4} + X_{21} x z^{n_5}  \nn \\
&=&  \phi_1 y + X_{13} z  + X_{31} + X_{12} x + X_{21} x z \quad \text{(for $n_2 = n_5 = 1$ and $n_i =0$ otherwise)} ~. \nn\\
\eea
The powers of $x, y, z$ in each term of $K$ give the coordinates of each point in the toric diagram.
We collect these points in the columns of the following $G_K$ matrix:
\bea
G_K = \left(
\begin{array}{ccccc}
 1&0 & 1&  0 & 0   \\
 0 & 0 & 0 &  0 &1   \\
 1 & 0 & 0  & 1 & 0 
\end{array}
\right)~.
\eea

\item {\bf The charge matrices.}  
From \eref{Kph3conxc}, the perfect matchings can therefore be taken as 
\bea
p_1 = X_{12}, \quad p_2 = X_{13}, \quad p_3 =   X_{21}, \quad p_4 = X_{31}, \quad p_5 = \phi_1~.
\eea
We see below that this choice of perfect matchings is precisely equal to the perfect matching of Phase I.
Since there is a one-to-one correspondence between the quiver fields and the perfect matchings, it follows that
\bea
Q_F=0~.
\eea  
According to the computation from \eref{QD}, we find that 
\bea
Q_D = (1,1,-1,-1,0)~. \label{QDph3conxc}
\eea  
Note that since the CS coefficient $k_1=0$, we can immediately identify the baryonic charges with the quiver charges under the gauge group 1, and hence arrive at \eref{QDph3conxc}.
The total charge matrix is then given by
\bea
Q_t = Q_D = (1,1,-1,-1,0)~,
\eea
which is identical to that of Phases I and II.  Hence, the $G'_t$ matrix is given by
\bea
G'_t = \left(
\begin{array}{ccccc}
 1 & 0 & 1 & 0 & 0 \\
 1 & 0 & 0 & 1 & 0 \\
 0 & 0 & 0 & 0 & 1
\end{array}
\right)~,
\eea
which is identical to \eref{Gptd2c}.  Thus, the toric diagram for this model is given by Figure \ref{f:torconxc}.  Thus, we have shown that the mesonic moduli space is indeed $\CC \times \BC$.
\end{itemize}
\noindent Note that the toric diagrams constructed from $G_K$ and $G'_t$ are the same up to a transformation 
{\footnotesize $\CT =\left( \begin{array}{ccc} 1&0&0\\0&0&1\\0&1&0 \end{array} \right)  \in GL(3, \BZ)$},
where we have $G_K = \CT \cdot G'_t$.

\paragraph{The moduli space.}  Since the $Q_F$ matrix is zero, the Master space is simply
\bea
\f_{\sD_2\sH_1} = \BC^5~.
\eea
From \eref{quoteFD}, the mesonic moduli space is given by
\bea
\CMm_{\sD_2\sH_1} = \f_{\sD_2\sH_1}//Q_D = \BC^{5}//(1,1,-1,-1,0) = \CC \times \BC ~, \label{mesonicd2h1}
\eea 
which is the same as Phases I and II, as expected.

\paragraph{The Hilbert series.}  From the charge matrices, it is clear that the global symmetry of this model is identical to that of Phase I, namely $SU(2)_1 \times SU(2)_2 \times U(1)_q \times U(1)_B \times U(1)_R$.  A consistent charge assignment to the perfect matchings is given by Table \ref{t:chargeconxc}.   It is easy to see that the Master space Hilbert series and the mesonic Hilbert series are given respectively by \eref{hsfflatd1c} and \eref{meshsd1c}.  From the plethystic logarithm \eref{mespld1c}, the generators are
\bea
M^{1}_{1} &=& X_{21}X_{12} = p_1 p_3 ~,\quad M^{2}_{1} = X_{21}X_{13} = p_2 p_3 ~,\quad M^{1}_{2}=X_{31}X_{12} = p_1 p_4~, \nn \\
 M^{2}_{2} &=& X_{21}X_{13} = p_2 p_4 ~,\quad \phi_{1} = p_5~.
\eea
Note that we require gauge invariance with respect to the gauge group 1, and so the indices corresponding to the gauge group 1 are contracted.
Among these generators, there is a relation which can be written as
\bea
\det ~ \! \! M = 0 \ .
\eea
Note that, in terms of the perfect matchings, the generators of this model are precisely the same as those of Phase I.

\subsection{A Comparison between Phases of the $\CC \times \BC$ Theory}
Here we make a comparison between phases of the $\CC \times \BC$ theory:
\begin{itemize}
\item {\bf Perfect matchings.} The perfect matchings of different phases are exactly the same (including the labels).  They are charged in the same way under the global symmetry according to Table \ref{t:chargeconxc}.

\item {\bf Generators.}  In terms of the perfect matchings, the generators of different phases are precisely the same.  These are summarised in Table \ref{t:comgenconxc}.

\begin{table}[h]
 \begin{center} 
  \begin{tabular}{|c||c|c|c|}
    \hline
    Perfect matchings & Generator of Phase I & Generator of Phase II & Generator of Phase III \\
    \hline
     $p_1 p_3$ & $X_{13}X^{1}_{32}$ & $X^{1}_{12}$ & $X_{21}X_{12}$ \\
     $p_2 p_3$ &  $X_{13}X^{2}_{32}$ & $X^{1}_{21}$  &$X_{21}X_{13}$ \\
     $p_1 p_4$ & $X_{23}X^{1}_{32}$ & $X^{2}_{12}$ &$X_{31}X_{12}$  \\
     $p_2 p_4$ & $X_{23}X^{2}_{32}$ & $X^{2}_{21}$  &$X_{21}X_{13}$ \\
     $p_5$ & $X_{21}$ & $\phi_1=\phi_2$& $\phi_{1}$ \\
     \hline
  \end{tabular}
  \end{center}
 \caption{A comparison between the generators of different phases of the $\CC \times \BC$ theory.  In terms of the perfect matchings, the generators of different phases are precisely the same.  In Phase I, we require gauge invariance with respect to the gauge group 3, and so the indices corresponding to the gauge group 3 are contracted.  In Phase III, we require gauge invariance with respect to the gauge group 1, and so the indices corresponding to the gauge group 1 are contracted.}
\label{t:comgenconxc}
\end{table}

\item {\bf Quiver fields.} The quiver fields of Phases I and III are the perfect matchings, whereas the quiver fields of Phase II are bilinears in perfect matchings (except the adjoint field which is linear in the perfect matching). 

\item {\bf Mesonic moduli space.} The mesonic moduli spaces of all phases are identical; they are $\CC \times \BC$.

\item {\bf Baryonic symmetries.} The baryonic symmetries of all phases are identical.  However, not all of them come from the same origin.  The baryonic symmetries of Phases I and III are induced by the D-terms, and each of them arises from one node of the quiver.  On the other hand, the baryonic symmetry of Phase II arises from the relation between perfect matchings.

\item {\bf Master space \& space of perfect matchings.} The Master spaces of Phases I and III and the space of perfect matchings in Phase II are identical; they are $\BC^5$.  Each of them is a combined baryonic and mesonic moduli space for one's own phase.  Note that for Phase II, the Master space is the mesonic moduli space.      

\item  {\bf Conclusion.} Different concepts like Master space, quiver fields, get different meaning in different phases.  Nevertheless, each object in one theory is mapped to the other, giving rise to a one-to-one correspondence.
\end{itemize}

\section{Phases of the $D_3$ Theory}
\subsection{Phase I: The Two-Double-Bonded Chessboard Model}
\begin{figure}[h]
\begin{center}
  \vskip 1cm
  \hskip -7cm
  \includegraphics[totalheight=4cm]{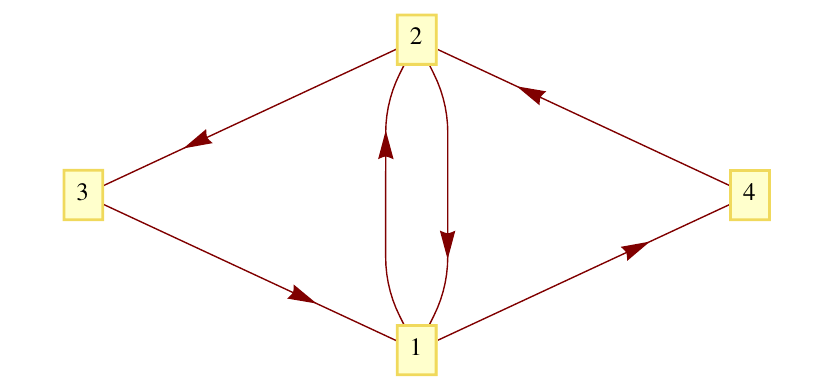}
    \vskip -4.5cm
  \hskip 8.4cm
  \includegraphics[totalheight=5cm]{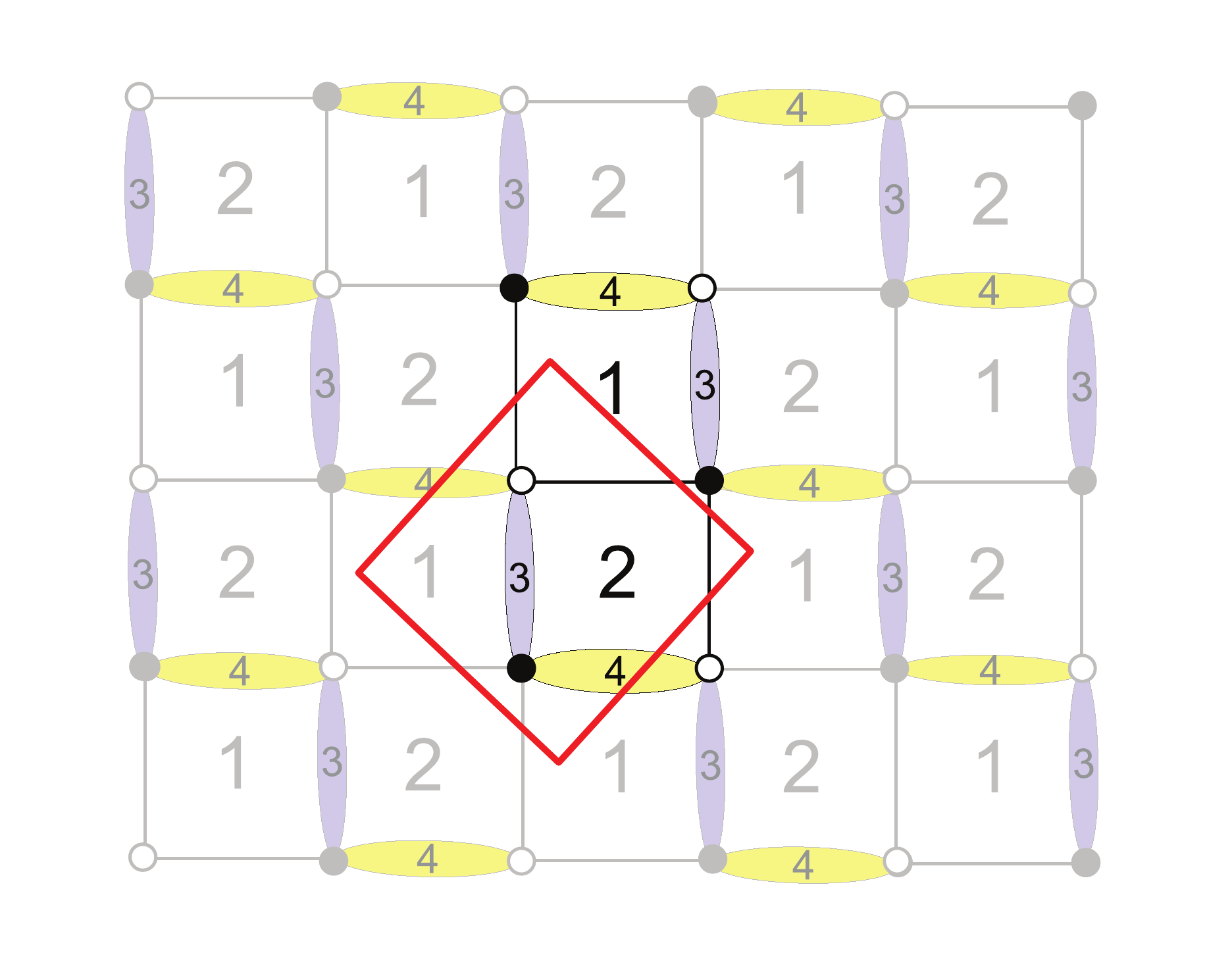}
 \caption{[Phase I of the $D_3$ theory] (i) Quiver diagram of the $\sD_2 \sC$ model.\ (ii) Tiling of the $\sD_2 \sC$ model.}
  \label{f:phase1D3}
\end{center}
\end{figure}

This model was studied before in \cite{Franco:2008um, taxonomy}.
The quiver diagram and tiling of this model (which we shall refer to as $\sD_2 \sC$) are drawn in Figure \ref{f:phase1D3}.
The superpotential of this model is given by
\bea
W = \tr \left( X_{14}X_{42}X_{21}X_{12}X_{23}X_{31} - X_{14}X_{42}X_{23}X_{31}X_{12}X_{21} \right)~.
\eea
We choose the CS levels to be $(k_1,k_2,k_3,k_4) = (1,1,-1,-1)$.

\begin{figure}[h]
\begin{center}
   \includegraphics[totalheight=6.0cm]{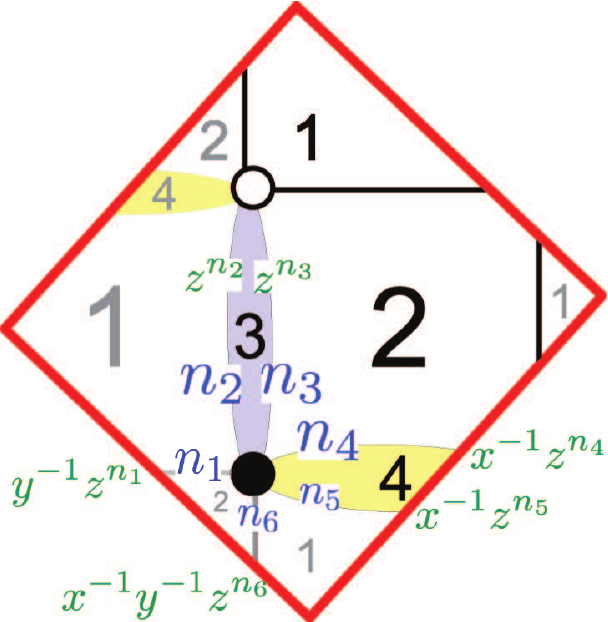}
 \caption{[Phase I of the $D_3$ theory] The fundamental domain of tiling for the $\sD_2 \sC$ model: Assignments of the integers $n_i$ to the edges are shown in blue and the weights for these edges are shown in green.}
  \label{f:fdph1d3}
\end{center}
\end{figure} 

\paragraph{The toric diagram.} We demonstrate two methods of constructing the toric diagram. 
\begin{itemize}
\item {\bf The Kasteleyn matrix.}
We assign the integers $n_i$ to the edges according to Figure \ref{f:fdph1d3}.  From \eref{kn}, we find that 
\bea
\text{Gauge group 1~:} \qquad k_1 &=& 1  = n_1 - n_2 +n_5 - n_6 ~, \nn \\
\text{Gauge group 2~:} \qquad k_2 &=& 1 = - n_1 + n_3 - n_4 +n_6  ~, \nn \\
\text{Gauge group 3~:} \qquad k_3 &=& -1 = n_2 - n_3~, \nn \\
\text{Gauge group 4~:} \qquad k_4 &=& -1 = n_4 - n_5~.
\eea  
We choose
\bea
n_3 = n_5 = 1,\quad n_i=0 \; \text{otherwise}~.
\eea
Since the fundamental domain contains only one white node and one black node, the Kasteleyn matrix is $1\times 1$ and, therefore, coincides with its permanent:
\bea \label{Kph2d3}
K &=& X_{31}z^{n_2} + X_{23}z^{n_3} +  X_{42} x^{-1} z^{n_4} + X_{14} w^{-1} z^{n_5} + X_{21} x^{-1} y^{-1} z^{n_6} + X_{12} y^{-1} z^{n_1} \nn \\
&=& X_{31} + X_{23}z +  X_{42} x^{-1}  + X_{14} x^{-1} z +  X_{21} x^{-1} y^{-1}  + X_{12} y^{-1} \nn \\ 
&& \text{(for $n_3 = n_5 =1$ and $n_i=0$)}~,
\eea
where the powers of $x, y, z$ in each term give the coordinates of each point in the toric diagram. 
We collect these points in the columns of the following $G_K$ matrix:
\bea
G_K = \left(
\begin{array}{cccccc}
 0 & -1 & 0 & -1 & -1 & 0 \\
 0 & 0 & -1 & -1 & 0 & 0 \\
 1 & 0 & 0 & 0 & 1 & 0
\end{array}
\right)~.
\eea

\item{\bf The charge matrices.}  From \eref{Kph2d3}, we can take the perfect matchings to be
\bea
p_1= X_{23}, \quad p_2 = X_{42}, \quad p_3 = X_{12}, \quad p_4 = X_{21}, \quad p_5 = X_{31}, \quad p_6 = X_{14}~.
\eea
Since there is a one-to-one correspondence between the perfect matchings and the quiver fields,
\bea
Q_F = 0~.
\eea
Since the number of gauge groups is $G=4$, there are $G-2 = 2$ baryonic charges coming from the D-terms.  From \eref{QD}, we find that the $Q_D$ matrix is given by
\bea
Q_D = \left(
\begin{array}{cccccc}
 1 & 0 & -1 & 1 & 0 & -1\\
 1 & 1 & 0 & 0 & -1 & -1 
\end{array}
\right)~. \label{qdph1d3}
\eea
The total charge matrix $Q_t$ therefore coincides with $Q_D$:
\bea
Q_t = \left(
\begin{array}{cccccc}
 1 & 0 & -1 & 1 & 0 & -1\\
 1 & 1 & 0 & 0 & -1 & -1 
\end{array}
\right)~. \label{qtph1d3}
\eea 
Hence, the $G'_t$ matrix is given by
\bea
G'_t =\left(
\begin{array}{cccccc}
  0 & 1 & 0 & 1 & 1 & 0\\
 0 & 0 & 1 & 1 & 0 & 0 \\
1 & 0 & 0 & 0 & 1 & 0
\end{array}
\right)~. \label{gtpph1d3}
\eea
Thus, we arrive at the toric diagram in Figure \ref{f:tord3}.
This is in fact the toric diagram of $D_3$ \cite{Hanany:2008fj}.

\begin{figure}[h]
\begin{center}
  \includegraphics[totalheight=3.5cm]{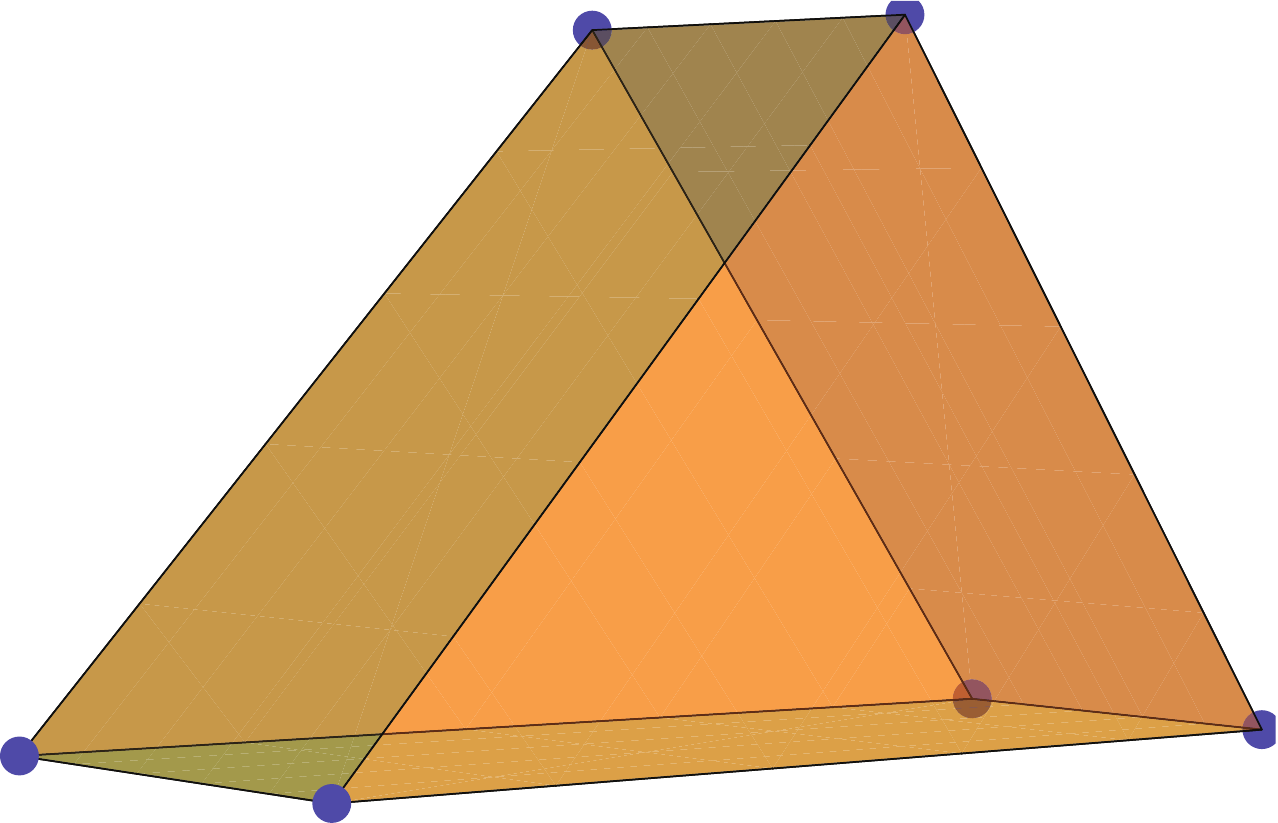}
 \caption{The toric diagram of the $D_3$ theory.}
  \label{f:tord3}
\end{center}
\end{figure}

\end{itemize}

\noindent Note that the toric diagrams constructed from $G_K$ and $G'_t$ are the same up to a transformation
{\footnotesize $\CT =\left( \begin{array}{ccc} -1&0&0\\ 0&-1&0 \\ 0&0&1 \end{array} \right)  \in GL(3, \BZ)$},
where we have $G_K = \CT \cdot G'_t$.


\paragraph{The global symmetry.}  
Since all columns of the $Q_t$ matrix are distinct, the symmetry of the mesonic moduli space is expected to be $U(1)^3 \times U(1)_R$. 
The presence of four mesonic $U(1)$ charges implies that there is a minimisation problem to be solved in order to determine which linear combination of these charges gives the right R-charge in the IR \cite{Hanany:2008fj}.   
Alternatively, we can use the symmetry argument as follows: The 6 perfect matchings are completely symmetric and the requirement of R-charge 2 to the superpotential divides $2$ equally among them, resulting in R-charge of $1/3$ per each.
 From Figure \ref{f:tord3}, there are 6 external points in the toric diagram. 
 From \eref{numberbary}, we thus have $6-4 =2$ baryonic charges, under which the perfect matchings are charged according to the $Q_D$ matrix.  
The global symmetry of this model is the product of mesonic and baryonic symmetries:  $U(1)^3 \times U(1)_R \times U(1)_{B_1} \times U(1)_{B_2}$.  A consistent charge assignment to the perfect matchings for this model is given in Table \ref{chargeph1d3}. 

\begin{table}[h]
 \begin{center}  
  \begin{tabular}{|c||c|c|c|c|c|c|c|}
  \hline
  \;& $U(1)_{1}$&$U(1)_{2}$&$U(1)_{3}$& $U(1)_R$& $U(1)_{B_1}$ & $U(1)_{B_2}$ & fugacity\\
  \hline \hline  
  $p_1$& $1$	& $1$  & 1 	& $1/3$	&1		&$1$ 	& $t q_1 q_2 q_3 b_1 b_2$ \\
  $p_2$& $-1$	& $-1$ & 0 	& $1/3$	&$0$	& 1 		& $t b_2/(q_1 q_2)$ \\
  $p_3$& $1$	& $0$  & $-1 $ 	& $1/3$	&$-1$	&0 		& $t q_1/(b_1 q_3)$ \\
  $p_4$& $-1$	& $0$  & $-1 $ 	& $1/3$	&$1$ 	&0 		& $t b_1/(q_1 q_3)$ \\
  $p_5$& $-1$	& $1$  & 1 	& $1/3$	&$0$	&$-1$ 	& $t q_3 q_2/(q_1 b_2)$\\
  $p_6$& $1$	& $-1$ & 0	& $1/3$	&$-1$	&$-1$	& $t q_1/(q_2 b_1 b_2)$\\
  \hline
  \end{tabular}
  \end{center}
\caption{Charges under the global symmetry of the $D_3$ theory. Here $t$ is the fugacity of R-charge and $q_1,q_2,q_3,b_1, b_2$ are the respectively fugacities of the $U(1)_{1}, U(1)_{2}, U(1)_3, U(1)_{B_1}, U(1)_{B_2}$ charges. }
\label{chargeph1d3}
\end{table}

\begin{figure}[h]
\begin{center}
  \includegraphics[totalheight=3.5cm]{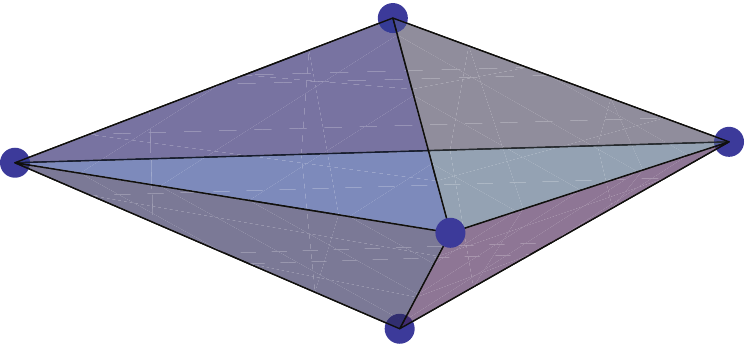}
 \caption{The lattice of generators of the $D_3$ theory.}
  \label{f:latd3}
\end{center}
\end{figure}

\paragraph{The Hilbert series.} Since the $Q_F$ matrix is zero, the Master space is simply 
\bea
\f_{\sD_2 \sC} = \BC^6~.  
\eea
The Hilbert series is given by
\bea
\gf_1(t,q_1,q_2,q_3,b_1,b_2; \sD_2 \sC) &=& \frac{1}{\left(1-t q_1 q_2 q_3 b_1 b_2\right)\left(1- \frac{t b_2}{q_1 q_2} \right)\left(1- \frac{t q_1}{b_1 q_3}\right)}  \times \nn \\
&&  \times \frac{1}{\left(1- \frac{t b_1}{q_1 q_3}\right)\left(1- \frac{t q_3 q_2}{q_1 b_2}\right)\left(1- \frac{t q_1}{q_2 b_1 b_2} \right)}~. \qquad
\label{masterd2c}
\eea
From \eref{quoteFD}, the mesonic moduli space is given by
\bea
\CMm_{\sD_2 \sC} =  \BC^{6}//Q_D ~, 
\eea 
Therefore, we can obtain the Hilbert series of the mesonic moduli space by integrating (\ref{masterd2c}) over the two baryonic fugacities $b_1$ and $b_2$:
\bea
\gm_1 (t,q_1,q_2,q_3; \sD_2 \sC) &=& \frac{1}{(2 \pi i)^2} \oint_{|b_1|=1} \frac{\ud b_1}{b_1} \oint_{|b_2|=1} \frac{\ud b_2}{b_2} ~ \gf_1(t,q_1,q_2,q_3,b_1,b_2; \sD_2 \sC) \nn \\
&=& \frac{1-t^6}{\left(1- \frac{t^2}{q_3^2}\right)\left(1-\frac{q_3 t^2}{q_1^2}\right)\left(1-q_1^2 q_3 t^2\right)\left(1-\frac{t^3}{q_1q_2^2 q_3}\right)\left(1-q_1 q_2^2 q_3t^3\right)}~. \qquad
\label{hsmesd3ph1}
\eea 
The unrefined Hilbert series is given by
\bea
\gm_1 (t,1,1,1;\sD_2 \sC) = \frac{1-t^6}{(1-t^3)^2(1-t^2)^3} = \frac{1+t^3}{(1-t^3)(1-t^2)^3}~.
\eea 
Since the pole at $t=1$ is of order 4 and the numerator is palindromic, it follows that the mesonic moduli space is a Calabi--Yau 4-fold which, in the literature, is usually referred to as $D_3$.  The plethystic logarithm of the mesonic Hilbert series is
\bea
\PL [\gm_1 (t,q_1,q_2,q_3;\sD_2 \sC)] =  \frac{t^2}{q_3^2}+\frac{q_3 t^2}{q_1^2}+q_1^2 q_3 t^2+\frac{t^3}{q_1q_2^2 q_3}+q_1 q_2^2 q_3t^3-t^6~. \label{plph1d3}
\label{plmesd3ph1}
\eea
Therefore, we see that the mesonic moduli space of this phase is a complete intersection generated by 
\bea
&& X_{23}X_{14} = p_1 p_6~, \quad X_{42}X_{31} = p_2 p_5~, \quad X_{12}X_{21} = p_3 p_4~, \nn \\ 
&& X_{23}X_{12}X_{31} = p_1 p_3 p_5~, \quad X_{42}X_{21}X_{14} = p_2 p_4 p_6~, \label{genph1d3}
\eea
subject to the relation
\bea
\left(X_{23}X_{14}\right)\left(X_{42}X_{31}\right)\left(X_{12}X_{21} \right) &=& \left(X_{23}X_{12}X_{31}\right)\left(X_{42}X_{21}X_{14}\right)~.
\eea
We can represent these generators \eref{genph1d3} in a lattice (Figure \ref{f:latd3}) by plotting the powers of the weights of the characters in \eref{plph1d3}.  Note that the lattice of generators is the dual of the toric diagram (nodes are dual to faces and edges are dual to edges): The toric diagram has 6 nodes, 9 edges and 5 faces, whereas the generators form a convex polytope that has 5 nodes, 9 edges and 6 faces.

\subsection{Phase II: The Two-Hexagon with One-Diagonal Model}
\begin{figure}[ht]
\begin{center}
  \vskip -0.5cm
  \hskip -7cm
  \includegraphics[totalheight=6.2cm]{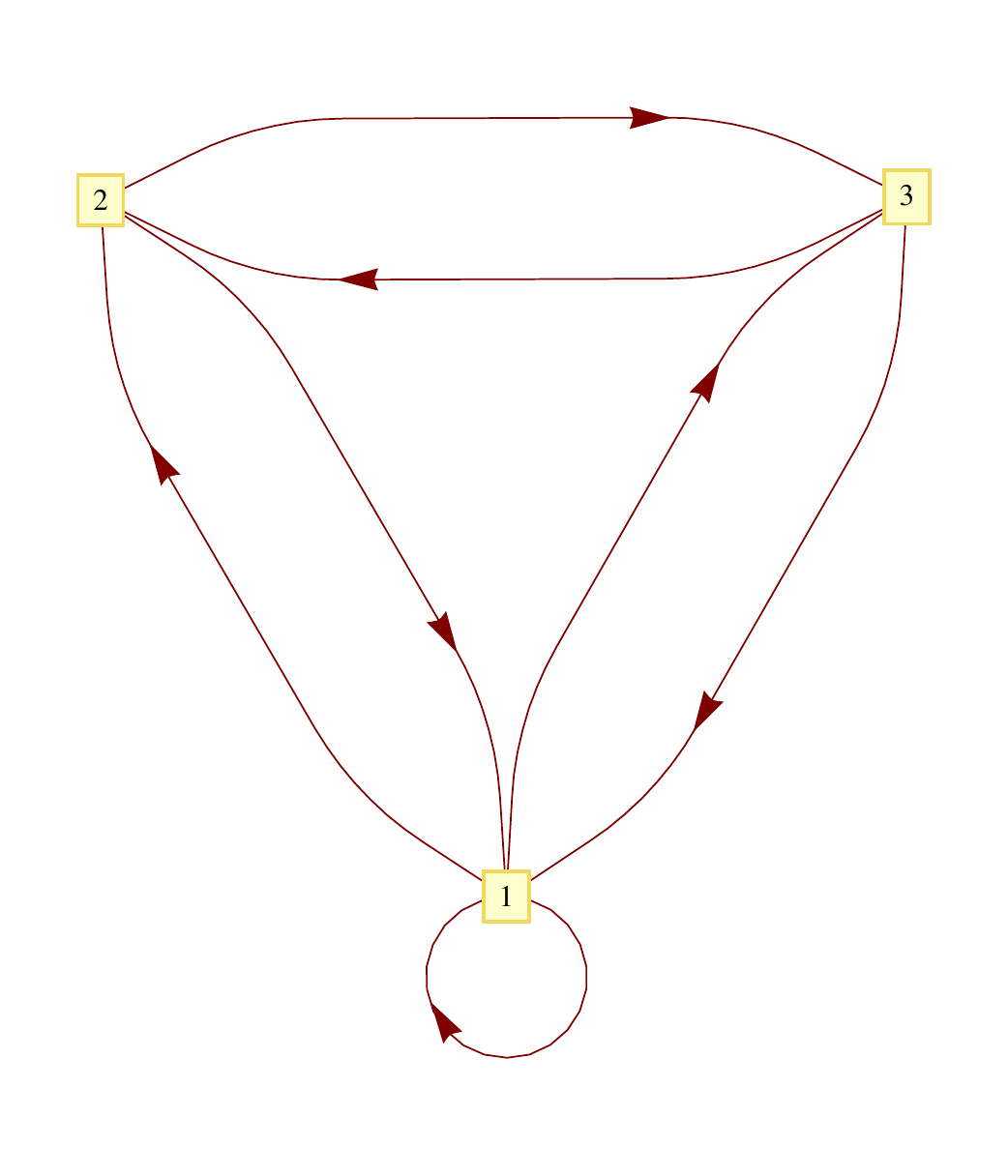}
    \vskip -5.5cm
  \hskip 8cm
  \includegraphics[totalheight=5.5cm]{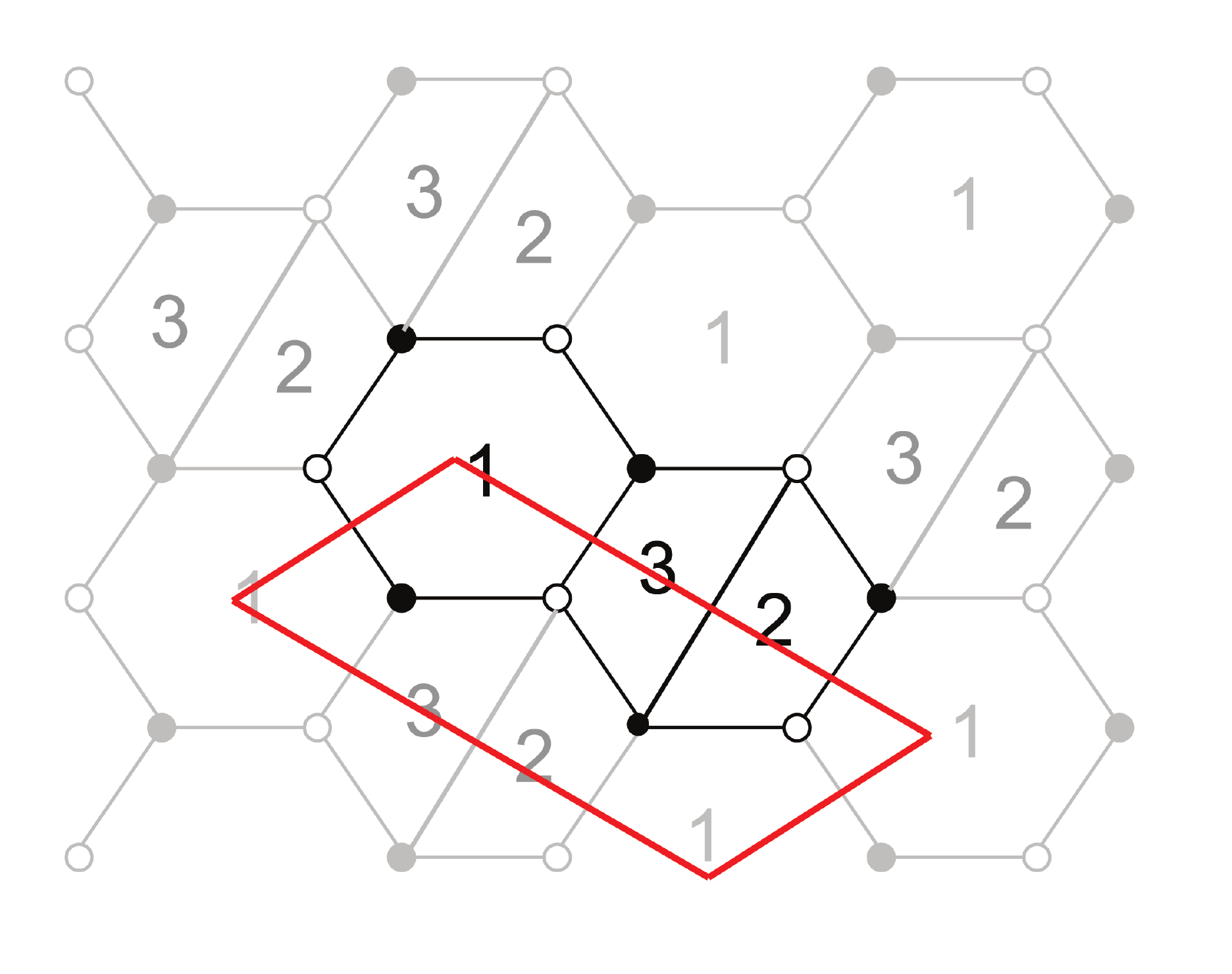}
 \caption{[Phase II of the $D_3$ theory]  (i) Quiver diagram for the $\sH_{2} \partial_1$ model.\ (ii) Tiling of the $\sH_{2} \partial_1$ model.}
  \label{f:phase2D3}
\end{center}
\end{figure}

The quiver diagram and tiling of this model (which we shall refer to as $\sH_{2} \partial_1$) are discussed in this context in \cite{Hanany:2008cd, Hanany:2008fj} and are given in Figure \ref{f:phase2D3}. Note that in 3+1 dimensions this tiling corresponds to the SPP model.
The superpotential is given by
\bea
W = \tr \left( X_{32}X_{23}X_{31}X_{13} - X_{23}X_{32}X_{21}X_{12}  - \phi_1 \left(X_{13}X_{31} - X_{12}X_{21}\right) \right) \ .
\eea
We choose the CS levels to be $k_1= 1,~k_2= -1,~k_3 = 0$.

\paragraph{The Master space.} From the superpotential, we find that the Master space is a reducible variety
$\f_{\sH_{2} \partial_1} = \firr{\sH_{2} \partial_1}~  \cup~L_{\sH_{2} \partial_1}$~, where
\bea
\firr{\sH_{2} \partial_1} &=& \mathbb{V} (X_{21}X_{12} - X_{31}X_{13}, \phi_1 - X_{23}X_{32})~, \nn \\
L_{\sH_{2} \partial_1} &=& \mathbb{V} (X_{13}, X_{31}, X_{12}, X_{21})~.
\eea 
We see that the coherent component is 
\bea
\firr{\sH_{2} \partial_1} = \CC \times \BC^2~, 
\eea
where the $\BC^2$ is parametrised by the fields $\{ \phi_1, X_{23}, X_{32} \}$ with the relation $\phi_1 = X_{23}X_{32}$ and the conifold singularity $\CC$ is described by the fields $\{ X_{12}, X_{21}, X_{13}, X_{31} \}$ with the relation $X_{21}X_{12} = X_{31}X_{13}$.
The linear component $L_{\sH_{2} \partial_1} = \BC^3$ is parametrised by the fields $\{ \phi_1, X_{23}, X_{32} \}$.  The intersection between these two components is
\bea
\firr{\sH_{2} \partial_1}~  \cap~L_{\sH_{2} \partial_1} = \BC^2~.
\eea

\begin{figure}[ht]
\begin{center}
   \includegraphics[totalheight=6.0cm]{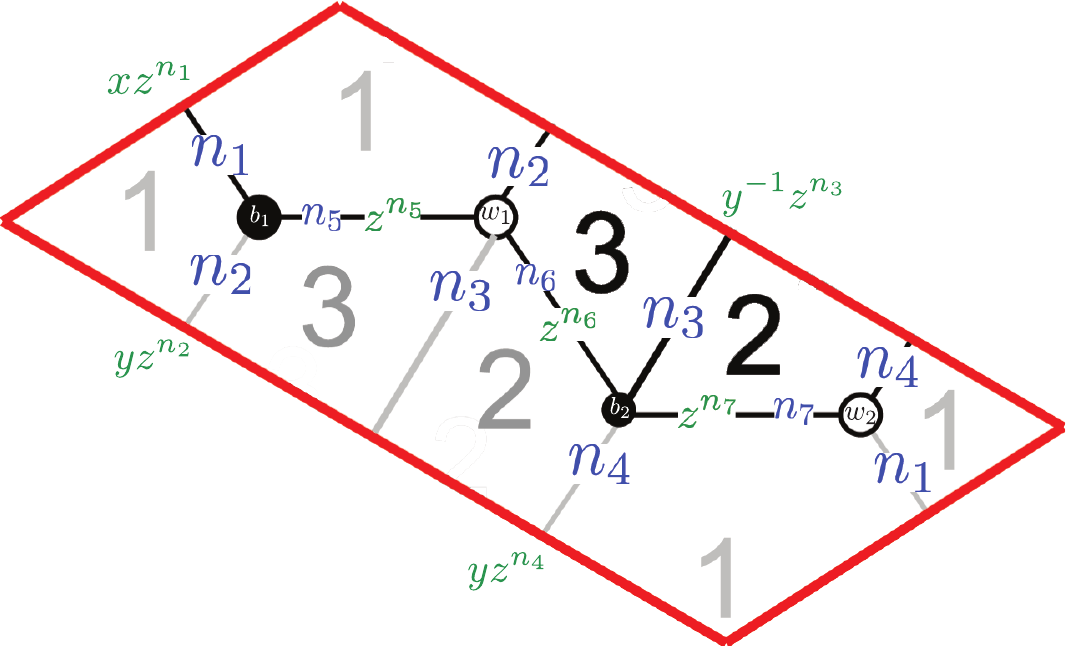}
 \caption{[Phase II of the $D_3$ theory]  The fundamental domain of tiling for the $\sH_{2} \partial_1$ model: Assignments of the integers $n_i$ to the edges are shown in blue and the weights for these edges are shown in green.}
  \label{f:fdphase2d3}
\end{center}
\end{figure}

\paragraph{The Kasteleyn matrix.}
We assign the integers $n_i$ to the edges according to Figure \ref{f:fdphase2d3}.  From \eref{kn}, we find that 
\bea
\text{Gauge group 1~:} \qquad k_1 &=& 1  = n_2 - n_4 - n_5 + n_7  ~, \nn \\
\text{Gauge group 2~:} \qquad k_2 &=& -1 =  n_3 + n_4 - n_6  - n_7 ~,  \nn \\
\text{Gauge group 3~:} \qquad k_3 &=& 0 =  - n_2 - n_3  + n_5 + n_6  ~.
\eea  
We choose
\bea
n_4= -1,\quad n_i=0 \; \text{otherwise}~.
\eea
The Kasteleyn matrix for this theory is
\be
K =   \left(
\begin{array}{c|cc}
& w_1 & w_2 \\
\hline
b_1& X_{13} y z^{n_2}+ X_{31} z^{n_5}  &   \  \phi_1 x z^{n_1}  \\
b_2 & X_{32} z^{n_6}+ X_{23} y^{-1}  z^{n_3}  & \  X_{21} y z^{n_4}+ X_{12} z^{n_7} 
\end{array}
\right)~.
\ee
The permanent of this matrix is given by
\bea  \label{permKph2d3}
\perm~ K &=& X_{31} X_{12} z^{n_5+n_7} + X_{13} X_{12} y z^{n_2+n_7} + X_{31} X_{21} y z^{n_4+n_5} +  \nn \\
&& + X_{13} X_{21} y^{2} z^{n_2+n_4} + \phi_1 X_{32} x z^{n_1+n_6} + \phi_1 X_{23} x y^{-1} z^{n_1+n_3} \nn \\
&=& X_{31} X_{12}  + X_{13} X_{12} y  + X_{31} X_{21} y z^{-1}  + X_{13} X_{21} y^{2} z^{-1}  + \phi_1 X_{32} x  + \nn\\
&& + \phi_1 X_{23} x y^{-1}  \qquad \text{(for $n_4= -1$ and $n_i= 0$ otherwise)} \  .
\eea

\paragraph{The perfect matchings.} From the permanent of the Kasteleyn matrix, we can write the perfect matchings as collections of fields as follows:
\bea
&& p_1 = \{X_{31}, X_{12}\},~ p_2 =  \{ X_{21}, X_{13} \} ,~p_3 = \{X_{23}, \phi_1 \}, \nn \\
&& p_4 = \{ X_{32},  \phi_1 \},~p_5 = \{X_{31} , X_{21} \},~ p_6 = \{ X_{12}, X_{13} \} \ . 
\eea
In turn, we find the parameterisation of fields in terms of perfect matchings: 
\bea
&& X_{31} = p_1 p_5, \quad X_{12} = p_1 p_6, \quad X_{21} = p_2 p_5, \nn \\
&& X_{13} = p_2 p_6, \quad X_{23} = p_3, \quad \phi_1 = p_3 p_4, \quad X_{32} = p_4~. \label{quivfieldph2d3}
\eea
This is summarised in the perfect matching matrix:
\beq
P=\left(\begin{array} {c|cccccc}
  \;& p_1&p_2&p_3&p_4&p_5&p_6\\
  \hline
  X_{31}&1&0&0&0&1&0 \\
  X_{12}&1&0&0&0&0&1\\
    X_{21}&0&1&0&0&1&0\\
  X_{13}&0&1&0&0&0&1\\
  X_{23}&0&0&1&0&0&0\\
    \phi_{1}&0&0&1&1&0&0 \\
  X_{32}&0&0&0&1&0&0
 \end{array}
\right).
\eeq
Basis vectors of the null space of $P$ are given in the rows of the charge matrix:
\bea
Q_F = (1,1,0, 0 -1,-1)~. \label{qfph2d3}
\eea
Hence, from \eref{relpm}, we see that the relations between the perfect matchings are given by
\bea
p_1+p_2 - p_5 - p_6 = 0 \ . \label{relph2d3}
\eea
Since the coherent component of the Master space is generated by the perfect matchings (subject to the relation \eref{relph2d3}), it follows that 
\bea
\firr{\sH_{2} \partial_1} = \BC^6//Q_F = \BC^6//(1,1,0, 0 -1,-1)~.  \label{quot2a}
\eea
Since the quotient $\BC^4//(1,1,-1,-1)$ is known to be conifold ($\CC$) and $\BC^2$ is parametrised by the remaining perfect matchings with charge 0, it follows that
\bea
\firr{\sH_{2} \partial_1}  = \CC \times \BC^2~.
\eea

\paragraph{The toric diagram.} We demonstrate two methods of constructing the toric diagram. 
\begin{itemize}
\item{\bf The charge matrices.}  
Since the number of gauge groups is $G=3$, there is $G-2 = 1$ baryonic charge, which we shall denote as $U(1)_{B_1}$, coming from the D-terms.  We collect the $U(1)_{B_1}$ charges of the perfect matchings in the $Q_D$ matrix:
\bea
Q_D = (1, 0, -1, 1, 0, -1)~.  \label{qdph2d3}
\eea 
Note that since the CS coefficient $k_3=0$, the $Q_D$ matrix \eref{qdph2d3} has been chosen such that the baryonic charge of each quiver field in \eref{quivfieldph2d3} coincides with the quiver charge under gauge group 3.
From \eref{qfph2d3} and \eref{qdph2d3}, the total charge matrix is given by
\bea
Q_t = \left(
\begin{array}{cccccc}
1& 0 & -1 & 1 & 0 & -1 \\
 1 & 1 & 0 & 0 & -1 & -1 
\end{array}
\right)~. 
\eea
Note that this is precisely the same as the $Q_t$ matrix \eref{qtph1d3} for Phase I. 
We thus obtain the same matrix $G'_t$ as for Phase I \eref{gtpph1d3}.
The toric diagram is therefore given by Figure \ref{f:tord3}.   

\item {\bf The Kasteleyn matrix.}
The powers of $x, y, z$ in each term of \eref{permKph2d3} give the coordinates of each point in the toric diagram.
We collect these points in the columns of the following $G_K$ matrix:
\bea
G_K = \left(
\begin{array}{cccccc}
 1 & 0 & 0 & 0 & 1 & 0 \\
 -1 & 1 & 1 & 2 & 0 & 0 \\
 0 & 0 & -1 & -1 & 0 & 0
\end{array}
\right)~.
\eea
Note that the toric diagrams constructed from $G_K$ and $G'_t$ are the same up to a transformation
{\footnotesize $\CT =\left( \begin{array}{ccc} 1&0&0\\-1&1&1\\0&-1&0 \end{array} \right)  \in GL(3, \BZ)$},
where we have $G_K = \CT \cdot G'_t$.
\end{itemize}

\paragraph{The baryonic charges.} 
From Figure \ref{f:tord3}, there are 6 external points in the toric diagram. From \eref{numberbary}, we thus have $6-4 =2$ baryonic charges.  One of them comes from the D-terms (as discussed above) and the other arises from the $Q_F$ matrix.  Let us donote the latter by $U(1)_{B_2}$. 

\paragraph{The global symmetry.}  Since all columns of the $Q_t$ matrix are distinct, the symmetry of the mesonic moduli space is expected to be $U(1)^3 \times U(1)_R$.  It was shown in \cite{Hanany:2008fj} that each perfect matching has an R-charge $1/3$. As discussed above, there are two baryonic charges $U(1)_{B_1}$ and $U(1)_{B_2}$.  The global symmetry of this model is the product of mesonic and baryonic symmetries: $U(1)^3 \times U(1)_R \times U(1)_{B_1} \times U(1)_{B_2}$.  The $U(1)_{B_1}$ and $U(1)_{B_2}$ charges of the perfect matchings can be read off respectively from the $Q_D$ and $Q_F$ matrices.  We present a consistent assignment of the charges in Table \ref{chargeph1d3}.

\paragraph{The Hilbert series.} From \eref{quot2a}, the Hilbert series of the coherent component can be obtained by integrating the $\BC^6$ Hilbert series over the baryonic fugacity $b_2$ corresponding to the $U(1)_{B_2}$ charge:
\bea
g^{\firr{}}_1 (t,q_1,q_2,q_3,b_1;\sH_{2} \partial_1) &=&  \frac{1}{2\pi i} \oint_{|b_2|=1} \frac{\ud b_2}{ b_2} ~  \frac{1}{\left(1-t q_1 q_2 q_3 b_1 b_2\right)\left(1- \frac{t b_2}{q_1 q_2} \right)\left(1- \frac{t q_1}{b_1 q_3}\right)}  \times \nn \\
&&  \times \frac{1}{\left(1- \frac{t b_1}{q_1 q_3}\right)\left(1- \frac{t q_3 q_2}{q_1 b_2}\right)\left(1- \frac{t q_1}{q_2 b_1 b_2} \right)} \nn\\
&=& \frac{\left(1-t^4 q_3^2 \right)}{\left(1-\frac{t^2}{b_1 q_2^2}\right) \left(1-\frac{t^2 q_3}{q_1^2}\right) \left(1- \frac{t q_1}{b_1 q_3}\right) \left(1-\frac{t b_1}{q_1 q_3}\right) \left(1-t^2 q_1^2 q_3\right) \left(1-t^2 b_1 q_2^2 q_3^2\right)}~. \label{hsmasterh2d1} \nn \\  
\eea
The unrefined Hilbert series is 
\bea
g^{\firr{}}_1 (t,1,1,1,1;\sH_{2} \partial_1) =  \frac{1-t^4}{(1-t^2)^4} \times \frac{1}{(1-t)^2} = \frac{1+t^2}{(1-t)^2 (1-t^2)^3} ~. 
\eea
Note that this is the Hilbert series of $\CC \times \BC^2$ and the space $\firr{\sH_{2} \partial_1}$ is 5 dimensional (which is the order of the pole at $t=1$).  Integrating \eref{hsmasterh2d1} over the baryonic fugacity $b_1$, we obtain the same result the mesonic Hilbert series \eref{hsmesd3ph1} for Phase I.  
Therefore, the plethystic logarithm is given by \eref{plmesd3ph1}.
We see that mesonic moduli space is a complete intersection and is generated by
\bea
&& X_{12} = p_1 p_6~, \quad X_{21} = p_2 p_5~, \quad \phi_1 = p_3 p_4~, \nn \\
&& X_{23}X_{31} = p_1 p_3 p_5~, \quad X_{13}X_{32} = p_2p_4p_6~,
\eea
Note that we require gauge invariance with respect to the gauge group 3, and so the indices corresponding to the gauge group 3 are contracted.
Among these generators, there is a relation:
\bea
(X_{23}X_{31})(X_{13}X_{32}) = X_{12}X_{21} \phi_1~.
\eea
Note that, in terms of the perfect matchings, the generators of this model are precisely the same as those of Phase I.

\subsection{Phase III: The Three Double-Bonded One-Hexagon Model}
\begin{figure}[h]
\begin{center}
  \hskip -7cm
  \includegraphics[totalheight=5cm]{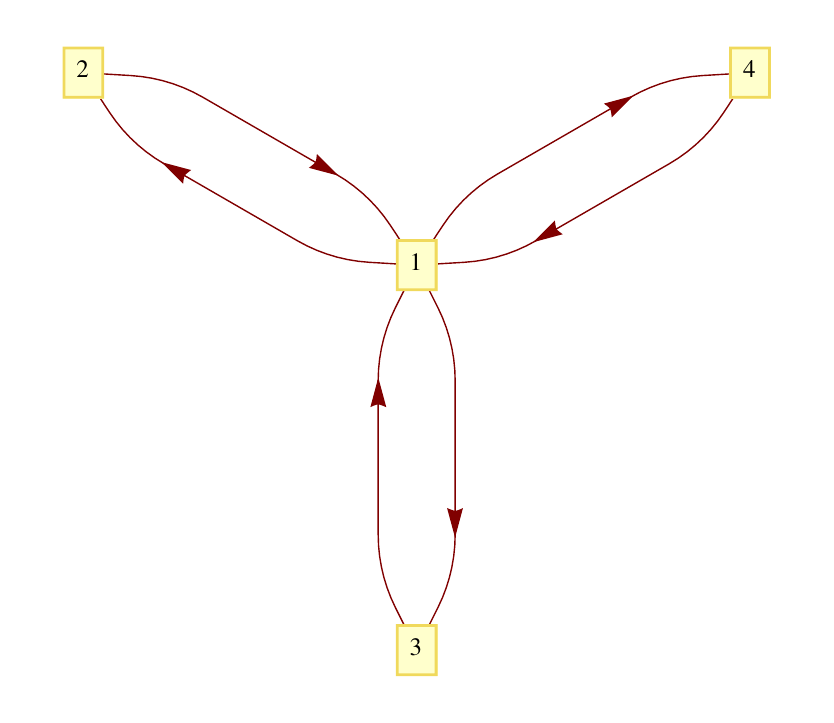}
    \vskip -5.0cm
  \hskip 8.4cm
  \includegraphics[totalheight=5cm]{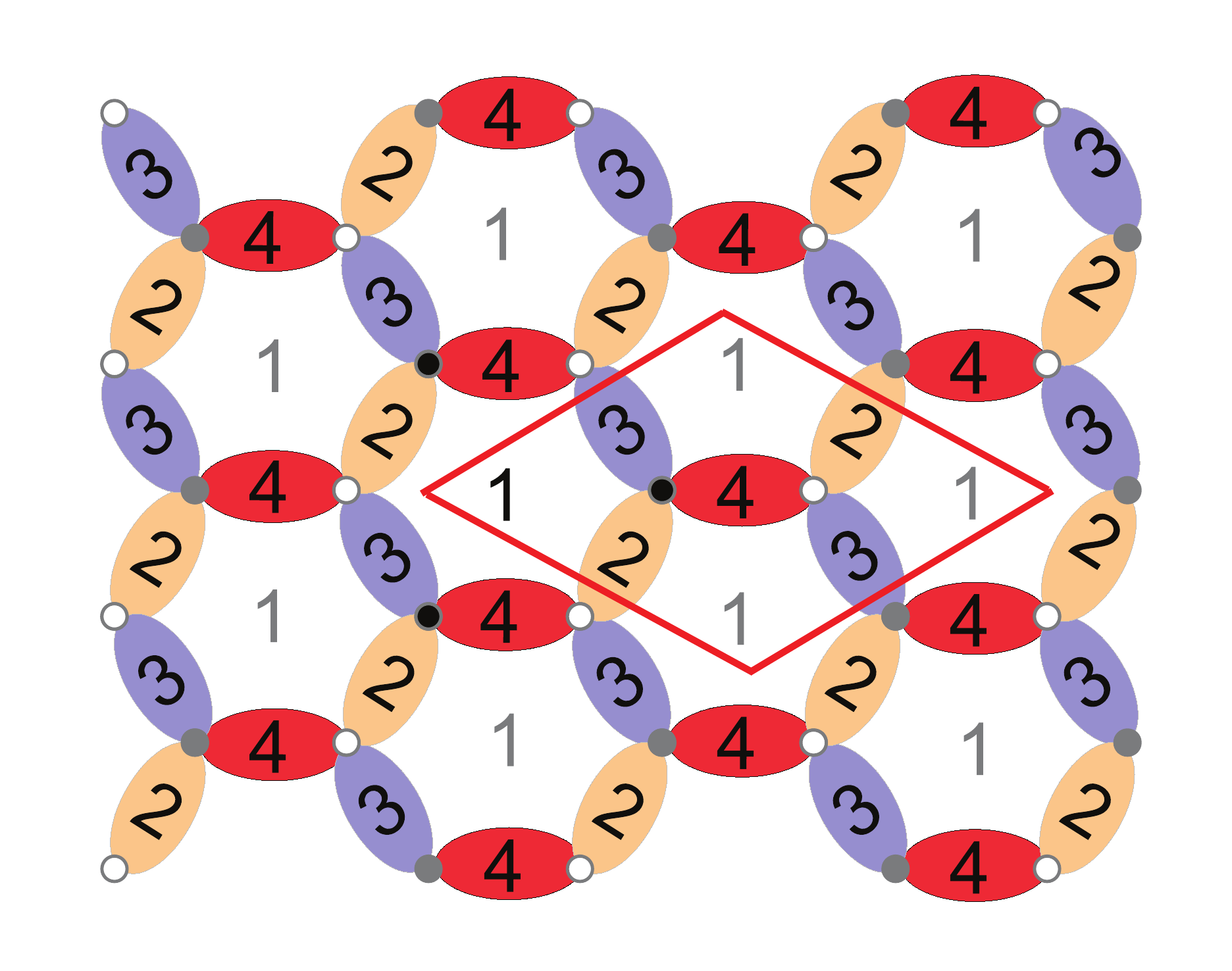}
 \caption{[Phase III of the $D_3$ theory]  (i) Quiver diagram of the $\sD_3\sH_1$ model.\qquad (ii) Tiling of the $\sD_3\sH_1$ model.}
  \label{f:ph3d3}
\end{center}
\end{figure}

This model (which we shall refer to as $\sD_3\sH_1$) was first introduced in \cite{taxonomy} as part of a classification procedure for all models that have 2 terms in the superpotential.  Its quiver diagram and tiling of this model are drawn in Figure \ref{f:ph3d3}.
The superpotential of this model is given by
\bea
W = \tr \left( X_{13}X_{31}X_{14}X_{41}X_{12}X_{21} - X_{14}X_{41}X_{13}X_{31}X_{12}X_{21} \right)~.
\eea
We choose the CS levels to be $(k_1,k_2,k_3,k_4) = (1,-1,1,-1)$.

\begin{figure}[h]
\begin{center}
   \includegraphics[totalheight=6.0cm]{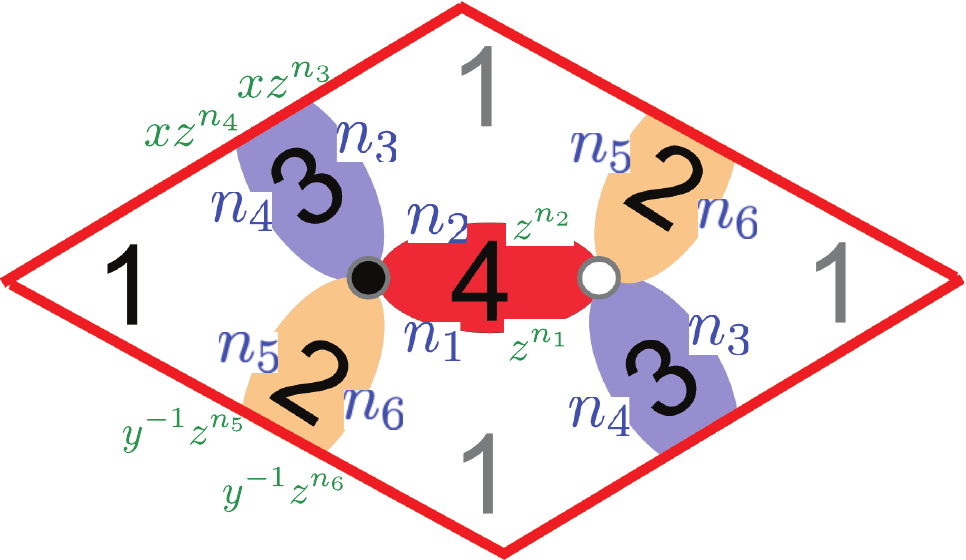}
 \caption{[Phase III of the $D_3$ theory]  The fundamental domain of tiling for the $\sD_3\sH_1$ model: Assignments of the integers $n_i$ to the edges are shown in blue and the weights for these edges are shown in green.}
  \label{f:fdph3d3}
\end{center}
\end{figure}

\paragraph{The toric diagram.} We demonstrate two methods of constructing the toric diagram. 
\begin{itemize}
\item {\bf The Kasteleyn matrix.}
We assign the integers $n_i$ to the edges according to Figure \ref{f:fdph3d3}.  From \eref{kn}, we find that 
\bea
\text{Gauge group 1~:} \qquad k_1 &=& 1  = n_1 - n_2 + n_3 - n_4 + n_5 - n_6 ~, \nn \\
\text{Gauge group 2~:} \qquad k_2 &=& -1 = - n_5 +n_6  ~, \nn \\
\text{Gauge group 3~:} \qquad k_3 &=& 1 = - n_3 + n_4 ~, \nn \\
\text{Gauge group 4~:} \qquad k_4 &=& -1 = -n_1 + n_2~.
\eea  
We choose
\bea
n_1 = n_4 = n_5 = 1,\quad n_i=0 \; \text{otherwise}~.
\eea
Since the fundamental domain contains only one white node and one black node, the Kasteleyn matrix is $1\times 1$ and, therefore, coincides with its permanent:
\bea \label{Kph3d3}
K &=& X_{14}z^{n_1} + X_{41}z^{n_2} +  X_{13} x z^{n_3} + X_{31} x z^{n_4} + X_{12} y^{-1} z^{n_5} + X_{21} y^{-1} z^{n_6} \nn \\
&=& X_{14}z + X_{41} +  X_{13} x  + X_{31} x z + X_{12} y^{-1} z + X_{21} y^{-1}  \nn \\ 
&& \text{(for $n_1 = n_4 = n_5 = 1$ and $n_i=0$)}~,
\eea
where the powers of $x, y, z$ in each term give the coordinates of each point in the toric diagram. 
We collect these points in the columns of the following $G_K$ matrix:
\bea
G_K = \left(
\begin{array}{cccccc}
 1 & 0 & 0 & 0 & 1 & 0 \\
 0 & 0 & -1 & -1 & 0 & 0 \\
 0 & 1 & 0 & 1 & 1 & 0
\end{array}
\right)~.
\eea

\item{\bf The charge matrices.}  From \eref{Kph3d3}, we can take the perfect matchings to be
\bea
p_1 = X_{41}, \quad p_2 = X_{12}, \quad p_3 = X_{13}, \quad p_4 = X_{31}, \quad p_5 = X_{14}, \quad p_6 = X_{21}~.
\eea
Since there is a one-to-one correspondence between the perfect matchings and the quiver fields, it follows that
\bea
Q_F = 0~.
\eea
Since the number of gauge groups is $G=4$, there are $G-2 = 2$ baryonic charges coming from the D-terms.  From \eref{QD}, we find that the $Q_D$ matrix is given by
\bea
Q_D = \left(
\begin{array}{cccccc}
 1 & 0 & -1 & 1 & 0 & -1 \\
 1 & 1 & 0 & 0 & -1 & -1
\end{array}
\right)~.
\eea
The total charge matrix $Q_t$ therefore coincides with $Q_D$:
\bea
Q_t = \left(
\begin{array}{cccccc}
 1 & 0 & -1 & 1 & 0 & -1 \\
 1 & 1 & 0 & 0 & -1 & -1
\end{array}
\right)~,
\eea 
Note that this is exactly the same as the $Q_t$ matrix \eref{qtph1d3} for Phase I.  Hence, the $G'_t$ matrix coincides with that of Phase I \eref{gtpph1d3}.
Thus, we arrive at the toric diagram in Figure \ref{f:tord3}. In this way, we have shown that the mesonic moduli space is indeed $D_3$.
\end{itemize}
\noindent Note that the toric diagrams constructed from $G_K$ and $G'_t$ are the same up to a transformation
{\footnotesize $\CT =\left( \begin{array}{ccc} 1&0&0\\ 0&-1&0 \\ 0&0&1 \end{array} \right)  \in GL(3, \BZ)$},
where we have $G_K = \CT \cdot G'_t$.

\paragraph{The global symmetry.}  Since all columns of the $Q_t$ matrix are distinct, the symmetry of the mesonic moduli space is expected to be $U(1)^3 \times U(1)_R$.   
The 6 perfect matchings are completely symmetric and the requirement of R-charge 2 to the superpotential divides $2$ equally among them, resulting in R-charge of $1/3$ per each.
As discussed above, there are two baryonic charges $U(1)_{B_1}$ and $U(1)_{B_2}$, under which the perfect matchings are charged according to the $Q_D$ matrix.  Thus, the global symmetry of this model is expected to be $U(1)^3 \times U(1)_R \times U(1)_{B_1} \times U(1)_{B_2}$, which is the same as in Phases I and II.  We emphasise that both baryonic charges arise from the $Q_D$ matrix, as for Phase II.  A consistent charge assignment to the perfect matchings for this model is given in Table \ref{chargeph1d3}. 

\paragraph{The Hilbert series.} Since the $Q_F$ matrix is zero, the Master space is simply 
\bea
\f_{\sD_3\sH_1} = \BC^6~.  
\eea
From \eref{quoteFD}, the mesonic moduli space is given by
\bea
\CMm_{\sD_3\sH_1} =  \BC^{6}//Q_D ~, 
\eea 
which is the same as Phase I, as expected.  Therefore, the Master space Hilbert series and the mesonic Hilbert series are the same as those of Phase I, and are given respectively by \eref{masterd2c} and \eref{hsmesd3ph1}.
Therefore, we see that the mesonic moduli space of this phase is a complete intersection generated by 
\bea
&& X_{41}X_{21} = p_1p_6~, \quad X_{12}X_{14} = p_2 p_5~, \quad X_{31}X_{13} = p_3p_4~, \nn \\
&& X_{13}X_{41}X_{14} = p_1p_3p_5~, \quad X_{31}X_{12}X_{21} = p_2p_4p_6~,
\eea
subject to the relation
\bea
\left(X_{41}X_{21} \right)\left(X_{12}X_{14}\right)\left(X_{31}X_{13}\right) &=& \left(X_{41}X_{13}X_{14}\right)\left(X_{12}X_{31}X_{21}\right)~.
\eea
Note that, in terms of the perfect matchings, the generators of this model are precisely the same as those of Phase I.

\subsection{A Comparison between Phases of the $D_3$ Theory}
Here we make a comparison between phases of the $D_3$ theory:
\begin{itemize}
\item {\bf Perfect matchings.} The perfect matchings of different phases are exactly the same (including the labels).  They are charged in the same way under the global symmetry according to Table \ref{chargeph1d3}.

\item {\bf Generators.}  In terms of the perfect matchings, the generators of different phases are precisely the same.  These are summarised in Table \ref{t:comgend3}.

\begin{table}[h]
 \begin{center}
  \begin{tabular}{|c||c|c|c|}
    \hline
    Perfect matchings & Generator of Phase I & Generator of Phase II & Generator of Phase III \\
    \hline
     $p_1 p_6$ & $X_{23}X_{14}$ 		& $X_{12}$ & $X_{41}X_{21}$ \\
     $p_2 p_5$ &  $X_{42}X_{31}$ 		& $X_{21}$  &$X_{12}X_{14}$ \\
     $p_3 p_4$ & $X_{12}X_{21}$ 		& $\phi_1$ &$X_{31}X_{13}$  \\
     $p_1 p_3  p_5$ & $X_{23}X_{12}X_{31}$ & $X_{23}X_{31}$  &$X_{13}X_{41}X_{14}$ \\
     $p_2 p_4 p_6$ & $X_{42} X_{21} X_{14}$ & $X_{13}X_{32}$& $X_{31} X_{12}X_{21}$ \\
     \hline
  \end{tabular}
  \end{center}
 \caption{A comparison between the generators of different phases of the $D_3$ theory.  In terms of the perfect matchings, the generators of different phases are precisely the same.  In Phase II, we require gauge invariance with respect to the gauge group 3, and so the indices corresponding to the gauge group 3 are contracted.}
\label{t:comgend3}
\end{table}

\item {\bf Quiver fields.} The quiver fields of Phases I and III are the perfect matchings, whereas some of the quiver fields of Phase II are bilinear and some are linear in perfect matchings.

\item {\bf Mesonic moduli space.} The mesonic moduli spaces of all phases are identical; they are $D_3$.

\item {\bf Baryonic symmetries.} The baryonic symmetries of all phases are identical.  However, not all of them have the same origin. The baryonic symmetries  for Phases I and III arises from the D-terms, whereas the baryonic symmetry for Phase II arises from the relation between perfect matchings as well as the D-terms (induced by one node of the quiver).

\item {\bf Master space \& space of perfect matchings.} The Master spaces of Phases I and III and the space of perfect matchings of Phase II are identical; they are $\BC^6$.  However, the Master spaces of Phases I and III are combined baryonic and mesonic moduli spaces, whereas the space of perfect matchings of Phase II is a combination of partial baryonic moduli space and the mesonic moduli space.
\end{itemize}

\section{Phases of the  $Q^{1,1,1} / \BZ_2$ Theory}
This theory was introduced in \cite{Hanany:2008cd, Hanany:2008fj} as a modified $\BF_0$ theory. In the following subsections, we examine two phases of this theory in details.

\subsection{Phase I: The Four-Square Model}
\begin{figure}[ht]
\begin{center}
  \hskip -7cm
  \includegraphics[totalheight=6.2cm]{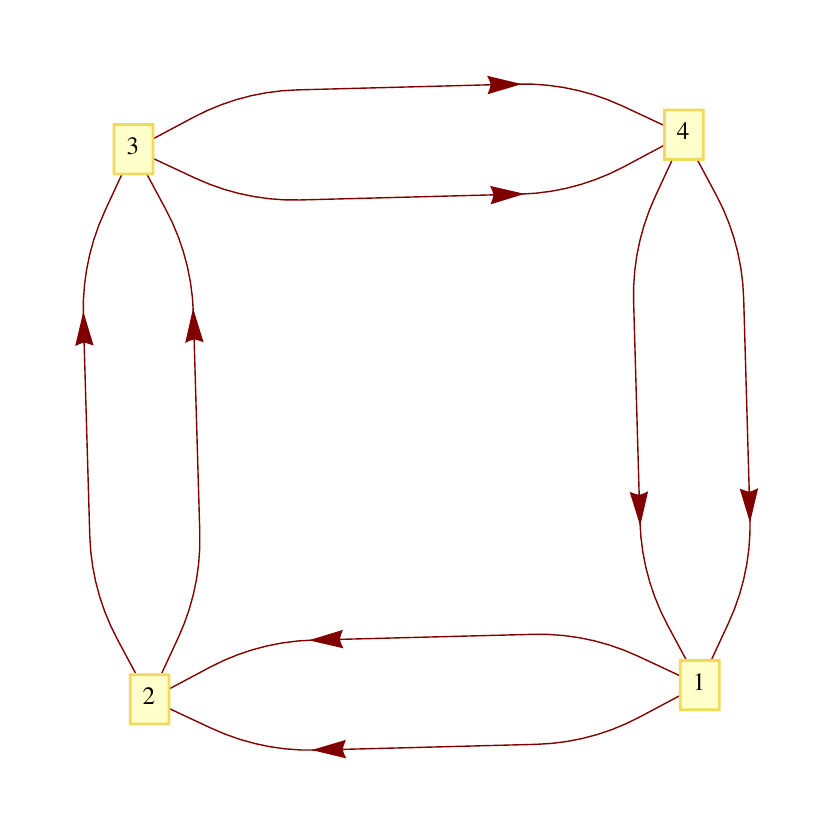}
    \vskip -5.5cm
  \hskip 8.9cm
  \includegraphics[totalheight=5cm]{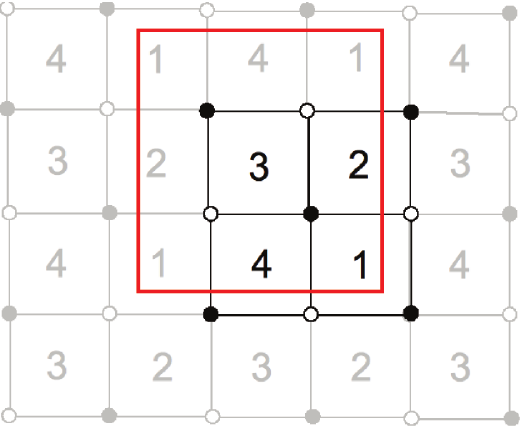}
 \caption{[Phase I of $Q^{1,1,1}/\BZ_2$]  (i) Quiver diagram for the $\mathscr{S}_4$ model. \ (ii) Tiling for the $\mathscr{S}_4$ model.}
  \label{f:phase1f0}
\end{center}
\end{figure}

This model (which we shall refer to as $\mathscr{S}_4$) has 4 gauge groups and bi-fundamental fields $X_{12}^i$, $X_{23}^i$, $X_{34}^i$ and $X_{41}^i$ (with $i=1,2$). The superpotential is given by
\bea
W = \epsilon_{ij} \epsilon_{pq} \tr(X_{12}^i X_{23}^p X_{34}^j X_{41}^q)~.
\eea
The quiver diagram and tiling are drawn in Figure \ref{f:phase1f0}.  Note that in 3+1 dimensions, these correspond to Phase I of the $\BF_0$ theory \cite{ Forcella:2008ng, master, Butti:2007jv}.
We choose the CS levels to be $k_1 = -k_2 = -k_3 = k_4=1$.

\paragraph{The Master space.}  A primary decomposition indicates that the Master space of this phase is a reducible variety and has 3 irreducible components \cite{master, Forcella:2008ng}: 
\bea
\f_{\mathscr{S}_4} = \firr{\mathscr{S}_4}~  \cup~L^1_{\mathscr{S}_4}~  \cup~L^2_{\mathscr{S}_4}~,
\eea
where
\bea
\firr{\mathscr{S}_4} &=& \mathbb{V} (X^1_{41}X^2_{23}-X^2_{41}X^1_{23} , X^1_{34}X^2_{12} - X^2_{34}X^1_{12})~, \nn \\
L^1_{\mathscr{S}_4} &=& \mathbb{V} (X^1_{23}, X^2_{23}, X^1_{41}, X^2_{41})~,  \nn \\
L^2_{\mathscr{S}_4} &=& \mathbb{V} (X^1_{34}, X^2_{34}, X^1_{12}, X^2_{12})~. 
\eea
We see that the coherent component is the product of two conifolds:
\bea
\firr{\mathscr{S}_4} = \CC \times \CC~,
\eea
and the linear components are simply copies of $\BC^4$:
\bea
L^i_{\mathscr{S}_4} = \BC^4 \qquad \text{(for $i =1,2$)}~.
\eea

\begin{figure}
\begin{center}
   \includegraphics[totalheight=8.0cm]{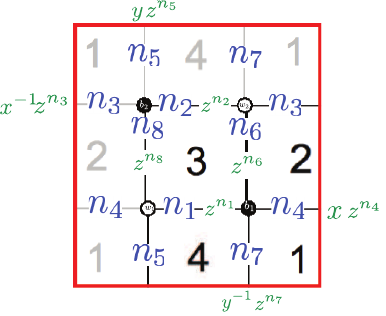}
 \caption{[Phase I of $Q^{1,1,1}/\BZ_2$] The fundamental domain of tiling for the $\mathscr{S}_4$ model: Assignments of the integers $n_i$ to the edges are shown in blue and the weights for these edges are shown in green.}
  \label{f:fdph1f0}
\end{center}
\end{figure}

\paragraph{The Kasteleyn matrix.} We assign the integers $n_i$ to the edges according to Figure \ref{f:fdph1f0}.  From \eref{kn}, we find that 
\bea
\text{Gauge group 1~:} \qquad k_1 &=& 1  =   n_3 + n_4 - n_5 - n_7 ~, \nn \\
\text{Gauge group 2~:} \qquad k_2 &=& -1 =    n_6 + n_8 - n_3 - n_4 ~, \nn \\
\text{Gauge group 3~:} \qquad k_3 &=& -1 =   n_1 + n_2 - n_6 - n_8 ~, \nn \\
\text{Gauge group 4~:} \qquad k_4 &=& 1 = - n_1 - n_2 + n_5 + n_7 ~.
\eea  
We choose
\bea
n_3 = - n_1 = 1,\quad n_i=0 \; \text{otherwise}~.
\eea

\noindent We can now construct the Kasteleyn matrix. The fundamental domain contains two black nodes and two white nodes and, therefore, the Kasteleyn matrix is a $2\times 2$ matrix:
\be
K =   \left(
\begin{array}{c|cc}
& w_1 & w_2 \\
\hline
b_1 & X^{1}_{34} z^{n_1} + X^{2}_{12} x z^{n_4} &\ X^{1}_{23} z^{n_6} + X^{2}_{41} y^{-1} z^{n_7}   \\
b_2 & X^{2}_{23} z^{n_8} + X^{1}_{41} y z^{n_5} &\ X^{2}_{34} z^{n_2} + X^{1}_{12} x^{-1} z^{n_3}  
\end{array}
\right) ~.
\ee
The permanent of this matrix is given by
\bea
\perm~K &=& X^{1}_{34} X^{2}_{34}  {z^{(n_1 + n_2)}} +  X^{1}_{12} X^{2}_{12} z^{(n_3 + n_4)}  + X^{1}_{34} X^{1}_{12} x^{-1} z^{(n_1 + n_3)} +  X^{2}_{34} X^{2}_{12} x z^{(n_2 + n_4)}\nn \\
&& X^{1}_{41} X^{1}_{23} y z^{(n_5 + n_6)} + X^{2}_{41} X^{2}_{23} y^{-1} z^{(n_7 + n_8)} +  X^{1}_{41} X^{2}_{41} z^{(n_5 + n_7)} + X^{1}_{23} X^{2}_{23} z^{(n_6 + n_8)} \nn \\
&=& X^{1}_{34} X^{2}_{34} z^{-1}+  X^{1}_{12} X^{2}_{12} z + X^{1}_{34} X^{1}_{12} x^{-1}  +  X^{2}_{34} X^{2}_{12} x +  X^{1}_{41} X^{1}_{23} y  + X^{2}_{41} X^{2}_{23}y^{-1} \nn \\
&& +  X^{1}_{41} X^{2}_{41}  + X^{1}_{23} X^{2}_{23} \qquad \text{(for $n_3 = - n_1 = 1,~ n_i=0 \; \text{otherwise}$)} ~. \label{permKph1f0}
\eea

\paragraph{The perfect matchings.} From \eref{permKph1f0}, we write the perfect matchings as collections of fields as follows:
\bea 
&& p_1 = \{ X^1_{34},  X^2_{34} \},\;\; p_2 = \{X^{1}_{12}, X^{2}_{12} \}, \;\; q_1 = \{X^{1}_{34}, X^{1}_{12} \}, \;\; q_2 = \{ X^{2}_{34},  X^2_{12} \}, \nn \\  
&& r_1=  \{ X^{1}_{41}, X^1_{23} \}, \;\; r_2 = \{ X^2_{41}, X^2_{23} \}, \;\; 
s_1 = \{  X^1_{41},  X^2_{41} \}, \; \; s_2 = \{ X^{1}_{23}, X^2_{23}  \} \ . \qquad
\eea
From \eref{permKph1f0}, we see that the perfect matchings $p_i, q_i, r_i$ correspond to the external points in the toric diagram, whereas the perfect matchings $s_i$ correspond to the internal point at the origin.
In turn, we find the parameterisation of fields in terms of perfect matchings:
\bea
&& X^1_{34} = p_1 q_1 , \quad X^2_{34} = p_1 q_2 , \quad X^1_{12} = p_2 q_1, \quad X^2_{12} = p_2 q_2, \nn \\
&& X^1_{41} = r_1 s_1, \quad X^1_{23} = r_1 s_2, \quad X^2_{41} = r_2 s_1, \quad X^2_{23} = r_2 s_2~.
\eea
This is summarised in the perfect matching matrix:
\beq
P=\left(\begin{array} {c|cccccccc}
  \;& p_1 & p_2 & q_1 & q_2 & r_1 & r_2 & s_1 & s_2\\
  \hline 
    X^{1}_{34}& 1&0&1&0&0&0&0&0\\
  X^{2}_{34}& 1&0&0&1&0&0&0&0\\
  X^{1}_{12}& 0&1&1&0&0&0&0&0\\
  X^{2}_{12}& 0&1&0&1&0&0&0&0\\
  X^{1}_{41}& 0&0&0&0&1&0&1&0\\
  X^{1}_{23}& 0&0&0&0&1&0&0&1\\
  X^{2}_{41}& 0&0&0&0&0&1&1&0\\
  X^{2}_{23}& 0&0&0&0&0&1&0&1
  \end{array}
\right).
\eeq
Basis vectors of the nullspace of $P$ are given in the rows of the charge matrix:
\be
Q_F =   \left(
\begin{array}{cccccccc}
1&1&-1&-1&0&0&0&0 \\
0&0&0&0&1&1&-1&-1
\end{array}
\right)~.  \label{qfph1q111z2}
\ee
Hence, from \eref{relpm}, we see that the relations between the perfect matchings are given by
\bea
p_1+p_2-p_3-p_4 &=& 0~, \nn \\ 
p_5+p_6-s_1-s_2 &=& 0~. \label{relf0I}
\eea
Since the coherent component $\firr{\mathscr{S}_4}$ of the Master space is generated by the perfect matchings (subject to the relation \eref{relf0I}), it follows that 
\bea
\firr{\mathscr{S}_4} = \BC^8//Q_F~.  \label{firrS4}
\eea

\paragraph{The toric diagram.} We demonstrate two methods of constructing the toric diagram. 
\begin{itemize}
\item{\bf The charge matrices.}  Since the number of gauge groups is $G=4$, there are $G-2 = 2$ baryonic charges coming from the D-terms.  We collect these charges of the perfect matchings in the $Q_D$ matrix:
\bea
Q_D = \left(
\begin{array}{cccccccc}
 1 & 1 & 0 & 0 & -1 & -1 & 0 & 0 \\
 0 & 0 & 0 & 0 & -1 & -1 & 2 & 0
\end{array}
\right)~.  \label{qdph1q111z2}
\eea
From \eref{qfph1q111z2} and \eref{qdph1q111z2}, the total charge matrix is given by
\bea
Q_t = \left(
\begin{array}{cccccccc}
 1 & 1 & 0 & 0 & -1 & -1 & 0 & 0 \\
 0 & 0 & 0 & 0 & -1 & -1 & 2 & 0 \\
 1&1&-1&-1&0&0&0&0 \\
 0&0&0&0&1&1&-1&-1
\end{array}
\right)~. \label{qtph1q111z2}
\eea
We obtain the matrix $G_t$ from \eref{Gt}, and after removing the first row, the columns give the coordinates of points in the toric diagram:  
\bea
G'_t = \left(
\begin{array}{cccccccc}
 0 & 0 & 0 & 0 & -1 & 1 & 0 & 0 \\
 0 & 0 & -1 & 1 & 0 & 0 & 0 & 0 \\
 -1 & 1 & 0 & 0 & 0 & 0 & 0 & 0
\end{array}
\right)~.
\eea
The toric diagram is drawn in Figure \ref{f:torq111z2}.  
\begin{figure}[h]
\begin{center}
  \includegraphics[totalheight=4cm]{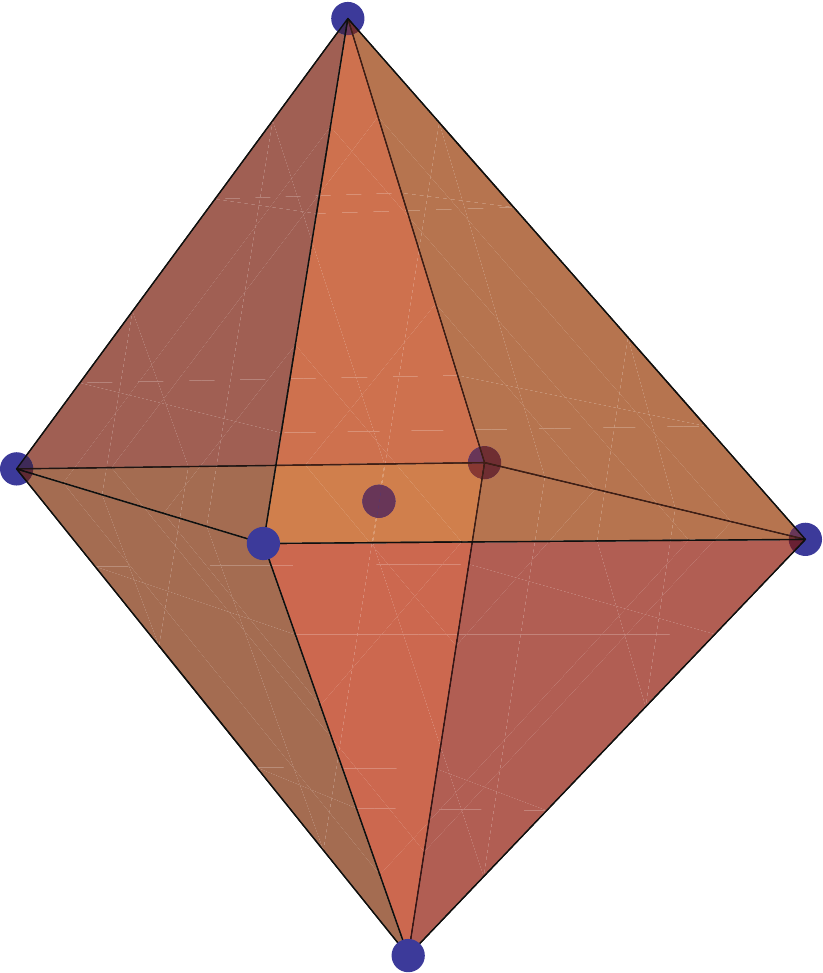}
 \caption{The toric diagram of the $Q^{1,1,1}/\BZ_2$ theory.}
  \label{f:torq111z2}
\end{center}
\end{figure}
Observe that there is an internal point (with multiplicity 2) in the toric diagram for this theory, whereas the toric diagram for the $Q^{1,1,1}$ theory is simply 6 corners of an octahedron without an internal point (see Appendix A of \cite{Hanany:2008fj}).
Comparing Figure \ref{f:torq111z2} with the 2d toric diagram of Phase I of $\BF_0$ theory \cite{master, Butti:2007jv}, we see that the CS levels split two of the four points at the centre of the 2d toric diagram  along the vertical axis into the two tips, and the rest remain at the centre of the octahedron.

\item{\bf The Kasteleyn matrix.}  The powers of $x, y, z$ in each term of \eref{permKph1f0} give the coordinates of each point in the toric diagram.  We collect these points in the columns of the following $G_K$ matrix:
\bea
G_K = \left(
 \begin{array}{cccccccc}
 0 & 0 & 0 & 0 & -1 & 1 & 0 & 0 \\
 0 & 0 & -1 & 1 & 0 & 0 & 0 & 0 \\
 -1 & 1 & 0 & 0 & 0 & 0 & 0 & 0
\end{array}
\right) = G'_t~.
\eea
Thus, the toric diagrams constructed from these two methods are indeed identical.
\end{itemize}

\paragraph{The baryonic charges.}  Since the toric diagram has 6 external points, this model has precisely $6-4 = 2$ baryonic charges which we shall denote by $U(1)_{B_1}, U(1)_{B_2}$.  From the above discussion, we see that they arise from the D-terms.  Therefore, the baryonic charges of the perfect matchings are given by the rows of the $Q_D$ matrix. 

\paragraph{The global symmetry.}  Since the $Q_t$ matrix has 3 pairs of repeated columns, it follows that the mesonic symmetry of this model is $SU(2)^3 \times U(1)_R$.   
Since $s_1$ and $s_2$ are the perfect matchings corresponding to internal points in the toric diagram, we assign to each of them a zero R-charge.  
The remaining 6 external perfect matchings are completely symmetric and the requirement of R-charge 2 to the superpotential divides 2 equally among them, resulting in R-charge of 1/3 per each.
The global symmetry of the theory is a product of mesonic and baryonic symmetries: $SU(2)^3 \times U(1)_R \times U(1)_{B_1} \times U(1)_{B_2}$.
In Table \ref{chargeph1f0}, we present a consistent way of assigning charges to the perfect matchings under these global symmetries.
\begin{table}[h]
 \begin{center}  
  \begin{tabular}{|c||c|c|c|c|c|c|c|}
  \hline
  \;& $SU(2)_{1}$&$SU(2)_{2}$&$SU(2)_{3}$&$U(1)_R$&$U(1)_{B_1}$&$U(1)_{B_2}$& fugacity\\
  \hline \hline  
  $p_1$& $1$&$0$&$0$&$1/3$&$1$&$0$& $t b_1 x_1$ \\
  $p_2$& $-1$&$0$&$0$&$1/3$&$1$&$0$ & $t b_1/x_1$ \\
  $q_1$& $0$&$1$&$0$&$1/3$&$0$&$0$  & $t x_2$\\
  $q_2$& $0$&$-1$&$0$&$1/3$&$0$&$0$& $ t / x_2$\\
  $r_1$& $0$&$0$&$1$&$1/3$&$-1$&$-1$& $ t x_3/(b_1 b_2)$\\
  $r_2$& $0$&$0$&$-1$&$1/3$&$-1$&$-1$& $ t / (x_3  b_1 b_2)$\\
  $s_1$& $0$&$0$&$0$&$0$&$0$&$2$&  $b_2^2$ \\
  $s_2$& $0$&$0$&$0$&$0$&$0$&$0$ & $1$ \\
  $\Blue s_3$& $0$&$0$&$0$&$0$&$0$&$0$ & $1$ \\ 
  \hline
  \end{tabular}
  \end{center} \Black
  \caption{Charges under the global symmetry of the $Q^{1,1,1}/\BZ_2$ theory. Here $t$ is the fugacity of R-charge, $x_1,x_2,x_3$ are weights of $SU(2)_{1}, SU(2)_{2}, SU(2)_3$, and $b_1, b_2$ are baryonic fugacities of $U(1)_{B_1}, U(1)_{B_2}$. Note that the perfect matching $s_3$ (represented in blue) does not exist in Phase I but exists in Phase II.}
\label{chargeph1f0}
\end{table}

\paragraph{The Hilbert series.} From \eref{firrS4}, we compute the Hilbert series of the coherent component of the Master space by integrating the Hilbert series of $\BC^8$ over the fugacities $z_1$ and $z_2$ associated with the $Q_F$ charges:
\bea
g^{\firr{}}_1 (t,x_1,x_2,x_3,b_1,b_2; \mathscr{S}_4) &=& \frac{1}{(2\pi i)^2} \oint \limits_{|z_1| =1} {\frac{\ud z_1}{z_1 }} \oint \limits_{|z_2| =1} {\frac{\ud z_2}{z_2}}  \frac{1}{\left(1- t b_1 z_1 x_1  \right)\left(1- \frac{t b_1 z_1}{x_1}\right)\left(1- \frac{t x_2}{z_1}\right)} \times \nn \\
&& \times \frac{1}{\left(1-\frac{t}{x_2 z_1}\right)\left(1- \frac{t x_3 z_2}{b_1 b_2} \right)\left(1-\frac{t z_2}{b_1 b_2 x_3}\right)\left(1- \frac{b_2^2}{z_2} \right)\left(1- \frac{1}{z_2} \right)} \nn \\
&=& \frac{\left(1-\frac{t^2}{b_1^2}\right)}{\left(1-\frac{t b_2}{b_1 x_3}\right)\left(1-\frac{t b_2 x_3}{b_1}\right)\left(1-\frac{t}{b_1 b_2 x_3}\right)\left(1- \frac{t x_3}{b_1 b_2}\right)} \times \nn \\
&& \times \frac{\left(1-t^4 b_1^2 \right)}{\left(1-\frac{t^2 b_1}{x_1 x_2}\right)\left(1-\frac{t^2 b_1 x_2}{x_1}\right)\left(1-\frac{t^2 b_1 x_1}{x_2}\right)\left(1-t^2 b_1 x_1 x_2\right)}~. \label{hsfirrs4}
\eea
The unrefined Hilbert series of the Master space can be written as:
\bea
g^{\firr{}}_1 (t,1,1,1,1,1; \mathscr{S}_4) = \frac{1-t^2}{\left(1-t\right)^4}\times \frac{1-t^4}{\left(1-t^2\right)^4}~. \label{hsfph1f0}
\eea
We see that this space is indeed the product of two conifolds.  The Hilbert series of the mesonic moduli space can be obtained by integrating \eref{hsfirrs4} over the two baryonic fugacities $b_1$ and $b_2$:
\bea
\gm_1 (t,x_1,x_2,x_3; \mathscr{S}_4) &=& \frac{1}{(2\pi i)^2}\oint  \limits_{|b_1| =1} {\frac{db_1}{b_1}}\oint  \limits_{|b_2| =1} {\frac{db_2}{b_2}} g^{\firr{}}_1 (t,x_1,x_2,x_3,b_1,b_2; \mathscr{S}_4)\nn \\
&=& \frac{P(t,x_1,x_2,x_3)}{\left(1 - t^6 x_1^2 x_2^2 x_3^2 \right)\left(1 - \frac{t^6 x_1^2 x_2^2}{x_3^2}\right)\left(1 - \frac{t^6 x_1^2 x_3^2}{x_2^2}\right)\left(1 - \frac{t^6 x_2^2 x_3^2}{x_1^2}\right)} \times \nn \\
&& \times \frac{1}{\left(1 - \frac{t^6 x_1^2}{x_2^2 x_3^2}\right)\left(1 - \frac{t^6 x_2^2}{x_1^2 x_3^2}\right)\left(1 - \frac{t^6 x_3^2}{x_1^2 x_2^2}\right)\left(1 - \frac{t^6}{x_1^2 x_2^2 x_3^2}\right)}\nn \\
&=& \sum^{\infty}_{n=0}[2n;2n;2n]t^{6n}~.
\label{meshsph1fo}
\eea
where $P(t,x_1,x_2,x_3)$ is a polynomial of degree $42$ in $t$ which is too long to present here. The unrefined Hilbert series of the mesonic moduli space can be written as:
\bea
\gm_1 (t,1,1,1; \mathscr{S}_4) = \frac{1+23t^6+23t^{12}+t^{18}}{\left(1-t^6\right)^4}~.
\eea
This indicates that the mesonic moduli space is a Calabi--Yau 4-fold, as expected.
The plethystic logarithm of the mesonic Hilbert series is given by 
\bea
\PL[\gm_1 (t,x_1,x_2,x_3;  \mathscr{S}_4)] &=& [2;2;2] t^6 - ([4;4;0] + [4;0;4] + [0;4;4] + [4;0;0] + [0;4;0] +   \nn\\
&& + [0;0;4] + [4;2;2] + [2;4;2] + [2;2;4] + [2;2;0] +  [2;0;2] + \nn \\
&& + [0;2;2] + 1) t^{12} + O(t^{18})~. 
\label{PLf0}
\eea

\begin{figure}[ht]
\begin{center}
  \includegraphics[totalheight=6.0cm]{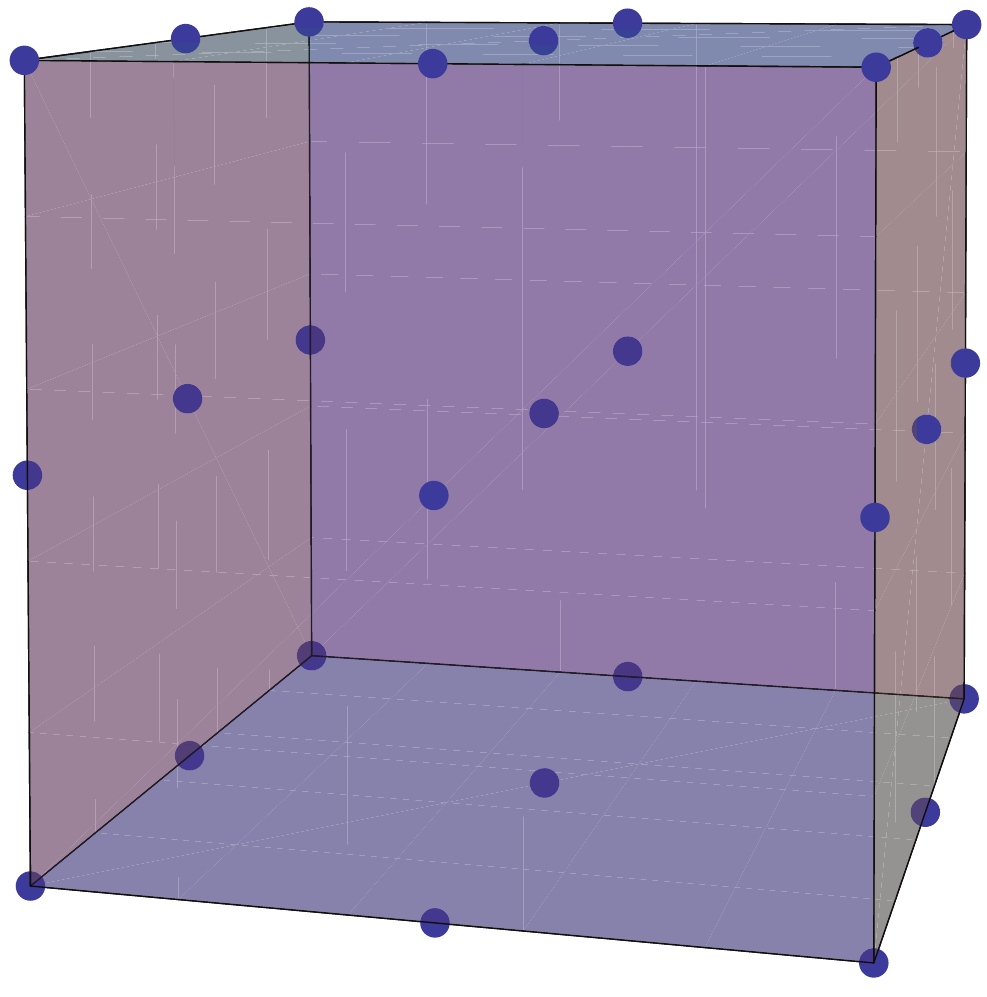}
 \caption{The lattice of generators of the $Q^{1,1,1}/\BZ_2$ theory.}
  \label{f:latq111z2}
\end{center}
\end{figure}

\paragraph{The generators.} Each of the generators can be written as a product of the perfect matchings:
\bea
p_i ~p_j ~q_k ~q_l ~r_m ~r_n ~s_1~ s_2~,  \label{genph1f0}
\eea
where the indices $i, j, k,  l, m, n$ run from 1 to 2.  
Since, for example, $p_i p_j$ has 3 independent components $p_1 p_1, ~ p_1 p_2, ~ p_2 p_2$, it follows that there are indeed 27 independent generators.
We can represent the generators in a lattice (Figure \ref{f:latq111z2}) by plotting the powers of the weights of the characters in \eref{PLf0}.  Note that the lattice of generators is the dual of the toric diagram (nodes are dual to faces and edges are dual to edges): The toric diagram has  6 nodes (external points), 12 edges and 8 faces, whereas the generators form a convex polytope that has 8 nodes (corners of the cube), 12 edges and 6 faces.

\paragraph{The $\BZ_2$ orbifold action.} It is interesting to compare the last equality of \eref{meshsph1fo} to the Hilbert series of the $Q^{1,1,1}$ theory, which is given by (A.7) of \cite{Hanany:2008fj}:
\bea
\gm_1(t, x_1, x_2, x_3; Q^{1,1,1}) &=& \sum^{\infty}_{n=0}[n;n;n]t^{3n}~. \label{hsq111}
\eea
This indicates that the $\mathscr{S}_4$ model is indeed the orbifold $Q^{1,1,1}/ \BZ_2$.  The reason is as follows.  As discussed in \cite{Hanany:2008qc}, under the $\BZ_2$ orbifold action, $t \rightarrow -t$ 
and we need to sum over both sectors, with $t$ and with $-t$.  Therefore, starting from \eref{hsq111} and applying the $\BZ_2$ action, we are left with the terms correponding to even $j$ and hence \eref{meshsph1fo}.

\subsection{Phase II: The Two-Square and Two-Octagon Model}
\begin{figure}[ht]
\begin{center}
  \hskip -5cm
  \includegraphics[totalheight=3.6cm]{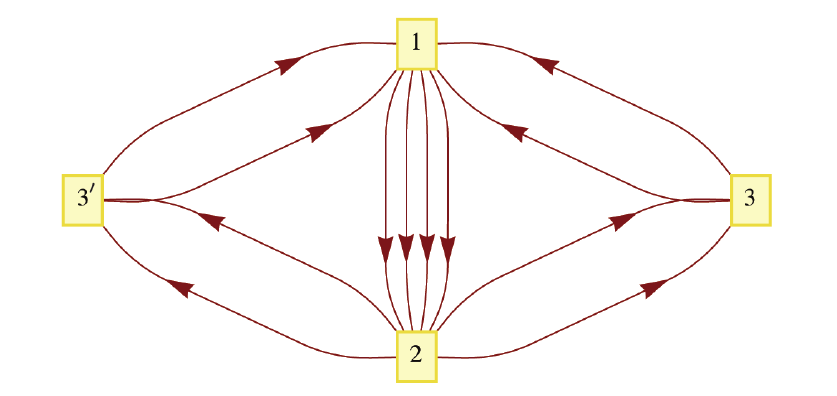}
    \vskip -3.5cm
  \hskip 8cm
  \includegraphics[totalheight=3.5cm]{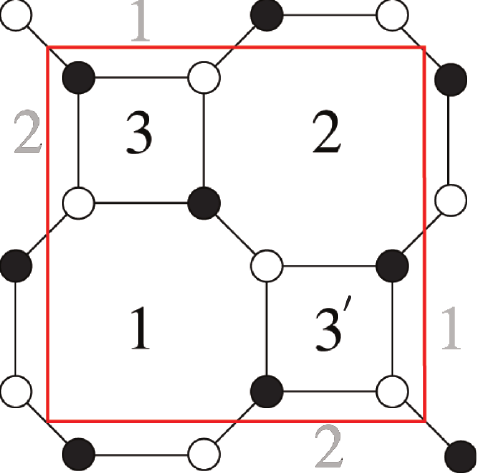}
 \caption{[Phase II of $Q^{1,1,1}/\BZ_2$]  (i) Quiver diagram for the $\mathscr{S}_2 \mathscr{O}_2$ model.\ (ii) Tiling for the $\mathscr{S}_2 \mathscr{O}_2$ model.}
  \label{f:phase2f0}
\end{center}
\end{figure}

This model, first studied in \cite{Hanany:2008fj}, (which we shall denote as $\mathscr{S}_2 \mathscr{O}_2$) has four gauge groups and bi-fundamental fields $X_{12}^{ij}$, $X_{23}^i$, $X_{23'}^i$, $X_{31}^i$ and $X^{i}_{3'1}$ (with $i,j=1,2$). From the features of this quiver gauge theory, this phase is also known as a \emph{three-block model} (see for example \cite{Benvenuti:2004dw}). The superpotential is given by 
\bea
W &=& \epsilon_{ij}\epsilon_{kl} \tr(X^{ik}_{12}X^{l}_{23} X^{j}_{31}) - \epsilon_{ij}\epsilon_{kl} \tr(X^{ki}_{12}X^{l}_{23'}X^{j}_{3'1})~.
\eea
The quiver diagram and tiling of this phase of the theory are given in Figure \ref{f:phase2f0}.  Note that in 3+1 dimensions, these quiver and tiling correspond to Phase II of the $\BF_0$ theory \cite{Forcella:2008ng, master}.  
We choose the CS levels to be $k_1 = k_2 = -k_3 = -k_{3'}=1$.
\\

\paragraph{The Master space.} A primary decomposition indicates that the Master space of this phase is a reducible variety and has 4 irreducible components \cite{Forcella:2008ng, master}: 
\bea
\f_{\mathscr{S}_2 \mathscr{O}_2} = \firr{\mathscr{S}_2 \mathscr{O}_2} ~  \cup~L^1_{\mathscr{S}_2 \mathscr{O}_2}~  \cup~L^2_{\mathscr{S}_2 \mathscr{O}_2} ~\cup~L^3_{\mathscr{S}_2 \mathscr{O}_2}~, 
\eea
where
\bea
\firr{\mathscr{S}_1 \mathscr{O}_1} &=& \mathbb{V} (X^{12}_{12}X^1_{3'1} - X^{11}_{12} X^2_{3'1}, X^2_{31}X^1_{3'1} - X^1_{31}X^2_{3'1}, X^{22}_{12}X^1_{3'1} - X^{21}_{12} X^2_{3'1},
X^2_{23}X^1_{23'} - X^1_{23}X^2_{23'}, \nn \\
&&
X^{21}_{12}X^1_{23'} - X^{22}_{12}X^2_{23'}, X^{22}_{12}X^1_{23'} - X^{12}_{12}X^2_{23'}, X^{21}_{12}X^{12}_{12} -  X^{22}_{12}X^{22}_{12}, X^1_{31}X^{12}_{12} - X^2_{31}X^{11}_{12}, \nn \\ 
&&
X^2_{31}X^2_{23} - X^2_{3'1}X^2_{23'}, X^1_{31}X^2_{23} - X^1_{3'1}X^2_{23'}, X^1_{23}X^{21}_{12} - X^2_{23}X^{11}_{12},X^1_{23}X^2_{31} - X^1_{23'}X^2_{3'1},\nn  \\
&& 
X^1_{31}X^{22}_{12} - X^2_{31}X^{21}_{12}, X^1_{23}X^{22}_{12} - X^2_{23}X^{12}_{12}, X^1_{23}X^1_{31} - X^1_{23'}X^1_{3'1} )~, \nn \\
L^1_{\mathscr{S}_1 \mathscr{O}_1} &=& \mathbb{V}(X^2_{23'}, X^2_{3'1}, X^1_{3'1}, X^1_{23'}, X^2_{23}, X^2_{31}, X^1_{31},X^1_{23})~, \nn \\
L^2_{\mathscr{S}_1 \mathscr{O}_1} &=& \mathbb{V}(X^2_{3'1}, X^1_{3'1}, X^{11}_{12}, X^{12}_{12}, X^{21}_{12}, X^2_{31}, X^{22}_{12}, X^1_{31})~, \nn \\
L^3_{\mathscr{S}_1 \mathscr{O}_1} &=& \mathbb{V}(X^2_{23'}, X^1_{23'},X^{11}_{12}, X^{12}_{12}, X^2_{23}, X^{21}_{12}, X^{22}_{12}, X^1_{23}) \ .
\eea
We see that the linear components are simply copies of $\BC^4$:
\bea
L^i_{\mathscr{S}_2 \mathscr{O}_2} = \BC^4 \qquad \text{(for $i =1,2, 3$)}~.
\eea

\begin{figure}[h]
\begin{center}
   \includegraphics[totalheight=6.5cm]{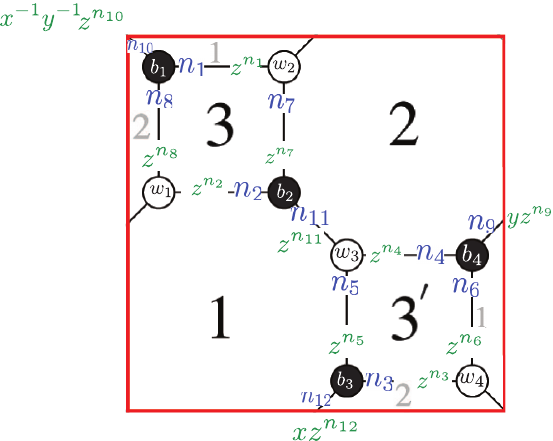}
 \caption{[Phase II of $Q^{1,1,1}/\BZ_2$]  The fundamental domain of tiling for the $\mathscr{S}_2 \mathscr{O}_2$ model: Assignments of the integers $n_i$ to the edges are shown in blue and the weights for these edges are shown in green.}
  \label{f:fdph2f0}
\end{center}
\end{figure}

\paragraph{The Kasteleyn matrix.}  We assign the integers $n_i$ to the edges according to Figure \ref{f:fdph2f0}.  From \eref{kn}, we find that 
\bea
\text{Gauge group 1~:} \qquad k_1 &=& 1  = - n_1 - n_2 - n_5 - n_6 +n_9 + n_{10} +n_{11} + n_{12}  ~, \nn \\
\text{Gauge group 2~:} \qquad k_2 &=& 1 =  n_3 + n_4 + n_7 +n_8 - n_9 -n_{10} - n_{11} - n_{12} ~, \nn \\
\text{Gauge group 3~:} \qquad k_3 &=& -1 =  n_1 + n_2 - n_7  - n_8 ~, \nn \\
\text{Gauge group 4~:} \qquad k_{3'} &=& -1 =  - n_3 - n_4 + n_5 + n_6 ~.
\eea  
We choose
\bea
n_2 = -1, \quad n_4 = 1,\quad n_i=0 \; ~ \text{otherwise}~.
\eea

\noindent We can now determine the Kasteleyn matrix.  Since the fundamental domain contains 4 black nodes and 4 white nodes, the Kasteleyn matrix is a $4\times 4$ matrix:
\be
K =   \left(
\begin{array}{c|cccc}
 \;& w_1&w_2&w_3&w_4 \\
  \hline 
b_1 & X^{2}_{23} z^{n_{8}} &\ X^{1}_{31} z^{n_{1}} &\ 0 &\ X^{21}_{12} x^{-1} y^{-1} {z^{n_{10}}} \\
b_2 & X^{2}_{31} z^{n_{2}} &\ X^{1}_{23} z^{n_{7}} &\ X^{12}_{12} z^{n_{11}} &\ 0  \\
b_3 & 0 &\ X^{22}_{12} x z^{n_{12}} &\ X^{1}_{3'1} z^{n_{5}} &\ X^{1}_{23'} z^{n_{3}}  \\
b_4 & X^{11}_{12} y z^{n_{9}} &\ 0 &\ X^{2}_{23'} z^{n_{4}} &\ X^{2}_{3'1} z^{n_{6}} 
\end{array}
\right) ~.
\ee
The permanent of this matrix is given by
\bea \label{permKph2f0}
\perm~K &=&  X^{1}_{31} X^{2}_{31} X^{1}_{3'1} X^{2}_{3'1} z^{(n_{1} + n_{2} + n_{5} + n_{6})} + X^{1}_{23'} X^{2}_{23'} X^{2}_{23} X^{1}_{23} z^{(n_3 + n_4 + n_7 + n_8)}  \nn \\
&&+ X^{1}_{3'1} X^{1}_{23} X^{11}_{12} X^{21}_{12} x^{-1} z^{(n_5 + n_7 + n_9 + n_{10})} + X^{2}_{3'1} X^{2}_{23} X^{12}_{12} X^{22}_{12} x {z^{(n_{11} + n_{12} + n_6 + n_8)}}\nn \\
&&+ X^{1}_{31} X^{1}_{23'} X^{11}_{12} X^{12}_{12} y z^{(n_1 + n_3 + n_9 + n_{11})} + X^{2}_{31} X^{2}_{23'} X^{21}_{12} X^{22}_{12} y^{-1} z^{(n_{2} + n_{4} + n_{10} + n_{12})} \nn \\
&&+ X^{1}_{31} X^{2}_{31} X^{1}_{23'} X^{2}_{23'} {z^{(n_{1} + n_2 + n_3 + n_4)}}+ X^{1}_{3'1} X^{2}_{3'1} X^{2}_{23} X^{1}_{23} z^{(n_{5} + n_6 + n_7 + n_8)}\nn \\
&&  + X^{11}_{12} X^{21}_{12} X^{12}_{12} X^{22}_{12} z^{(n_9 + n_{10} + n_{11} + n_{12})} \nn \\
&=& X^{1}_{31} X^{2}_{31} X^{1}_{3'1} X^{2}_{3'1} z^{-1}  +  X^{1}_{23'} X^{2}_{23'} X^{2}_{23} X^{1}_{23} z + X^{1}_{3'1} X^{1}_{23} X^{11}_{12} X^{21}_{12} x^{-1} + X^{2}_{3'1} X^{2}_{23} X^{12}_{12} X^{22}_{12} x  \nn \\
&&+ X^{1}_{31} X^{1}_{23'} X^{11}_{12} X^{12}_{12} y  + X^{2}_{31} X^{2}_{23'} X^{21}_{12} X^{22}_{12} y^{-1} + X^{1}_{31} X^{2}_{31} X^{1}_{23'} X^{2}_{23'} + X^{1}_{3'1} X^{2}_{3'1} X^{2}_{23} X^{1}_{23}  \nn \\
&&  + X^{11}_{12} X^{21}_{12} X^{12}_{12} X^{22}_{12} \qquad \text{(for $n_2 = -1, ~ n_4 = 1,~ n_i=0 \; ~ \text{otherwise}$)} ~.
\eea

\paragraph{The perfect matchings.} We summarise the correspondence between the quiver fields and the perfect matchings in the $P$ matrix as follows: 
\beq
P=\left(\begin{array} {c|ccccccccc}
  \;& p_1 & p_2 &q_1&q_2&r_1&r_2&s_1&s_2&s_3\\
  \hline
  X^{1}_{31}  & 1 & 0 & 0 & 0 & 1 & 0 & 1 & 0 & 0 \\
  X^{2}_{31}  & 1 & 0 & 0 & 0 & 0 & 1 & 1 & 0 & 0 \\
  X^{1}_{23'}  & 0 & 1 & 0 & 0 & 1 & 0 & 1 & 0 & 0 \\
  X^{2}_{23'}  & 0 & 1 & 0 & 0 & 0 & 1 & 1 & 0 & 0 \\
  X^{1}_{3'1}  & 1 & 0 & 1 & 0 & 0 & 0 & 0 & 1 & 0 \\
  X^{2}_{3'1}  & 1 & 0 & 0 & 1 & 0 & 0 & 0 & 1 & 0 \\
  X^{1}_{23}  & 0 & 1 & 1 & 0 & 0 & 0 & 0 & 1 & 0 \\
  X^{2}_{23}  & 0 & 1 & 0 & 1 & 0 & 0 & 0 & 1 & 0 \\
  X^{11}_{12} & 0 & 0 & 1 & 0 & 1 & 0 & 0 & 0 & 1 \\
  X^{21}_{12} & 0 & 0 & 1 & 0 & 0 & 1 & 0 & 0 & 1 \\
  X^{12}_{12} & 0 & 0 & 0 & 1 & 1 & 0 & 0 & 0 & 1\\
  X^{22}_{12} & 0 & 0 & 0 & 1 & 0 & 1 & 0 & 0 & 1 
  \end{array}
\right)~.
\eeq
From \eref{permKph2f0}, we see that the perfect matchings $p_i, q_i, r_i$ correspond to the external points in the toric diagram, whereas the perfect matchings $s_i$ correspond to the internal point at the origin.
Basis vectors of the null space of $P$ are given in the rows of the charge matrix:
\be
Q_F =   \left(
\begin{array}{ccccccccc}
 1 & 1 & 0 & 0 & 0 & 0 & -1 & -1 & 0\\
 0 & 0 & 1 & 1 & 0 & 0 & 0 & -1 & -1 \\
 0 & 0 & 0 & 0 & 1 & 1 & -1 & 0 & -1
 \end{array}
\right)~. \label{qfph2f0}
\ee
Hence, from \eref{relpm}, we see that the relations between the perfect matchings are given by
\bea
p_1 + p_2 - s_1 - s_2 &=& 0~, \nn\\
q_1 + q_2 - s_2 - s_3 &=& 0~, \nn \\ \
r_1 + r_2 - s_1 - s_3 &=& 0~. \label{relf0ii}
\eea
Since the coherent component of the Master space is generated by the perfect matchings (subject to the relations \eref{relf0ii}), it follows that 
\bea
\firr{\mathscr{S}_2 \mathscr{O}_2} = \BC^9//Q_F~.  \label{firrph2f0}
\eea

\paragraph{The toric diagram.} We demonstrate two methods of constructing the toric diagram. 
\begin{itemize}
\item{\bf The charge matrices.}    Since the number of gauge groups is $G=4$, there are $G-2 = 2$ baryonic charges coming from the D-terms.  We collect these charges of the perfect matchings in the $Q_D$ matrix:
\bea
Q_D = \left(
\begin{array}{ccccccccc}
 1 & 1 & 0 & 0 & -1 & -1 & 0 & 0 & 0 \\
 0 & 0 & 0 & 0 & -1 & -1 & 2 & 0 & 0
\end{array}
\right)~. \label{qdph2f0}
\eea
From \eref{qfph2f0} and \eref{qdph2f0}, the total charge matrix is given by
\bea
Q_t = \left(
\begin{array}{ccccccccc}
 1 & 1 & 0 & 0 & -1 & -1 & 0 & 0 & 0 \\
 0 & 0 & 0 & 0 & -1 & -1 & 2 & 0 & 0\\
  1 & 1 & 0 & 0 & 0 & 0 & -1 & -1 & 0\\
 0 & 0 & 1 & 1 & 0 & 0 & 0 & -1 & -1 \\
 0 & 0 & 0 & 0 & 1 & 1 & -1 & 0 & -1
\end{array}
\right)~.
\eea
We obtain the matrix $G_t$ from \eref{Gt}, and after removing the first row, the columns give the coordinates of points in the toric diagram:  
\bea
G'_t = \left(
\begin{array}{ccccccccc}
 0 & 0 & 0 & 0 & -1 & 1 & 0 & 0 & 0 \\
 0 & 0 & -1 & 1 & 0 & 0 & 0 & 0 & 0 \\
 -1 & 1 & 0 & 0 & 0 & 0 & 0 & 0 & 0
\end{array}
\right)~.
\eea
We see that the toric diagram is given by Figure \ref{f:torq111z2}, with three degenerate internal points at the centre.  Comparing Figure \ref{f:torq111z2} with the 2d toric diagram of Phase II of $\BF_0$ theory \cite{master, Butti:2007jv}, we see that the CS levels split two of the five points at the centre of the 2d toric diagram  along the vertical axis into the two tips, and the rest remain at the centre of the octahedron.

\item {\bf The Kasteleyn matrix.} The powers of $x, y, z$ in each term of the permanent of the Kasteleyn matrix give the coordinates of each point in the toric diagram.  We collect these points in the columns of the following $G_K$ matrix:
\bea
G_K = \left(
\begin{array}{ccccccccc}
 0 & 0 & 0 & 0 & -1 & 1 & 0 & 0 & 0 \\
 0 & 0 & -1 & 1 & 0 & 0 & 0 & 0 & 0 \\
 -1 & 1 & 0 & 0 & 0 & 0 & 0 & 0 & 0
\end{array}
\right) = G'_t~.
\eea
Thus, the toric diagrams constructed from these two methods are indeed identical.
\end{itemize}

\paragraph{The baryonic charges.}
Since the toric diagram has 6 external points, this model has precisely $6-4 = 2$ baryonic charges which we shall denote by $U(1)_{B_1}, U(1)_{B_2}$.  From the above discussion, we see that they arise from the D-terms.  Therefore, the baryonic charges of the perfect matchings are given by the rows of the $Q_D$ matrix. 

\paragraph{The global symmetry.} From the $Q_t$ matrix, the charge assignment breaks the symmetry of the space of perfect matchings to $SU(2)^3\times U(1)_R$. 
Since $s_1, s_2, s_3$ are the perfect matchings corresponding to internal points in the toric diagram, we assign to each of them a zero R-charge.  
The remaining 6 external perfect matchings are completely symmetric and the requirement of R-charge 2 to the superpotential divides 2 equally among them, resulting in R-charge of 1/3 per each. 
The global symmetry of the theory is a product of mesonic and baryonic symmetries: $SU(2)^3 \times U(1)_R \times U(1)_{B_1} \times U(1)_{B_2}$.
In Table \ref{chargeph1f0}, we give a consistent charge assignment for the perfect matchings under the global symmetries.

\paragraph{The Hilbert series.} From \eref{firrph2f0}, we compute the Hilbert series of the coherent component of the Master space by integrating the Hilbert series of $\BC^{9}$ over the fugacities $z_1, z_2, z_3$ associated with the $Q_F$ charges:
\bea
g^{\firr{}}_1 (t,x_1,x_2,x_3,b_1,b_2; \mathscr{S}_2 \mathscr{O}_2) &=& \frac{1}{(2 \pi i)^3} \oint \limits_{|z_1|=1} {\frac{dz_1}{z_1}} \oint \limits_{|z_2|=1} {\frac{dz_2}{z_2}}\oint \limits_{|z_3|=1} { \frac{dz_3}{z_3}}  \frac{1}{\left(1- t b_1 z_1 x_1\right)\left(1-\frac{t b_1 z_1}{x_1}\right)}  \times \nn \\
&& \times \frac{1}{\left(1- t x_2 z_2\right) \left(1-\frac{t z_2}{x_2}\right)\left(1-\frac{t x_3 z_3}{b_1b_2}\right)\left(1-\frac{t z_3}{x_3 b_1 b_2}\right)\left(1-\frac{b_2^2}{z_1 z_3}\right)}  \times \nn \\
&& \times \frac{1}{\left(1-\frac{1}{z_1 z_2}\right)\left(1- \frac{1}{z_2z_3} \right)}~.
\eea
The unrefined Hilbert series of the Master space can be written as:
\bea
g^{\firr{}}_1 (t,1,1,1,1,1;\mathscr{S}_2 \mathscr{O}_2) &=& \frac{1+6t^2+6t^4+t^6}{\left(1-t^2\right)^6}~. \label{hsfph2f0}
\eea
Integrating the Hilbert series of the Master space over the baryonic fugacities gives the Hilbert series of the mesonic moduli space:
\bea
\gm_1 (t,x_1,x_2,x_3; \mathscr{S}_2 \mathscr{O}_2) &=& \frac{1}{(2\pi i)^2}\oint \limits_{|b_1|=1} {\frac{db_1}{b_1}}\oint \limits_{|b_2|=1} {\frac{db_2}{b_2}} g^{\firr{}}_1 (t,x_1,x_2,x_3,b_1,b_2; \mathscr{S}_2 \mathscr{O}_2) \nn \\
&=& \frac{P(t,x_1,x_2,x_3)}{\left(1 - t^6 x_1^2 x_2^2 x_3^2 \right)\left(1 - \frac{t^6 x_1^2 x_2^2}{x_3^2}\right)\left(1 - \frac{t^6 x_1^2 x_3^2}{x_2^2}\right)\left(1 - \frac{t^6 x_2^2 x_3^2}{x_1^2}\right)} \times \nn \\
&& \times  \frac{1}{\left(1 - \frac{t^6 x_1^2}{x_2^2 x_3^2}\right)\left(1 - \frac{t^6 x_2^2}{x_1^2 x_3^2}\right)\left(1 - \frac{t^6 x_3^2}{x_1^2 x_2^2}\right)\left(1 - \frac{t^6}{x_1^2 x_2^2 x_3^2}\right)} \nn \\
&=& \sum^{\infty}_{j=0}[2j;2j;2j]t^{6j}~. \label{meshsph2f0}
\eea
where $P(t,x_1,x_2,x_3)$ is a polynomial of order $42$ in $t$ mentioned in \eref{meshsph1fo}. 
This precisely identical to the Hilbert series \eref{meshsph1fo} of the mesonic moduli space of Phase I .   

\paragraph{The generators.} Each of the generators can be written as a product of the perfect matchings:
\bea
p_i ~p_j ~q_k ~q_l ~r_m ~r_n ~s_1~ s_2 ~s_3~,  \label{genph2f0}
\eea
where the indices $i, j, k,  l, m, n$ run from 1 to 2.  
Since, for example, $p_i p_j$ has 3 independent components $p_1 p_1, ~ p_1 p_2, ~ p_2 p_2$, it follows that there are indeed 27 independent generators.
Note that the generators of this model are identical to those of Phase I, apart from a factor of the internal perfect matching $s_3$.

\paragraph{Discussion.} The toric diagram and the Hilbert series \eref{meshsph2f0} confirms that the mesonic moduli space of this model is indeed $Q^{1,1,1}/\BZ_2$.  However, from \eref{hsfph1f0} and \eref{hsfph2f0}, we see that the Master spaces of the two phases are different.  Since the mesonic and baryonic symmetries of the two phases are identical, it remains an open question why the Master spaces, which are expected to be the combined baryonic and mesonic moduli space, of the two phases are different.  This situation was also encountered in \cite{Forcella:2008ng}, where two phases of the $\mathbb{F}_0$ were studied. There, it was found that the Hilbert series of the two phases are different unless the fugacities associated with the anomalous charges are set to 1.


\acknowledgments
We are indebted to Alastair Craw, Alastair King, Ed Segal, Angel Uranga and Alberto Zaffaroni for valuable discussions.  
J.~D.~ would like to thank the STFC for his studentship, Rak-Kyeong Seong for his invaluable help during the Summer of 2008 and Jennifer Forrester for her kindness and support.
N.~M.~ is grateful to the $26^\text{th}$ Winter School in Theoretical Physics at the IAS of Jerusalem for hospitality during the writing of this paper, as well as to Ofer Aharony, Oren Bergman and Alexander Shannon for useful discussions.  He would like to express his deep gratitude towards his family for the warm encouragement and support, as well as towards the DPST project and the Royal Thai Government for funding his research.  
G.~T.~ wants to express his deep gratitude to his family for the great support during the preparation of this work, as well as to Elisa Rebessi for her unique sweetness and intelligence, which are always splendid sources of encouragement for his life.

\appendix
\section{Permanent of the Kasteleyn Matrix and Coordinates of the Points in the Toric Diagram} \label{permkastel}
In this Appendix, we show that the permanent of the Kasteleyn matrix indeed gives rise to coordinates of the points in the toric diagram.
Given a Kasteleyn matrix $K(x,y,z)$, its permanent is given by \eref{permk}:
\bea
\perm~K = \sum_{\alpha=1}^c  p_{\alpha} x^{u_\alpha}y^{v_\alpha}z^{w_\alpha}~, \label{perma}
\eea
where $u_\alpha, v_\alpha, w_\alpha$ are given by
\bea \label{uvw}
u_\alpha = \sum_{e_i \in E_x} \mathrm{sign}_x(e_i)P_{i\alpha}~,\quad
v_\alpha = \sum_{e_i \in E_y} \mathrm{sign}_y(e_i)P_{i\alpha}~,\quad
w_\alpha &= \sum_{i} n_i P_{i\alpha}~,
\eea
where $E_x$ and $E_y$ denote the set of edges crossing the horizontal and vertical boundary of the fundamental domain, and $\mathrm{sign}_l(e_i)$ denotes the sign arising from the edge $e_i$ crossing the fundamental domain in the $l$ direction.  The powers of $x, y, z$ in \eref{perma} are collected in each column of the $G_K$ matrix as follows: 
\beq
G_K = 
\left( \begin{array}{ccccc}  u_1& u_2& u_3&\ldots & u_c \\ v_1& v_2&  v_3&\ldots & v_c \\ w_1& w_2& w_3&\ldots & w_c \end{array} \right)~.
\label{e:gk}
\eeq

We would like to prove that the rows of $G_K$ are elements of the nullspace of $Q_t$.  It follows immediately from the forward algorithm that $(u_\alpha, v_\alpha, w_\alpha)$ are coordinates of the points in the toric diagram.  From the definition \eref{Qtdef} of the $Q_t$ matrix, it is equivalent to proving that the rows of $G_K$ are in the nullspace of both $Q_F$ and $Q_D$:
\bea
Q_F \cdot G^t_K = 0~,\qquad Q_D\cdot G^t_K = 0~.
\label{kercond2}
\eea

Let us first prove the first equation of (\ref{kercond2}). 
From \eref{uvw}, we see that $u, v, w$ are linear combinations of the rows of the matrix $P$.  
According to \eref{relpm}, the latter live in the nullspace of $Q_F$.
Therefore, all of the three rows of $G_K$ are indeed elements of the nullspace of $Q_F$.
Thus, we have proven the first equation of \eref{kercond2}. 

Now let us prove the second equation of (\ref{kercond2}).   
Using the definition \eref{QD} of the $Q_D$ matrix and \eref{uvw}, we find that
\bea
[Q_D \cdot (u_1~ u_2~ \ldots~ u_c)^t]_l &=& \sum_{\alpha = 1}^c \sum_{e_i\in E_x} \mathrm{sign}_x (e_i)\; [\ker(C) \cdot \widetilde{Q}]_{l\alpha} (P^t)_{\alpha i}  \nn \\
&=& \sum^{G}_{a=1}\sum_{e_i\in E_x} \mathrm{sign}_x (e_i)[\ker(C)]_{l a}\;(d)_{a i}~, \label{QDu}
\eea
where we have used \eref{qtilde} in the last equality.
At this point, we follow the line of arguments in \cite{Franco:2006gc}. Every face of the tiling is crossed by the $x$ boundary of the fundamental domain over an even number of edges. Every edge which gets intersected by the $x$ boundary transforms either in the fundamental or in the antifundamental representation of the gauge group associated with the face $a$. Let us consider two edges, $e_i, e_j \in E_x$, of the face $a$. Then we have that $d_{ai}/d_{aj}=1$ or $-1$ if they are separated by an odd or even number of edges respectively. On the other hand, $\mathrm{sign}_x(e_i)/\mathrm{sign}_x(e_j)=1$ or $-1$ if the edges are separated by an even or odd number of edges. Hence, 
\bea
 \frac{\mathrm{sign}_x(e_i)\; d_{ai}}{\mathrm{sign}_x(e_j)\; d_{aj}} = -1~. 
\eea
Therefore, from \eref{QDu}, we find that
\bea
[Q_D \cdot (u_1~ u_2~ \ldots~ u_c)^t]_l  = \sum^{G}_{a=1} [\ker(C)]_{l a} \left(\sum_{e_i\in E_x} \mathrm{sign}_x(e_i)\;d_{a i} \right)=0~.
\eea
Similarly, we have
\bea
[Q_D \cdot (v_1~ v_2~ \ldots~ v_c)^t]_l  = \sum^{G}_{a=1} [\ker(C)]_{l a} \left(\sum_{e_i\in E_x} \mathrm{sign}_y(e_i)\;d_{a i} \right)=0~.
\eea
Thus, the first and second rows of $G_K$ are elements of the nullspace of $Q_D$.
Let us now consider the third row:
\bea
[Q_D \cdot (w_1~ w_2~ \ldots~ w_c)^t]_l   &=& \sum_{\alpha=1}^c \sum_{i} n_i  \left[ \ker(C) \cdot \widetilde{Q} \right]_{l\alpha} (P^t)_{\alpha i}  \nn \\
&=& \sum^{G}_{a=1}\sum_{i} n_i[ \ker(C)]_{l a}\;d_{a i} \nn \\
&=& \sum^{G}_{a=1} [ \ker(C)]_{l a} k_a  \nn \\
&=& 0~.
\eea
where we have used \eref{kn} in the third equality and the definition \eref{C} of $C$ in the last equality.
Therefore, the third row of $G_K$ is an element of the nullspace of $Q_D$.  Hence, we have proven the second equation of (\ref{kercond2}).  Thus, we have shown that the rows of $G_K$ are elements of the nullspace of $Q_t$.




 \newpage
%
%

\end{document}